\documentclass[a4paper,12pt,useAMS]{emulateapj}
\usepackage{apjfonts,graphicx,graphics}

\usepackage{natbib}
 
\newcommand{\simgt}{\lower.5ex\hbox{$\; \buildrel > \over \sim \;$}}
\newcommand{\simlt}{\lower.5ex\hbox{$\; \buildrel < \over \sim \;$}}

\def\btheta{\mbox{\boldmath $\theta$}}

\def\bx{\mbox{\boldmath $x$}}

\def\bu{\mbox{\boldmath $u$}}
\def\bd{\mbox{\boldmath $d$}}

\def\singlebond{\@makechembond\@ne}
\def\doublebond{\@makechembond\tw@}
\def\triplebond{\@makechembond\thr@@}

\lefthead{Umetsu et al.}

\righthead{Mass and Hot Baryons in Massive Galaxy Clusters}

\begin{document}

\title{
Mass and Hot Baryons in Massive Galaxy Clusters from 
Subaru Weak Lensing and 
AMiBA 
SZE Observations
\altaffilmark{1}}

\author{
Keiichi Umetsu\altaffilmark{2,3},
Mark Birkinshaw\altaffilmark{4},
Guo-Chin Liu\altaffilmark{2,5}, 
Jiun-Huei Proty Wu\altaffilmark{6,3},
Elinor Medezinski\altaffilmark{7},
Tom Broadhurst\altaffilmark{7},
Doron Lemze\altaffilmark{7},
Adi Zitrin\altaffilmark{7},
Paul T. P. Ho\altaffilmark{2,8},
Chih-Wei Locutus Huang\altaffilmark{6,3}, 
Patrick M. Koch\altaffilmark{2}, 
Yu-Wei Liao\altaffilmark{6,3}, 
Kai-Yang Lin\altaffilmark{2,6},
Sandor M. Molnar\altaffilmark{2}, 
Hiroaki Nishioka\altaffilmark{2},
Fu-Cheng Wang\altaffilmark{6,3},
Pablo Altamirano\altaffilmark{2},
Chia-Hao Chang\altaffilmark{2},
Shu-Hao Chang\altaffilmark{2},
Su-Wei Chang\altaffilmark{2},
Ming-Tang Chen\altaffilmark{2},
Chih-Chiang Han\altaffilmark{2},
Yau-De Huang\altaffilmark{2},
Yuh-Jing Hwang\altaffilmark{2},
Homin Jiang\altaffilmark{2},
Michael Kesteven\altaffilmark{9},
Derek Y. Kubo\altaffilmark{2},
Chao-Te Li\altaffilmark{2},
Pierre Martin-Cocher\altaffilmark{2},
Peter Oshiro\altaffilmark{2},
Philippe Raffin\altaffilmark{2},
Tashun Wei\altaffilmark{2},
Warwick Wilson\altaffilmark{9}
}

\altaffiltext{1}
 {Based in part on data collected at the Subaru Telescope,
  which is operated by the National Astronomical Society of Japan}
\altaffiltext{2}
 {Institute of Astronomy and Astrophysics, Academia Sinica,
  P.~O. Box 23-141, Taipei 10617, Taiwan}
\altaffiltext{3}
 {Leung center for Cosmology and Particle Astrophysics, National Taiwan University, Taipei 10617, Taiwan}
\altaffiltext{4}
  {Department of Physics, University of Bristol, Tyndall Avenue, Bristol BS8 1TL, UK.}
\altaffiltext{5}{Department of Physics, Tamkang University, 251-37 Tamsui, Taipei County,
 Taiwan} 
\altaffiltext{6}{Department of Physics, National Taiwan University,
 Taipei 10617, Taiwan} 
\altaffiltext{7}
  {School of Physics and Astronomy, Tel Aviv University, Tel Aviv 69978, Israel
   }
\altaffiltext{8}
 {Harvard-Smithsonian Center for Astrophysics, 60 Garden
Street, Cambridge, MA 02138, USA}
\altaffiltext{9}{Australia Telescope National Facility, P.O.Box 76, 
 Epping NSW 1710, Australia}


\begin{abstract}

We present a multiwavelength analysis of a sample of four hot ($T_X>8\,
 {\rm keV}$) X-ray galaxy clusters (A1689, A2261, A2142, and A2390)
 using joint AMiBA Sunyaev-Zel'dovich effect (SZE) and Subaru weak
 lensing observations, combined with published X-ray temperatures, to
 examine the distribution of mass and the intracluster medium (ICM) in
 massive cluster environments. 
Our observations show that A2261 is very similar to A1689 in terms of
lensing properties.
 Many tangential arcs are visible around A2261, with an effective
 Einstein radius $\sim 40\arcsec$ (at $z\sim 1.5$), which when combined 
 with our weak lensing measurements implies a mass profile well fitted 
 by an NFW model with a high concentration $c_{\rm vir}\sim 10$, 
 similar to A1689 and to other massive clusters. 
The cluster A2142 shows complex mass substructure, 
and displays a shallower profile $(c_{\rm vir}\sim 5)$, 
consistent with detailed X-ray observations which
imply recent interaction.  
The AMiBA map of A2142 exhibits 
an SZE feature
associated with mass
substructure lying ahead of 
the sharp north-west edge of the X-ray core
suggesting a pressure increase in the ICM. 
For A2390 we obtain highly elliptical mass and ICM distributions at all
 radii, consistent with other X-ray and strong lensing work.
Our cluster gas fraction measurements, free from the hydrostatic
 equilibrium assumption, are overall in good agreement with published
 X-ray and SZE observations, with the sample-averaged gas fraction of
 $\langle f_{\rm gas}(<r_{200})\rangle=0.133\pm 0.027$, for our sample
 with $\langle M_{\rm  vir}\rangle =(1.2\pm  0.1)\times 10^{15}M_\odot
 h^{-1}$.  
When compared to the cosmic baryon fraction $f_b=\Omega_{b}/\Omega_{m}$
constrained by the WMAP  5-year data,
this indicates $\langle f_{\rm gas,200}\rangle/f_b=0.78\pm 0.16$, 
i.e., $(22 \pm 16)$\% of the
baryons are missing from the hot phase of clusters.
\end{abstract}



\section{Introduction}\label{sec:intro}

Clusters of galaxies, the largest virialized systems known,
are key tracers of the matter distribution in the large scale structure
of the Universe. 
In the standard picture of cosmic structure formation, clusters are
mostly composed of dark matter (DM)
as indicated by a great deal of observational evidence, 
with the added assumptions that DM is non relativistic (cold) and
collisionless, referred to as CDM.
Strong evidence for substantial DM in clusters comes from
multiwavelength studies of interacting clusters 
\citep{2002ApJ...567L..27M},
in which weak gravitational lensing of background galaxies enables us
to directly map the distribution of gravitating matter in merging
clusters regardless of the physical/dynamical state of the system
\citep{2006ApJ...648L.109C,Okabe&Umetsu08}.
 The bulk of the baryons in clusters, 
   on the other hand, reside in the X-ray emitting intracluster medium
   (ICM), 
   where the X-ray surface brightness traces the
   gravitational mass dominated by DM. 
   The remaining baryons are in the form of luminous galaxies and faint 
   intracluster light 
   \citep{Fukugita+1998,Gonzalez+2005_ICL}.
   Since rich clusters represent high density peaks in the primordial
   fluctuation field, their baryonic mass fraction and its redshift
   dependence can in principle  be used to constrain the background
   cosmology
   \citep[e.g.,][]{Sasaki1996,2002MNRAS.334L..11A,2004MNRAS.353..457A,Allen+2008_fgas}.
   In particular, the gas mass to total mass ratio (the gas fraction) in
   clusters can be used to place a lower limit on the cluster 
   baryon fraction, which is expected to match the cosmic baryon
   fraction, $f_b\equiv \Omega_b/\Omega_m$.  
   However, non-gravitational processes associated with cluster
   formation, such as radiative gas cooling and AGN feedback,
   would break the self-similarities in cluster properties,
   which can cause the gas fraction to acquire some mass dependence 
   \citep{Bialek+2001,2005ApJ...625..588K}.

The deep gravitational potential wells of massive clusters
generate weak shape distortions of the images of background
sources due to differential deflection of light rays, resulting in a
systematic distortion pattern around the centers of 
massive clusters, known as weak gravitational lensing 
\citep[e.g.,][]{1999PThPS.133...53U,BS2001}.
In the past decade, weak lensing has become a powerful,
reliable measure to map the distribution of matter in clusters,
dominated by invisible DM,
without requiring any assumption about the physical/dynamical state of
the system 
\citep[e.g.,][]{2006ApJ...648L.109C,Okabe&Umetsu08}.
Recently, cluster weak
lensing has been used to examine the form of DM density profiles
\cite[e.g.,][]{BTU+05,BUM+2008,Mandelbaum+2008,UB2008},  
aiming for an observational test of the equilibrium density profile of
DM halos and the scaling relation between halo mass and concentration,
predicted by $N$-body simulations in the standard Lambda Cold Dark
Matter ($\Lambda$CDM) model
\citep{2007ApJS..170..377S,2008arXiv0803.0547K}.
Observational results show that the form of lensing profiles in relaxed
clusters is
consistent with a continuously steepening density profile with
increasing radius, well described by the general NFW model
\citep{1997ApJ...490..493N}, 
expected for collisionless CDM halos.

The Yuan-Tseh Lee Array for Microwave Background Anisotropy 
\citep{Ho_AMiBA}
is a platform-mounted interferometer array of up
to 19 elements 
operating at $3\,$mm wavelength, specifically designed to study the
structure of the  cosmic microwave background (CMB) radiation.
In the course of early AMiBA operations
we conducted Sunyaev-Zel'dovich effect (SZE)
observations at 94$\,$GHz towards six massive Abell clusters
with the 7-element compact array \citep{Wu_AMiBA}.
At 94$\,$GHz, the SZE signal is a temperature decrement in the CMB sky, and
is a measure of the thermal gas pressure
in the ICM integrated along the line 
of sight \citep{1999PhR...310...97B,1995ARA&A..33..541R}.  
Therefore it is rather insensitive to the cluster core as compared 
with the X-ray data, allowing us to trace the 
distribution of the ICM out to large radii.

This paper presents
a multiwavelength analysis of four nearby massive
clusters in the AMiBA sample, A1689, A2261, A2142, and A2390, for which
high-quality deep Subaru images are available for accurate weak lensing
measurements. 
This AMiBA lensing sample  represents a subset of the high-mass clusters
that can be selected by their high ($T_X > 8\,$keV) gas temperatures
\citep{Wu_AMiBA}.
Our joint weak lensing and SZE observations, combined 
with supporting X-ray information available in the published literature,
will allow us
to constrain the cluster gas fractions without the assumption of
hydrostatic equilibrium 
\citep{Myers+1997,2005astro.ph..6065U},
complementing X-ray based studies.
Our companion papers complement details of the instruments, system
performance and verification, 
observations and data analysis, and early science results from AMiBA.
\citet{Ho_AMiBA} describe 
the design concepts and specifications of the AMiBA telescope. 
Technical aspects of the instruments are described in
\citet{Chen_AMiBA} and \citet{Koch_mount}. Details of the first SZE
observations and data analysis are presented in \citet{Wu_AMiBA}.
\citet{Nishioka_AMiBA} assess the integrity of AMiBA data with several
statistical tests.
\citet{Lin_AMiBA} discuss the system performance and verification.
\citet{Liu_AMiBA} examine the levels of contamination from foreground
sources and the primary CMB radiation. 
\citet{Koch_AMiBA} present a measurement of the Hubble constant,
$H_0$,
from AMiBA SZE and X-ray data.
\citet{Huang_AMiBA} discuss cluster scaling relations between AMiBA SZE
and X-ray observations.

The paper is organized as follows.  
We briefly summarize in \S2 the basis of cluster SZE and weak lensing.
In \S3 we present a concise summary of the AMiBA target clusters
and observations.
In \S4 we describe our weak lensing analysis of Subaru imaging data, 
and derive lensing distortion and mass
profiles for individual clusters.
In \S5 
we examine and compare
cluster ellipticity and orientation profiles on
mass and ICM structure in the Subaru weak lensing and AMiBA SZE
observations. 
In \S6 we present our cluster models and method for measuring 
cluster gas fraction profiles from joint weak-lensing and SZE
observations, combined with published X-ray temperature measurements; 
we then derive cluster gas fraction profiles, and constrain
the sample-averaged gas fraction profile for our massive AMIBA-lensing
clusters.
Finally, a discussion and summary are given in \S7.

Throughout this paper, we adopt a
concordance $\Lambda$CDM cosmology with 
$\Omega_{m}=0.3$, $\Omega_{\Lambda}=0.7$, and
$h\equiv H_0/(100\, {\rm km\, s^{-1}\, Mpc^{-1}})=0.7$.
Cluster properties are determined at 
the virial radius $r_{\rm vir}$
and radii $(r_{200}, r_{500}, r_{2500})$, 
corresponding to overdensities $(200, 500, 2500)$ relative to the
critical density of the universe at the cluster redshift.

\section{Basis of Cluster Sunyaev-Zel'dovich Effect and Weak Lensing}
\label{sec:basis}

\subsection{Sunyaev-Zel'dovich Effect}
\label{subsec:sze}

We begin with a brief summary of the basic equations of the thermal SZE.
Our notation here closely follows the standard notation of
\citet{1995ARA&A..33..541R}. 

The SZE is a spectral distortion of the CMB radiation
resulting from the inverse Compton scattering of cool CMB photons 
by the hot ICM. 
The non-relativistic form
of the spectral change was obtained by Sunyaev-Zel'dovich (1972)
from the Kompaneets equation in the non-relativistic limit.
The change in the CMB intensity $I_{\rm CMB}$ due to the SZE
is written
in terms of its spectral function $g$ and of the integral of the electron
pressure along the line-of-sight as 
\citep{1995ARA&A..33..541R,1999PhR...310...97B,Carlstrom+2002_SZE}:
\begin{equation}
\label{eq:SZE}
\Delta I_{\rm SZE}(\nu) = I_{\rm norm}\, g[x(\nu)] y(\btheta),
\end{equation}
where 
$x(\nu)$ is the dimensionless frequency, $x\equiv h\nu/(k_{\rm
B}T_{\rm CMB})\approx 1.66 (\nu/94\,{\rm GHz})$, with 
$k_{\rm B}$ being the Boltzmann constant and
$T_{\rm CMB}=2.725\,{\rm K}$ being the CMB
temperature at the present-day epoch, 
$I_{\rm norm}=(2h/c^3)\,(k_{\rm B}T_{\rm CMB}/h)^2\simeq 2.7\times
10^8\,{\rm Jy}\,{\rm sr}^{-1}$, 
and $y(\btheta)$ is the
Comptonization parameter defined as
\begin{equation}
\label{eq:y}
y=\int_{-r_{\rm max}}^{+r_{\rm max}}\!dl\,\sigma_{th}
 n_e \left(\frac{k_B T_e}{m_e c^2}\right)  =  
\frac{\sigma_{th}}{m_e c^2}\int_{-r_{\rm max}}^{+r_{\rm max}}\!dl\,
\frac{\rho_{\rm gas}}{\mu_e m_p} k_BT_{\rm  gas},
\end{equation}
where $\sigma_{th}$, $m_e$, $c$, and $\mu_e$ 
are the Thomson cross section, the
electron mass, the speed of light, and the mass per electron in units of
proton mass $m_p$, respectively; 
for a fully ionized H-He
plasma, $\mu_e=2/(1+X)\simeq 1.14$, with $X$ being the Hydrogen
primordial abundance by mass fraction, $X\simeq 0.76$;
$r_{\rm max}$ is the cutoff
radius for an isolated cluster (see \S\ref{subsec:szedata}).
The SZE spectral function $g(x)$ is expressed as
\begin{equation}
\label{eq:sze_gfunc}
g(x) = 
g_{\rm NR}(x)
\left[
1+\delta_{\rm SZE}(x,T_{\rm gas})
\right],
\end{equation}
where $g_{\rm NR}(x)$ is the thermal spectral function in the
non-relativistic limit (Sunyaev \& Zel'dovich
1972),
\begin{equation}
g_{\rm NR}(x) = \frac{x^4 e^x}{(e^x-1)^2}\left(
x\frac{e^x+1}{e^x-1}-4
\right),
\end{equation}
which is zero at the cross-over frequency $x_0\simeq 3.83$, or 
$\nu_0=217\,$GHz,
and $\delta_{\rm SZE}(x,T_{\rm gas})$ is the relativistic correction 
\citep{Challinor+Lasenby1998_SZE,Itoh+1998_SZE}.
The fractional intensity decrease due to the SZE 
with respect to the primary CMB 
is maximized at $\nu\sim 100\,$GHz
\citep[see Figure 1 of][]{2002ApJ...577..555Z}, 
which is well matched to the observing frequency range $86$--$102\,$GHz
of AMiBA.
At the central frequency $\nu_c=94\,$GHz of AMIBA, $g(x)\simeq -3.4$.
For our hot X-ray clusters with $T_X=8$--$10\,{\rm keV}$,
the relativistic correction to the thermal SZE 
is 6--7$\%$ at $\nu_c=94\,{\rm GHz}$.


\subsection{Cluster Weak Lensing}
\label{subsec:wl}

Weak gravitational lensing is responsible for the weak shape-distortion
and magnification of the images of background sources due to the
gravitational field of intervening foreground clusters of galaxies
and large scale structures in the universe.
The deformation of the image can be described by the $2\times 2$
Jacobian matrix $\cal{A}_{\alpha\beta}$
($\alpha,\beta=1,2$) of the lens mapping.
The Jacobian ${\cal A}_{\alpha\beta}$ is real and symmetric, so that
it can be decomposed as
\begin{eqnarray}
\label{eq:jacob}
{\cal A}_{\alpha\beta} &=& (1-\kappa)\delta_{\alpha\beta}
 -\Gamma_{\alpha\beta},\\
\Gamma_{\alpha\beta}&=&
\left( 
\begin{array}{cc} 
+{\gamma}_1   & {\gamma}_2 \\
 {\gamma}_2  & -{\gamma}_1 
\end{array} 
\right),
\end{eqnarray}
where 
$\delta_{\alpha\beta}$ is Kronecker's delta,
$\Gamma_{\alpha\beta}$ is the trace-free, symmetric shear matrix
with $\gamma_{\alpha}$ being the components of 
spin-2
complex gravitational
shear $\gamma:=\gamma_1+i\gamma_2$,
describing the anisotropic shape distortion,
and $\kappa$ is the 
lensing convergence responsible for the 
trace-part of the Jacobian matrix, describing the isotropic area
distortion.
In the weak lensing limit where $\kappa,|\gamma|\ll 1$, 
$\Gamma_{\alpha\beta}$ induces a quadrupole anisotropy of the 
background image, which can be observed from ellipticities 
of background galaxy images.
The flux  magnification due to gravitational lensing
is given by the inverse Jacobian
determinant,
\begin{equation}
\label{eq:mu}
\mu = 
\frac{1}{{\rm det}{\cal A}}
=
\frac{1}{(1-\kappa)^2-|\gamma|^2},
\end{equation}
where we assume subcritical lensing, i.e., 
${\rm det}{\cal A}(\btheta)>0$.

The lensing convergence is expressed as a line-of-sight projection
of the matter density contrast $\delta_m=(\rho_m-\bar{\rho})/\bar{\rho}$ 
out to the source plane ($s$)
weighted by certain combination of co-moving angular diameter
distances $r$
\citep[e.g.,][]{2000ApJ...530..547J},
\begin{eqnarray}
\label{eq:kappa}
&&\kappa =
\frac{3H_0^2\Omega_m}{2c^2}
\int_0^{\chi_s}\!d\chi\, 
{\cal G}(\chi,\chi_s)
\frac{\delta_m}{a} 
\equiv \int\!d\Sigma_m\,\Sigma_{\rm crit}^{-1},\\
\label{eq:gdistance}
&&{\cal G}(\chi,\chi_s)=
\frac{r(\chi)r(\chi_s-\chi)}{r(\chi_s)},
\end{eqnarray}
where $a$ is the cosmic scale factor,
$\chi$ is the co-moving distance,
$\Sigma_m$ is the surface mass density of matter, $\Sigma_m
=\int\!d\chi \, a(\rho_m-\bar{\rho})$,
with respect to the cosmic mean density $\bar{\rho}$, and
$\Sigma_{\rm crit}$ 
is the critical surface mass density for gravitational lensing,
\begin{equation} 
\label{eq:sigmacrit}
\Sigma_{\rm crit} = \frac{c^2}{4\pi G}\frac{D_{s}}{D_d D_{ds}}
\end{equation}
with $D_s$, $D_d$, and $D_{ds}$ being the (proper) angular diameter distances
from the observer to the source, from the observer to the deflecting
lens, and from the lens to the source, respectively.
For a fixed background cosmology and a lens redshift $z_d$,
$\Sigma_{\rm crit}$ is a function of background source redshift
$z_s$.
For a given mass distribution $\Sigma_m(\btheta)$, the lensing signal is
proportional to the angular diameter distance ratio,
$D_{ds}/D_s$.

In the present weak lensing study we aim to reconstruct
the dimensionless surface mass density $\kappa$ 
from weak lensing distortion and magnification data.
To do this, we utilize the relation between the gradients of 
$\kappa$ and $\gamma$
\citep{1995ApJ...439L...1K,2002ApJ...568...20C}, 
\begin{equation}
\label{eq:local}
\triangle \kappa (\btheta)
= \partial^{\alpha}\partial^{\beta}\Gamma_{\alpha\beta}(\btheta)
= 2\hat{\cal D}^*\gamma(\btheta)
\end{equation}
where 
$\hat{\cal D}$ is the complex differential operator
$\hat{\cal D}=(\partial_1^2-\partial_2^2)/2+i\partial_1\partial_2$.
The Green's function for the two-dimensional Poisson equation is
$\triangle^{-1}(\btheta,\btheta')=\ln|\btheta-\btheta'|/(2\pi)$,
so that equation (\ref{eq:local}) can be solved to yield the following
non-local relation between $\kappa$ and $\gamma$ (Kaiser \& Squires 1993):
\begin{equation}
\label{eq:gamma2kappa}
\kappa(\btheta) = 
\frac{1}{\pi}\int\!d^2\theta'\,D^*(\btheta-\btheta')\gamma(\btheta')
\end{equation}
where $D(\btheta)$ is the complex kernel defined as 
\begin{equation}
\label{eq:kerneld}
D(\btheta) = \frac{\theta_2^2-\theta_1^2-2i\theta_1\theta_2}{|\theta|^4}.
\end{equation}
Similarly, the spin-2 shear field can be expressed in terms of the
lensing convergence as
\begin{equation} 
\label{eq:kappa2gamma}
\gamma(\btheta) = 
\frac{1}{\pi}\int\!d^2\theta'\,D(\btheta-\btheta')\kappa(\btheta').
\end{equation} 
Note that adding a constant mass sheet to $\kappa$ in equation
(\ref{eq:kappa2gamma}) does not change the 
shear field
$\gamma(\btheta)$ which is observable in the weak lensing limit, 
leading to the so-called {\it mass-sheet degeneracy}
 (see eq.~[\ref{eq:invtrans}])
based solely on shape-distortion measurements
\citep[e.g.,][]{BS2001,1999PThPS.133...53U}.
In general, the observable quantity is not the 
gravitational shear $\gamma$ but the {\it reduced} shear,
\begin{equation}
\label{eq:redshear}
g=\frac{\gamma}{1-\kappa}
\end{equation}
in the subcritical regime where ${\rm det}{\cal A}>0$
(or $1/g^*$ in the negative parity region with ${\rm det}{\cal A}<0$). 
We see that the reduced shear $g$ is invariant under the following
global transformation:
\begin{equation}
\label{eq:invtrans}
\kappa(\btheta) \to \lambda \kappa(\btheta) + 1-\lambda, \ \ \ 
\gamma(\btheta) \to \lambda \gamma(\btheta)
\end{equation}
with an arbitrary scalar constant $\lambda\ne 0$ 
\citep{1995A&A...294..411S}.

\section{AMiBA Sunyaev-Zel'dovich Effect Observations}
\label{sec:amiba}
 


\subsection{AMIBA Telescope} 
 
The AMiBA is a dual channel 86--102$\,$GHz (3-mm wavelength)
interferometer array 
of up to 19-elements with dual polarization capabilities sited at
3396$\,$m on Mauna-Loa, Hawaii
(latitude: $+19.5^\circ$, longitude: $-155.6^\circ$)
\footnote{http://amiba.asiaa.sinica.edu.tw/}.
AMiBA is equipped with 4-lag analog, broadband (16$\,$GHz bandwidth
centered at 94$\,$GHz) correlators which output a set of 4 real-number
correlation signals \citep{Chen_AMiBA}.
These four degrees of freedom (dof) correspond to two complex visibilities
in two frequency channels.
The frequency of AMiBA operation was chosen to take advantage of 
the optimal frequency window at
3$\,$mm, where the fractional decrement in the SZE intensity relative to
the primary CMB is close to its maximum (see
\S \ref{subsec:sze}) and
contamination by the Galactic synchrotron emission, dust foregrounds,
and the population of cluster/background radio sources is minimized
\citep[see for detailed contamination analysis,][]{Liu_AMiBA}.
This makes AMiBA a unique CMB/SZE interferometer, 
and also complements the wavelength coverage of other
existing and planned CMB instruments:
interferometers such as 
AMI at 15$\,$GHz
\citep{AMI2001},
CBI at 30$\,$GHz 
\citep{Padin2001_CBI_1st,Padin2002_CBI,Mason2003_CBI,Pearson2003_CBI},
SZA at 30 and 90$\,$GHz
\citep{SZA2008},
and VSA\footnote{http://astro.uchicago.edu/sza/} at 34$\,$GHz 
\citep{Watson2003_VSA};
bolometer arrays such as
ACT,\footnote{http://www.hep.upenn.edu/act/act.html}  
APEX-SZ\footnote{http://bolo.berkeley.edu/apexsz}
\citep{Halverson2008_APEX},
and SPT.\footnote{http://pole.uchicago.edu}

In the initial operation of AMiBA, we
used seven 0.6$\,$m (0.58$\,$m to be precise) 
Cassegrain antennas 
\citep{Koch2006_0.6m}
co-mounted on a  
6$\,$m hexapod platform in a hexagonal close-packed configuration
\citep[see][]{Ho_AMiBA}.
At each of the frequency channels centered at about 90 and 98$\,$GHz,
this compact configuration provides 21 simultaneous baselines 
with three baseline lengths of $d=0.61$, 1.05, and 1.21$\,$m,
corresponding to angular multipoles $l=2\pi \sqrt{u^2+v^2} 
(=2\pi d/\lambda)$ of
$l\approx 1194,2073,2394$ at $\nu_c = 94\, {\rm GHz}$. 
This compact 7-element array is sensitive to multipole range
$800 \simlt l \simlt 2600$. 
With 0.6-m antennas, the instantaneous field-of-view of
AMiBA is about $23'$ FWHM \citep{Wu_AMiBA},
and its angular resolution
ranges from $2'$ to $6'$ depending on
the configuration and weighting scheme. 
In the compact configuration,
the angular resolution of AMiBA is about $6'$ FWHM using {\it natural
weighting} (i.e., inverse noise variance weighting).
The point source sensitivity is estimated to be $\sim 63\,$mJy 
\citep{Lin_AMiBA}
in 1$\,$hour of on-source integration in 2-patch main-trail/lead
differencing observations, where the overall noise
level is increased by a factor of $\sqrt{2}$ due to the differencing.

\subsection{Initial Target Clusters}
\label{subsec:targets}

The AMiBA lensing sample, A1689, A2142, A2261, A2390,
is a subset of the AMiBA cluster sample \citep[see][]{Wu_AMiBA},
composed of four massive
clusters at relatively low redshifts of $0.09 \simlt z \simlt 0.23$
with the median redshift of $\bar{z}\approx 0.2$.
The sample size is simply limited by the availability of high quality
Subaru weak lensing data.
A1689 is a relaxed, round system, and is
one of the best studied clusters for lensing work 
\cite[e.g.,][]{BTU+05,2007ApJ...668..643L,UB2008,BUM+2008}.
A2261 is a compact cluster with a regular X-ray morphology.
A2142 is a merging cluster with two sharp X-ray surface brightness
discontinuities in the cluster core 
\citep{2000ApJ...541..542M,Okabe&Umetsu08}.
A2390 shows an elongated morphology both in
the X-ray emission and strong-lensing mass distributions
\citep{2001MNRAS.324..877A,1998ApJ...499L.115F}.
Table \ref{tab:amiba_target} summarizes the physical properties of the four
target clusters in this multiwavelength study.

In 2007, AMiBA with the seven small antennas (henceforth AMiBA7) was
in the science verification phase. 
For our initial observations, we therefore
selected those target clusters observable from Mauna~Loa
during the observing period 
that were known to have strong SZEs at relatively low redshifts ($0.1\simlt z
\simlt 0.3$) 
from previous experiments, such as 
OVRO observations at 30$\,$GHz \citep{Mason2001_Hubble},
BIMA/OVRO observations at 30$\,$GHz
\citep{Grego2001_SZE,Reese2002_SZE},
VSA observations at 34$\,$GHz
\citep{Lancaster+05_VSA},
and SuZIE II observations at 145, 221, and 355$\,$GHz
\citep{Benson2004_SuZIE}.
The targeted redshift range allows the target clusters to be resolved
by the
$6'$ resolution of AMiBA7,
allowing us to derive useful measurements of cluster SZE
profiles for our multiwavelength studies.
At redshifts of $z\simlt 0.3$ (0.2),
the angular resolution of AMiBA7 corresponds to
$\simlt  560\,{\rm kpc} h^{-1}$ ($\sim 400\,{\rm kpc} h^{-1}$) in radius,
which is $\simlt 30$--$40\%$ ($\sim 20$--$30\%$)
of the virial radius 1.5--2$\,{\rm Mpc}h^{-1}$ of massive
clusters. 
The requirement of being SZE strong is to ensure reliable SZE
measurements at 3$\,$mm with AMiBA7.
We note that AMiBA and SZA are the only SZE instruments measuring
at 3$\,$mm, but complimentary in their baseline coverage.
With sensitivities of 20--30$\,$mJy$\,$beam$^{-1}$
typically achieved in 2-patch differencing observations
in 5--10 hours of net on-source integration
\citep{Wu_AMiBA}, we would expect $\simgt 5\sigma$ detections of
SZE fluxes $\simgt 100$--$150\,$mJy at 3~mm.
Finally, our observing period (April--August 2007)
limited the range of right ascension (RA) of targets,\footnote{The
elevation limit of AMiBA is $30^\circ$.}
since we restricted our science
observations to nights
(roughly 8pm to 8am), where we would expect high gain stability
because the ambient temperature varies slowly and little
\citep{Nishioka_AMiBA}.
The SZE strong clusters in our AMiBA sample are likely to have
exceedingly deep potential wells, and indeed our AMiBA sample represents
a class of hot X-ray clusters with observed X-ray temperatures exceeding
$8\,{\rm keV}$ (see Table \ref{tab:amiba_target}).
We note that this may affect the generality of the results presented in
this study.  
A main-trail/lead differencing scheme has been used in our targeted
cluster observations where the trail/lead (blank) field is
subtracted from the main (cluster) field. 
This differencing scheme sufficiently removes 
contamination from ground spillover and electronic DC offset in the
correlator output \citep{Wu_AMiBA}. 
A full description of AMiBA observations and 
analysis of the initial six target clusters, including the
observation strategy,
analysis methodology,
calibrations, 
and map-making,
can be found in \citet{Wu_AMiBA,Wu_AMiBA_MPLA}.


\section{Subaru Weak Lensing Data and Analysis}
\label{sec:subaru}

In this section we present a technical description of our weak
lensing distortion analysis of the AMiBA lensing sample 
based on Subaru data.
The present work on A1689 is based on the same Subaru images
as analyzed in our earlier work of 
\citet{BTU+05} and \citet{UB2008}, but our improved color selection of
the red background has increased the sample size by $\sim 16\%$
(\S\ref{subsec:color}).
This work on A2142 is based on the same Subaru images as
in \citet{Okabe&Umetsu08}, but our inclusion of blue, as well as red,
galaxies has increased the sample size by a factor of $4$ (cf.  Table 6
of Okabe \& Umetsu 2008), leading to a significant improvement of our
lensing measurements.
For A2261 and A2390 we present our new
weak lensing analysis based on Suprime-Cam imaging data
retrieved from the Subaru archive, SMOKA.
The reader only interested in the main result may skip directly to
\S\ref{subsec:2dmap}.

\subsection{Subaru Data and Photometry}
\label{subsec:data}

We analyze deep images of four high mass clusters in the AMiBA sample
taken by the wide-field camera Suprime-Cam 
\citep[$34^\prime\times 27^\prime$;][]{2002PASJ...54..833M}
at the prime-focus of the 8.3m Subaru telescope. 
The clusters were observed deeply in two optical passbands each
with seeing in the co-added images
ranging from $0.55\arcsec$ to $0.88\arcsec$ (see Table \ref{tab:lensdata}). 
For each cluster we select an optimal combination of two filters that
allows for an efficient separation of cluster/background galaxies based
on color-magnitude correlations (see Table \ref{tab:lensdata}).
We use either $R_{\rm c}$ or $i'$ band for our weak lensing
measurements (described in \S\ref{subsec:shear}) for which the instrumental
response, sky background and seeing conspire to provide the best-quality
images. 
 The standard pipeline reduction software for Suprime-Cam 
\citep{2002AJ....123...66Y}
is used for flat-fielding, instrumental distortion
correction, differential refraction, 
sky subtraction and
stacking. Photometric catalogs are constructed from stacked and
matched images using SExtractor 
\citep{1996A&AS..117..393B}, and used for our color selection of
background galaxies (see \S\ref{subsec:color}).

\subsection{Weak Lensing Distortion Analysis}
\label{subsec:shear}

We use the IMCAT package developed by N. Kaiser\footnote{http://www.ifa.hawaii.edu/\~kaiser/imcat}
to perform object
detection, photometry and shape measurements, following the formalism
outlined in  \citet[][KSB]{1995ApJ...449..460K}.
Our analysis pipeline is implemented based on the procedures described
in 
\citet{2001A&A...366..717E}
and on verification tests with STEP1 data of mock
ground-based observations \citep{2006MNRAS.368.1323H}.
The same analysis pipeline has been used in
\citet{UB2008}, \citet{Okabe&Umetsu08}, and \citet{BUM+2008}.

\subsubsection{Object Detection}
\label{subsubsec:detection}

Objects are first detected as local peaks in the image by 
using the IMCAT hierarchical peak-finding
algorithm {\it hfindpeaks}
which for each object yields object parameters such as a peak position
($\bx$), 
an estimate of the object size ($r_g$), the significance of
the peak detection ($\nu$).
The local sky level and its gradient are 
measured around each  object from the mode of pixel values 
on a circular annulus defined by
inner and outer radii of 
$16\times r_g$ and $32\times r_g$.
In order to avoid contamination in the background estimation
by bright neighboring stars and/or
foreground galaxies,
all pixels within $3\times r_g$ of another object are excluded from the
mode calculation. Total fluxes and half-light radii ($r_h$) are then
 measured on sky-subtracted images using a circular aperture of
 radius $3\times r_g$ from the object center. Any pixels within
 $2.5\times r_g$ of another object are excluded from the aperture.
The aperture magnitude is then calculated from the measured total
flux and a zero-point magnitude.
Any objects with positional  differences between the
peak location and the weighted-centroid 
greater than $|\bd|=0.4$ pixels are excluded from
the catalog.

Finally, bad objects such as spikes, saturated stars, and noisy
detections need to be removed from the weak lensing analysis.
We removed from our detection catalog 
extremely small or large objects with $r_g<1$ or $r_g>10$ pixels, 
objects with low detection significance, $\nu<7$
\citep[see][]{2001A&A...366..717E}, 
objects with large raw ellipticities, $|e|>0.5$ (see \S\ref{subsubsec:shape}), 
noisy detections with unphysical negative fluxes, and
objects containing more than $10$ bad pixels, ${\it nbad}>10$.

\subsubsection{Weak Lensing Distortion Measurements}
\label{subsubsec:shape}

To obtain an estimate of the reduced shear,
$g_{\alpha}=\gamma_{\alpha}/(1-\kappa)$ ($\alpha=1,2$), 
we measure using the {\it getshapes} routine in IMCAT
the image ellipticity 
$e_{\alpha} = \left\{Q_{11}-Q_{22}, Q_{12} \right\}/(Q_{11}+Q_{22})$ 
from the weighted quadrupole moments of the surface brightness of
individual galaxies defined in the above catalog,
\begin{equation}
\label{eq:Qij}
Q_{\alpha\beta} = \int\!d^2\theta\,
 W({\theta})\theta_{\alpha}\theta_{\beta} 
I({\btheta})
\ \ \ (\alpha,\beta=1,2)
\end{equation} 
where $I(\btheta)$ is the surface brightness distribution of an object,
$W(\theta)$ is a Gaussian window function matched to the size of the
object, and the object center is chosen as the coordinate origin.
In equation (\ref{eq:Qij}) the maximum radius of integration is
chosen to be $\theta_{\rm max}=4r_g$.

\begin{figure}[!htb]
 \begin{center}
    \includegraphics[width=180mm,angle=0]{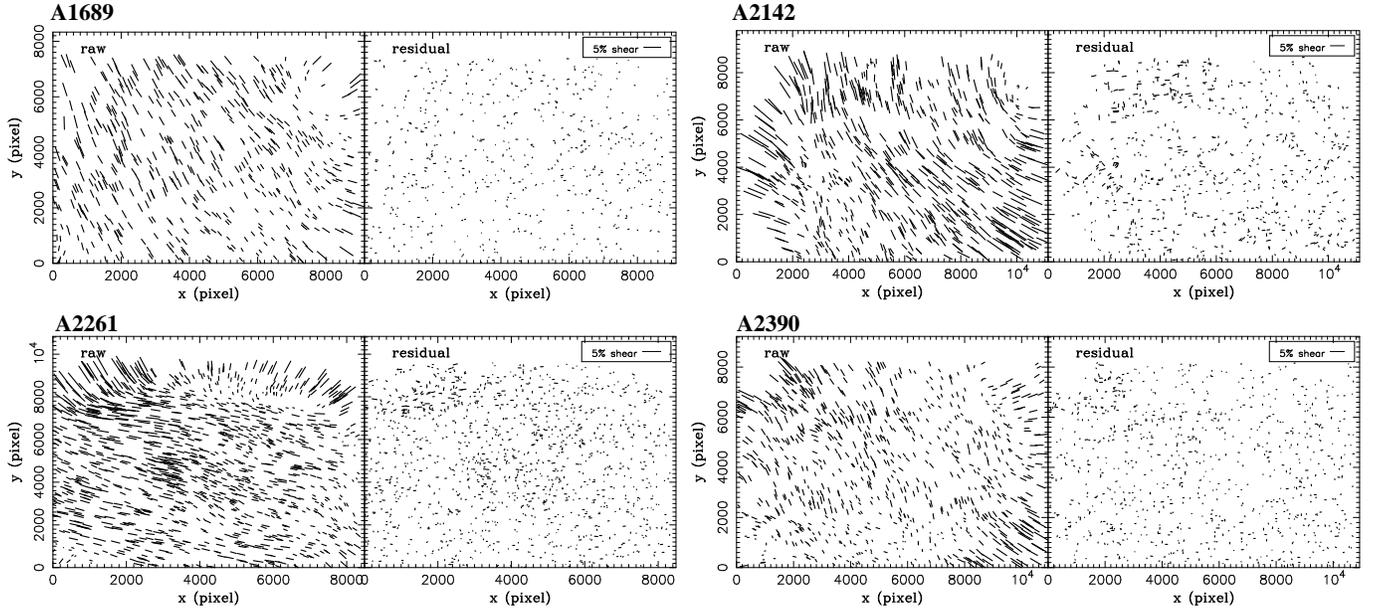}
 \end{center}
\caption{
\label{fig:anisopsf1}
The quadrupole PSF anisotropy field for individual clusters
as measured from 
stellar ellipticities before and after the PSF anisotropy correction.
For each cluster field,
the left panel shows the raw ellipticity field of stellar objects,
and the right panel shows the residual ellipticity field after
the PSF anisotropy correction.
The orientation of the sticks indicates the position angle of
the major axis of stellar ellipticity, whereas the length is
 proportional to the modulus of stellar ellipticity. A stick with the
 length of $5\%$ ellipticity is indicated in the top right of the right
 panel. 
} 
\end{figure}

\begin{figure}[!htb]
 \begin{center}
    \includegraphics[width=160mm,angle=0]{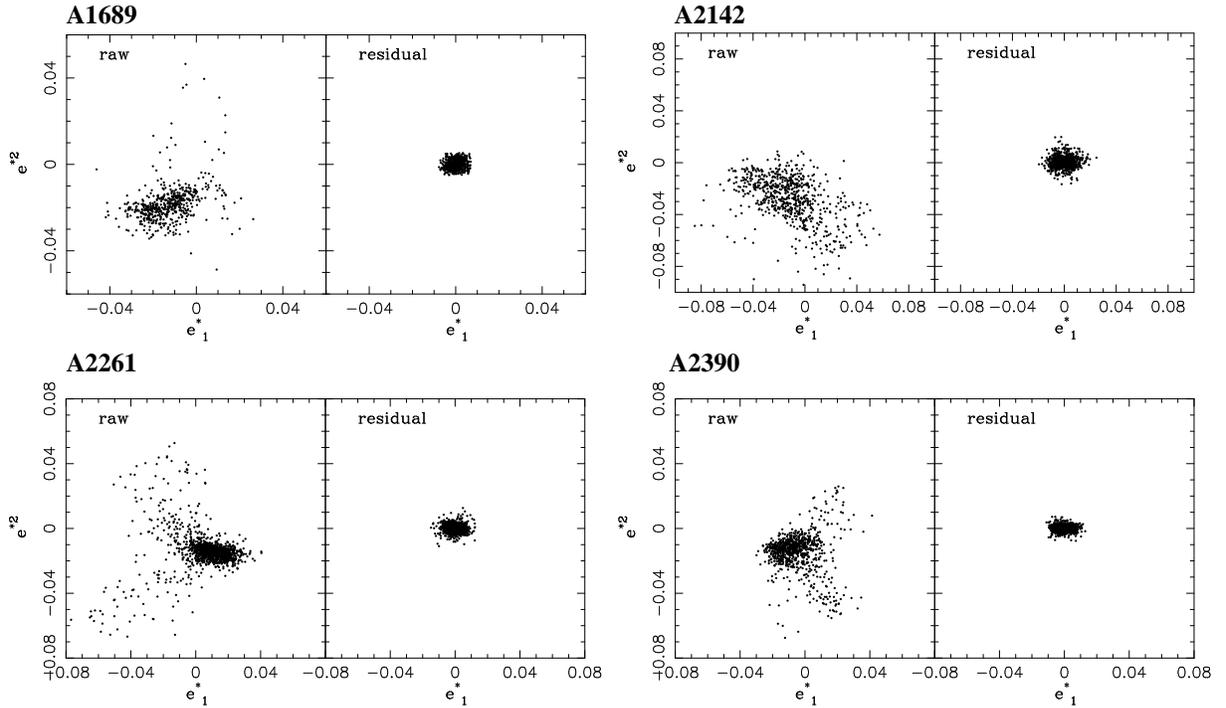}
 \end{center}
\caption{
\label{fig:anisopsf2}
Stellar ellipticity distributions before and after the PSF anisotropy 
correction for individual clusters. 
For each cluster field,
the left panel shows the raw ellipticity components 
$(e_1^*,e_2^*)$ of stellar objects, and the right panel shows
the residual ellipticity components $(\delta e_1^*, \delta e_2^*)$
after the PSF anisotropy correction.
} 
\end{figure}

Firstly the PSF anisotropy needs to be corrected using the star images
as references:
\begin{equation}
e'_{\alpha} = e_{\alpha} - P_{sm}^{\alpha \beta} q^*_{\beta} 
\label{eq:qstar}
\end{equation}
where $P_{sm}$ is the {\it smear polarizability} tensor  
(which is close to diagonal), and
$q^*_{\alpha} = (P_{sm}^*)^{-1}_{\alpha \beta}e_*^{\beta}$ 
is the stellar anisotropy kernel.
We select bright, unsaturated foreground stars 
identified in a branch
of the half-light radius vs. magnitude diagram
to measure $q^*_{\alpha}$.
In order to obtain a smooth map of $q^*_{\alpha}$ which is used in
equation (\ref{eq:qstar}), we divided the co-added mosaic image 
(of $\sim 10{\rm K} \times 8{\rm K}$ pixels)
into rectangular blocks.
The block length is based on the coherent
scale of PSF anisotropy patterns, and 
is typically $2{\rm K}\,$pixels.
In this way the PSF anisotropy in
individual blocks can be well described by fairly low-order
polynomials.
We then fitted the $q^*$ in each block independently with
second-order bi-polynomials, $q_*^{\alpha}(\btheta)$, in
conjunction with iterative outlier rejection on each
component of the residual:
$\delta e^*_\alpha =
e^*_{\alpha}-(P_{sm}^*)^{\alpha\beta}q^*_{\beta}(\btheta)$.   
The final stellar sample contains typically 500--1200 stars.
Uncorrected ellipticity components of stellar objects have on average 
a mean offset (from a value of zero) of 1--2$\%$ with a few $\%$ of rms, 
or variation of PSF
across the 
data field \cite[see, e.g.,][]{UB2008,Okabe&Umetsu08}.
On the other hand,  the mean residual stellar ellipticity
$\overline{\delta e^*_\alpha}$
after correction is less than or about $10^{-4}$, 
with the standard error on this
measurement, a few $\times 10^{-4}$.
We show in Figure \ref{fig:anisopsf1}
the quadrupole PSF anisotropy fields
as measured from stellar ellipticities before and after the anisotropic
PSF correction for our target clusters.
Figure \ref{fig:anisopsf2} shows
the distributions of stellar ellipticity
components before and after the PSF anisotropy correction.
In addition, we adopt a conservative magnitude 
limit $m < 25.5$--$26.0$ ABmag, depending on the depth of the data for
each cluster, to avoid systematic errors in the shape measurement
\cite[see][]{UB2008}.
From the rest of the object
catalog, we select objects with 
$r_h > \overline{r_h^*} + \sigma(r_h^*)$ pixels
as a magnitude-selected weak lensing galaxy sample, 
where $\overline{r_h^*}$
is the median value of stellar half-light radii $r_h^*$,
corresponding to half the median width of the circularized PSF
over the data field, and $\sigma(r_h^*)$ is the rms dispersion of
$r_h^*$.

Second, we need to correct image ellipticities for
the isotropic smearing effect 
caused by atmospheric seeing and the window function used for the shape
measurements. The pre-seeing reduced shear $g_\alpha$ can be
estimated from 
\begin{equation}
\label{eq:raw_g}
g_{\alpha} =(P_g^{-1})_{\alpha\beta} e'_{\beta}
\end{equation}
with the {\it pre-seeing shear polarizability} tensor
$P^g_{\alpha\beta}$ defined as
\cite{1998ApJ...504..636H},
\begin{equation}
\label{eq:Pgamma}
P^g_{\alpha\beta} = P^{sh}_{\alpha\beta}- 
\left[
P^{sm} (P^{sm*})^{-1} P^{sh*}
\right]_{\alpha\beta}
\approx 
 P^{sh}_{\alpha\beta}-  P^{sm}_{\alpha\beta} 
\frac{{\rm tr}[P^{sh*}]}{{\rm tr}[P^{sm*}]}
\end{equation}
with $P^{sh}$ being the {\it shear polarizability} tensor;
In the second equality we have used a trace approximation to the stellar
shape tensors, $P^{sh*}$ and $P^{sm*}$.
To apply equation (\ref{eq:raw_g}) the quantity ${\rm tr}[P^{sh*}]/{\rm
tr}[P^{sm*}]$ must be known for each of the galaxies with different
sizescales. 
Following
\citet{1998ApJ...504..636H}, we recompute 
the stellar shapes $P^{sh*}$ and $P^{sm*}$ in a range of filter scales
$r_g$ spanning that of the galaxy sizes ($r_g=[1,10]\,$pixels).
At each filter scale $r_g$, 
the median 
 $\langle {\rm tr}[P^{sh*}]/{\rm tr}[P^{sm*}]\rangle$ 
over the stellar sample is calculated, and used in equation 
(\ref{eq:Pgamma}) as an estimate of  ${\rm tr}[P^{sh*}]/{\rm
tr}[P^{sm*}]$.
Further, we adopt a scalar correction scheme, namely
\begin{equation}
(P_g)_{\alpha\beta}
=\frac{1}{2}{\rm tr}[P_g]\delta_{\alpha\beta}\equiv
P^{\rm s}_{g}\delta_{\alpha\beta}
\end{equation}
\citep{2001A&A...366..717E,1998ApJ...504..636H,UB2008}.
In order to suppress artificial effects due to the noisy
$P_g^{\rm s}$ estimated for individual galaxies, we apply filtering 
to raw $P_g^{\rm s}$ measurements. 
We compute for each object 
a median value of $P_g^{\rm s}$ among $N$-neighbors in the size 
and magnitude plane to define the object parameter space: 
firstly,
for each object,
$N$-neighbors with raw $P_g^{\rm s}>0$
are identified in the size ($r_g$) and magnitude 
plane; 
the median value of $P_g^{\rm s}$ is then used as the smoothed
$P_g^{\rm s}$ for the object, $\langle P_g^{\rm s}\rangle$,
and the variance $\sigma^2_{g}$ 
of $g=g_1+ig_2$ is calculated using equation (\ref{eq:raw_g}).
The dispersion $\sigma_g$ is used as an rms error of the shear estimate
for individual galaxies.
We take $N=30$.
Finally, we use the estimator 
$g_{\alpha} = e'_{\alpha}/\left< P_g^{\rm s}\right>$
for the reduced shear.

\subsection{Background Selection}
\label{subsec:color}


It is crucial in the weak lensing analysis
to make a secure selection of background galaxies in order
to minimize contamination by {\it unlensed}
cluster/foreground galaxies and hence to make an accurate determination 
of the cluster mass profile;
otherwise dilution of the distortion signal arises from
the inclusion of unlensed galaxies, particularly at small radius
where the cluster is relatively dense
\citep{BTU+05,Medezinski+07}.
This dilution effect is 
simply to reduce the strength of
the lensing signal when averaged over a local ensemble of 
galaxies, 
in proportion to the fraction of unlensed galaxies
whose orientations are randomly distributed, thus diluting the lensing
signal relative to the reference background level derived from
the background population 
\cite{Medezinski+07}.

To separate cluster members from the background
and hence minimize the weak lensing dilution,
 we follow an objective
background selection method developed by \citet{Medezinski+07}
and \citet{UB2008}.
We select red galaxies with colors redder than the color-magnitude
sequence of cluster E/S0 galaxies. The sequence forms a well defined
line in object color-magnitude space
due to the richness and relatively low redshifts of our
clusters. These red galaxies are expected to lie in the background by
virtue of $k$-corrections which are greater than for the red cluster
sequence galaxies;
This has been convincingly demonstrated spectroscopically by
\citet{Rines+Geller2008}.
We also include blue galaxies falling far from
the cluster sequence to minimize cluster contamination.

Figure \ref{fig:dilution} shows for each cluster 
the mean distortion strength 
averaged over a wide radial range of
$\theta=[1',18']$
as a function of color limit,
done separately for the blue ({\it left})
and red ({\it right}) samples, where the color boundaries
for the present analysis are indicated
by vertical dashed
lines for respective color samples. 
Here we do not apply area weighting to enhance the effect of dilution in
the central region \cite[see][]{UB2008}.
A sharp drop in the lensing signal is seen 
when the cluster red sequence starts to contribute significantly,
thereby reducing the mean lensing signal.
Note that 
the background populations do not need to be complete
in any sense but should simply be well defined and contain only
background.
For A1689, the weak lensing signal
in the blue sample is systematically lower than that of the red sample, 
so that blue galaxies in A1689 are excluded from the present
analysis, as was done in \citet{UB2008}; on the other hand, 
our improved color selection for the red sample has led to a 
$\sim 16\%$ increase of red galaxies.
In the present study we use for A2142 the same Subaru images as analyzed
by \citet{Okabe&Umetsu08},
but we have improved significantly our
lensing measurements by including blue, as well as red, galaxies,
where the sample size has been  increased by a factor of 4.

An estimate of the background depth is required when converting the
observed lensing signal into physical mass units, 
because the lensing signal depends on the source redshifts in proportion
to $D_{ds}/D_s$.
The mean depth is sufficient for our purposes
as the variation of the lens distance ratio, $D_{ds}/D_s$, is
slow for our sample because the clusters are at relatively low
redshifts ($z_d\sim 0.1-0.2$) compared to the redshift range of the
background galaxies. 
We estimate the mean depth $\langle D_{ds}/D_s\rangle$ of the
combined red+blue background galaxies
by applying our color-magnitude selection to Subaru multicolor
photometry of the HDF-N region 
\citep{Capak+04_HDFN}
or the COSMOS deep field  
\citep{Capak+07_COSMOS},
depending on the availability of filters.
The fractional uncertainty in the mean depth $\langle D_{ds}/D_s\rangle$
for the red galaxies is typically $\sim 3\%$,
while it is about $5\%$ for the blue galaxies. 
It is useful to define the distance-equivalent source redshift
$z_{s,D}$ \citep{Medezinski+07,UB2008}
defined as
\begin{equation}
\label{eq:zD}
\bigg\langle \frac{D_{ds}}{D_s} \bigg\rangle_{z_s} =
 \frac{D_{ds}}{D_s}\bigg|_{z_s=z_{s,D}}.
\end{equation}
We find $z_{s,D}=0.70^{+0.06}_{-0.05}, 0.95^{+0.79}_{-0.30},
0.98^{+0.24}_{-0.16}, 1.00^{+0.25}_{-0.16}$ for A1689, A2142, A2261, and
A2390, respectively. 
For the nearby cluster A2142 at $z\simeq 0.09$, a precise knowledge of
the source redshift is not critical at all for lensing work.
The mean surface number density ($n_g$) of the combined blue+red sample, the
blue-to-red fraction of background galaxies (B/R), the estimated mean
depth $\langle D_{ds}/D_s\rangle$, and the effective source redshift 
$z_{s,D}$ are listed in Table \ref{tab:lensdata}.

\subsection{Weak Lensing Map-Making}
\label{subsec:2dmap}

Weak lensing measurements of the gravitational shear field can be used
to reconstruct the underlying projected mass density field.
In the present study, we will use the dilution-free, color-selected
background sample (\S\ref{subsec:color}) both for the 2D mass
reconstruction and the lens profile measurements\footnote{Okabe \&
Umetsu (2008) used the magnitude-selected galaxy 
sample in their map-making of nearby merging clusters
to increase the background sample
size, while the dilution-free red background sample was used in their
lensing mass measurements.}.

Firstly, we pixelize distortion data of background galaxies
into a regular grid of pixels using a Gaussian 
$w_g(\theta)\propto \exp[-\theta^2/\theta_f^2]$ with $\theta_f = {\rm
FWHM}/\sqrt{4\ln{2}}$. 
Further we incorporate in the pixelization
a statistical weight $u_g$ for an individual galaxy, so that the
smoothed estimate of the reduced shear field at an angular position
$\btheta$ is written as
\begin{equation}
\label{eq:smshear}
\bar{g}_{\alpha}(\btheta) = \frac{\sum_i w_g(\btheta-\btheta_i) u_{g,i}
g_{\alpha,i}}{\sum_i 
w_g(\btheta-\btheta_i) u_{g,i}}
\end{equation}
where $g_{\alpha,i}$ is 
the reduced shear estimate of the $i$th galaxy
at angular position $\btheta_i$, 
and $u_{g,i}$ is the statistical weight of $i$th galaxy
taken as the inverse variance,
$u_{g,i}=1/(\sigma_{g,i}^2+\alpha^2)$, 
with $\sigma_{g,i}$ being the rms error for
the shear estimate of $i$th galaxy (see \S~\ref{subsubsec:shape})
and $\alpha^2$ being the softening constant variance 
\citep{2003ApJ...597...98H}.
We choose $\alpha=0.4$, which is a typical value of 
the mean rms $\bar{\sigma}_g$ over the background sample.
The case with $\alpha=0$ corresponds to an inverse-variance weighting.
On the other hand,
the limit  $\alpha\gg \sigma_{g,i}$ yields a uniform weighting.  
We have confirmed that our results are insensitive to the 
choice of $\alpha$ 
(i.e., inverse-variance or uniform weighting)
with the adopted smoothing parameters.
The error variance for the smoothed shear $\bar{g}=\bar{g}_1+i\bar{g}_2$ 
(\ref{eq:smshear}) is
then given as
\begin{equation}
\label{eq:smshearvar}
\sigma^2_{\bar{g}}(\btheta) = 
\frac{\sum_i w_{g,i}^2 u_{g,i}^2 \sigma^2_{g,i}}
{ \left( \sum_i w_{g,i} u_{g,i} \right)^2}
\end{equation} 
where $w_{g.i}=w_g(\btheta-\btheta_i)$ and 
we have used $\langle g_{\alpha,i}\, g_{\beta,j}\rangle = 
(1/2)\sigma_{g,i}^2\delta^{\rm K}_{\alpha\beta}\delta^{\rm K}_{ij}$
with $\delta^{\rm K}_{\alpha\beta}$ and $\delta_{ij}^{\rm K}$ being the
Kronecker's delta.

We then invert the pixelized reduced-shear field (\ref{eq:smshear}) to
obtain the lensing convergence field $\kappa(\btheta)$ using equation
(\ref{eq:gamma2kappa}). 
 In the map-making we assume 
linear shear in the weak-lensing limit, that is,
$g_{\alpha}=\gamma_{\alpha}/(1-\kappa) \approx \gamma_{\alpha}$.
We adopt the Kaiser \& Squires inversion method
\citep{1993ApJ...404..441K}, which
makes use of the 2D Green function in an infinite space
(\S\ref{subsec:wl}).
In the linear map-making process, the pixelized shear field is weighted
by the inverse of the variance (\ref{eq:smshearvar}).
Note that this
weighting scheme corresponds to using only the diagonal part of the
noise covariance matrix,
$N(\btheta_i,\btheta_j) = \langle
\overline{\Delta g}(\btheta_i)\overline{\Delta g}(\btheta_j)\rangle$,
which is only an approximation of the actual inverse noise weighting in
the presence of pixel-to-pixel correlation due to non-local Gaussian 
smoothing.  
In Table \ref{tab:lensdata} we list 
the rms noise level in the reconstructed $\kappa(\btheta)$
field for our sample of target clusters. 
For all of the clusters, the smoothing scale $\theta_f$
is taken to be $\theta_f= 1\arcmin$
($\theta_{\rm FWHM}\simeq 1.665\arcmin$), which is larger than 
the Einstein radius for our background galaxies. Hence our weak lensing
approximation here is valid in all clusters.

In Figure \ref{fig:maps} 
we show, for the four clusters,
2D maps of the lensing convergence
$\kappa(\btheta)=\Sigma_m(\btheta)/\Sigma_{\rm crit}$
reconstructed from the Subaru distortion data (\S \ref{subsec:2dmap}), 
each with the corresponding gravitational shear field overlaid.
Here the resolution of the $\kappa$ field is $\sim 1.665\arcmin$ in FWHM
for all of the four clusters. 
The side length of the displayed region is $22\arcmin$, corresponding
roughly to the instantaneous field-of-view of AMiBA ($\simeq 23\arcmin$ in
FWHM).
In the absence of higher-order effects, weak lensing only induces
curl-free $E$-mode distortions, responsible for tangential shear
patterns, while the $B$-mode lensing signal is expected to vanish.
For each case, a prominent mass peak is visible in the cluster center,
around which the lensing distortion pattern is clearly tangential.

Also shown in Figure \ref{fig:maps} are contours 
of the AMiBA flux density due to the thermal SZE
obtained by \citet{Wu_AMiBA}. The resolution of AMiBA7 is about
$6'$ in FWHM (\S\ref{sec:amiba}).
The AMiBA map of A1689 reveals a bright and compact structure in the
SZE, similar to the compact and round mass distribution 
reconstructed from the Subaru distortion data.
A2142 shows an extended structure in the SZE 
elongated along the northwest-southeast direction, consistent with the
direction 
of elongation of the X-ray halo, with its general cometary appearance 
\citep{2000ApJ...541..542M}.   
In addition, A2142 shows 
a slight excess in SZE signals
located $\sim 10\arcmin$ northwest of the cluster center,
associated with mass substructure seen in our lensing
$\kappa$ map (Figure \ref{fig:maps});
This slight excess SZE 
 appears extended for a couple of synthesized beams, although the
 per-beam significance level is marginal ($2-3\sigma$).
\citet{Okabe&Umetsu08} showed that this northwest mass substructure is
also associated with a slight excess of cluster sequence
galaxies, lying $\sim 5\arcmin$ ahead of 
the northwest edge of the central X-ray gas core.
On the other hand, no X-ray counterpart to
the northwest substructure was found in the X-ray images from Chandra
and XMM-Newton observations \citep{Okabe&Umetsu08}.
A2261 shows a filamentary mass structure with unknown redshift,
extending to the west of the cluster core 
\citep{2008ApJS..174..117M},
and likely background structures
which coincide with redder galaxy concentrations (see
\S\ref{subsubsec:kappa} for details). 
Our AMiBA and Subaru observations show a compact
structure both in mass and ICM.
The elliptical mass distribution in A2390 agrees well with the shape
seen by AMiBA in the thermal SZE, and
 is also consistent with other X-ray and strong  lensing work.
A quantitative comparison between the AMiBA SZE and Subaru lensing maps
will be given in \S\ref{sec:comparison}.

\subsection{Cluster Lensing Profiles}
\label{subsec:profile}

\subsubsection{Lens Distortion}
\label{subsubsec:gt}

The spin-2 shape distortion of an object 
due to gravitational lensing
is described by
the complex reduced shear, $g=g_1+i g_2$ (see equation [\ref{eq:redshear}]),
 which is coordinate dependent.
For a given reference point on the sky, one can instead 
form coordinate-independent
quantities, 
the tangential distortion $g_+$ and the $45^\circ$ rotated component,
from linear combinations of the distortion coefficients
$g_1$ and $g_2$ as
\begin{equation}
g_+ = -(g_1 \cos 2\phi + g_2\sin 2\phi), \ \ 
g_{\times} = -(g_2 \cos 2\phi - g_1\sin 2\phi),
\end{equation}
where $\phi$ is the position angle of an object with respect to
the reference position, and the uncertainty in the $g_+$ and
$g_{\times}$ 
measurement 
is $\sigma_+ = \sigma_{\times } = \sigma_{g}/\sqrt{2}\equiv \sigma$ 
in terms of the rms error $\sigma_{g}$ for the complex shear measurement.
In practice, the reference point is taken to be the cluster center,
which is well
determined by the locations of the brightest cluster galaxies. 
To improve the statistical significance of the distortion measurement,
we calculate the weighted average of 
$g_+$ and $g_\times$, and its weighted error,
as
\begin{eqnarray}
\label{eq:gt}
\langle g_+(\theta_m)\rangle &=& \frac{\sum_i u_{g,i}\, g_{+,i}}
{\sum_i u_{g,i}},\\
\langle g_\times(\theta_m)\rangle &=& \frac{\sum_i u_{g,i}\, g_{\times,i}}
{\sum_i u_{g,i}},\\
\sigma_{+}(\theta_m) &=& \sigma_{\times}(\theta_m)=\sqrt{
\frac{\sum_i u^2_{g,i}\sigma^2_i}{\left(\sum_i u_{g,i}\right)^2},
}
\end{eqnarray}
where the index $i$ runs over all of the objects located within 
the $m$th annulus with a median radius of $\theta_m$, and
$u_{g,i}$ is the inverse variance weight for $i$th object,
$u_{g,i}=1/(\sigma_{g,i}^2+\alpha^2)$,
softened with $\alpha=0.4$
(see \S\ref{subsec:2dmap}).

Now we assess cluster lens-distortion profiles from 
the color-selected background galaxies (\S \ref{subsec:color}) for the four
clusters,
in order to examine the form of the underlying cluster mass profile and to  
characterize cluster mass properties.   
In the weak lensing limit ($\kappa,|\gamma| \ll 1$), the azimuthally
averaged tangential distortion profile $\langle g_+(\theta)\rangle$
(eq. [\ref{eq:gt}]) is related to the projected mass density profile
\citep[e.g.,][]{BS2001} as
\begin{equation} 
\label{eq:gt2kappa}
\langle g_+(\theta)\rangle \simeq 
\langle \gamma_+ (\theta)\rangle
=\bar{\kappa}(<\theta)-\langle\kappa(\theta) \rangle,
\end{equation}
where $\langle \cdots \rangle$ denotes the azimuthal average,
and $\bar{\kappa}(<\theta)$ is the mean convergence within a circular
aperture of radius $\theta$ defined as
$\bar{\kappa}(<\theta)=(\pi\theta^2)^{-1}\int_{|\btheta'|\le
\theta}\!d^2\theta'\,\kappa(\btheta')$. Note that equation
(\ref{eq:gt2kappa})  holds for an arbitrary mass distribution.
With the assumption of quasi-circular symmetry in the
projected mass distribution, one can express the tangential distortion
as $\langle g_+(\theta)\rangle \simeq 
[\bar{\kappa}(<\theta)-\langle\kappa(\theta)\rangle]/
[1-\langle\kappa(\theta)\rangle]$ in the non-linear but sub-critical
(${\rm det}{\cal A}(\btheta)>0$) regime.

Figure \ref{fig:gprof} shows the azimuthally-averaged radial profiles of 
the tangential distortion, $\langle g_+\rangle$ ($E$ mode), and the
$45^\circ$-rotated 
component,  $\langle g_\times \rangle$ ($B$ mode).
Here the presence of $B$ modes can be used to check for systematic
errors. 
For each of the clusters, the observed $E$-mode signal 
is significant at the $12$--$16\sigma$ level out to the limit of our
data ($\theta \sim   20^\prime$).
The significance
level of the  $B$-mode detection is
about $2.5\sigma$ for each cluster, which is about a
factor of $5$ smaller than $E$-mode. 

The measured $g_+$ profiles are compared with two representative cluster
mass models, 
namely the NFW model and the singular isothermal sphere (SIS) model.
Firstly, the NFW universal density profile has a two-parameter
functional form as 
\begin{eqnarray}
\label{eq:NFW}
 \rho_{\rm NFW}(r)= \frac{\rho_s}{(r/r_s)(1+r/r_s)^2}, 
\end{eqnarray}
where $\rho_s$ is a characteristic inner density, and $r_s$ is a
characteristic inner radius.
The logarithmic gradient $n\equiv d\ln\rho(r)/d\ln r$
 of the NFW density profile flattens
continuously towards the center of mass, with a flatter central slope
$n=-1$ and a steeper outer slope ($n\to -3$ when $r\to \infty$)
than a purely isothermal structure ($n=-2$).
A useful index, the concentration, compares
the virial radius,
$r_{\rm vir}$, to $r_s$ of the NFW profile, $c_{\rm vir}=r_{\rm
vir}/r_s$. 
We specify the NFW model with the halo virial mass $M_{\rm vir}$ and the
concentration $c_{\rm vir}$ instead of $\rho_s$ and $r_s$.\footnote{We
assume the cluster redshift $z_d$ is equal to the cluster virial
redshift.}
We employ the radial dependence of the NFW lensing profiles,
$\kappa_{\rm NFW}(\theta)$ and $\gamma_{+,{\rm NFW}}(\theta)$,
given by
\citet{1996A&A...313..697B} and \citet{2000ApJ...534...34W}.
Next, the SIS density profile is given by
\begin{equation}
\rho_{\rm SIS}(r)=\frac{\sigma_v^2}{2\pi Gr^2},
\end{equation}
where $\sigma_v$ is the one-dimensional isothermal velocity dispersion
of the SIS halo. The lensing profiles for the SIS model, obtained by
projections of the three-dimensional mass distribution, are found to be 
\begin{equation}
\kappa_{\rm SIS}(\theta)=\gamma_{+,{\rm
 SIS}}(\theta)=\frac{\theta_E}{2\theta}, 
\end{equation}
where $\theta_E$ is the Einstein radius defined by
$\theta_E\equiv 4\pi(\sigma_v/c)^2D_{ds}/D_s$.

Table \ref{tab:model} lists the best-fitting parameters for
these models,
together with the predicted Einstein radius $\theta_{\rm E}$ for a
fiducial source at $z_s=1.5$, corresponding roughly to the median
redshifts of our blue background galaxies.
For a quantitative comparison of the models, we introduce 
as a measure of the {\it goodness-of-fit}
the significance probability $Q(\nu/2,\chi^2/2)$ to find 
by chance a value of $\chi^2$ as poor as the observed value for a given
number of dof, $\nu$ 
\citep[see \S 15.2 in][]{press1992}.\footnote{
Note that a $Q$ value greater
than 0.1 indicates a satisfactory agreement between the data and the
model; if $Q\simgt 0.001$, the fit may be acceptable, e.g. in a case
that the measurement errors are moderately underestimated; if $Q\simlt
0.001$, the model may be called into question.}
We find with our best-fit NFW models $Q$-values of
$Q\simeq 0.50, 0.95, 0.36$, and $0.80$,
and with our best-fit SIS models
$Q\simeq 0.28, 5.0\times 10^{-6}, 0.37$, and $0.87$,
for A1689, A2142, A2261, and A2390, respectively.
Both models provide
statistically acceptable fits for A1689, A2261, and A2390.
For our lowest-$z$ cluster A2142, the curvature in the $g_+$ profile is
pronounced, and a  SIS model for A2142 is
strongly ruled out by the Subaru distortion data alone,
where the minimum $\chi^2$ is $\chi^2_{\rm min} = 39$ with 8 dof.

\subsubsection{Lens Convergence}
\label{subsubsec:kappa}

Although the lensing distortion is directly observable,
the effect of substructure on the gravitational shear is non-local.
Here we examine the lens convergence ($\kappa$) profiles
using the shear-based 1D reconstruction method developed by
\citet{UB2008}. See Appendix \ref{app:1dmass} for details of the
reconstruction method.

In Figure \ref{fig:kprof} we show, 
for the four clusters,
model-independent $\kappa$ profiles
derived using the shear-based 1D reconstruction method, 
together with predictions from the best-fit NFW models for
the $\kappa(\theta)$ and $g_+(\theta)$ data. The substructure
contribution to $\kappa(\theta)$ is local, whereas the inversion
from the observable distortion to $\kappa$ involves a non-local process.
Consequently the 1D inversion method requires a boundary
condition specified in terms of the mean $\kappa$ value
within an outer annular region (lying out to $18'$--$19'$). 
We determine this value for each cluster using the best-fit NFW
model for the $g_+$ profile (Table \ref{tab:model}).

We find that the two sets of best-fit NFW parameters are in excellent
agreement for all except A2261: For A2261, the best-fit values of
$c_{\rm vir}$ from the $g_+$ and $\kappa$ profiles are 
are in poorer agreement.
From Figures \ref{fig:maps} and 
\ref{fig:kprof} we see that the NFW fit to the $g_+$ profile of A2261 is
affected 
by the presence of mass structures at outer radii,
$\theta\simeq  4'$ and $10'$,
resulting in a slightly shallower profile ($c_{\rm vir}\simeq  6.4$)
than in the $\kappa$ analysis.
It turns out that 
these mass structures are associated with galaxy overdensities
whose mean colors are redder than the cluster sequence for A2261
at $z=0.224$,
$\Delta(V-R_{\rm c})\equiv
(V-R_{\rm c})-(V-R_{\rm c})_{A2261}\sim +0.6$, and hence
they are likely to be physically unassociated
 background objects.
The NFW fit to $\kappa(\theta)$ yields a steeper
profile with a high concentration, $c_{\rm vir}\simeq  10.2$, 
which implies a large Einstein angle of $\theta_{\rm E}\simeq 37\arcsec$
at $z_s=1.5$ (Table \ref{tab:model}). This 
is in good agreement with our preliminary strong-lensing model 
(Zitrin et al. in preparation)
based on the method by \citet{2005ApJ...621...53B},
in which 
the deflection field
is constructed based on the smoothed cluster light distribution
to predict the appearance and positions of
counter images of background galaxies.
This model is refined  as new multiply-lensed images are identified 
in deep Subaru $VR_{\rm c}$ and CFHT/WIRCam $JHK_s$ images, 
and incorporated to improve the cluster mass model.
Figure \ref{fig:a2261} shows the tangential critical curve predicted for
a background source at $z_s\sim 1.5$, overlaid on 
the Subaru $V+R_{\rm c}$ pseudo-color image
in the central $6.7'\times 6.7'$ region of A2261.
The predicted critical curve is a nearly circular Einstein ring,
characterized by an effective radius of $\theta_{\rm E}\sim 40\arcsec$
\citep[see][]{Oguri+Blandford}.
This motivates us to further improve the statistical constraints on the
NFW model by combining the outer lens convergence profile with 
the observed constraint on the inner Einstein radius.
A joint fit of the NFW profile to the $\kappa$ 
profile and the inner Einstein-radius constraint with $\theta_{\rm
E}=40\arcsec \pm 4\arcsec$ ($z_s=1$)
tightens the constraints on the NFW parameters (see \S5.4.2 of Umetsu
\& Broadhurst 2008):
$M_{\rm vir}=1.25^{+0.17}_{-0.16}\times 10^{15}M_\odot h^{-1}$ and $c_{\rm
vir}=11.1^{+2.2}_{-1.9}$;
This model yields an Einstein radius of 
$\theta_E= (40\pm 11)\arcsec$ at $z_s=1.5$. 
In the following analysis 
we will adopt this as our primary mass model of A2261.

For the strong-lensing cluster A1689, 
more detailed lensing constraints are available from
joint observations with the high-resolution {\it Hubble Space Telescope} 
(HST) Advanced Camera for Surveys (ACS) and 
the wide-field Subaru/Suprime-Cam
\citep{BTU+05,UB2008}.
In \citet{UB2008} we combined all possible
lensing measurements, namely, the ACS strong-lensing profile of
\citet{BTU+05} and Subaru weak-lensing distortion and magnification data,
in a full two-dimensional treatment,
 to achieve the maximum possible lensing precision.
Note, the combination of distortion and magnification data breaks the
mass-sheet degeneracy (see eq.~[\ref{eq:invtrans}])
inherent in all reconstruction methods based on
distortion information alone \citep{1996ApJ...464L.115B}.
It was found that the joint ACS and Subaru data, covering a wide range
of radii
from 10 up to 2000$\,$kpc$\,h^{-1}$, are well approximated by a single NFW
profile
with $M_{\rm vir}=(1.5\pm 0.1 ^{+0.6}_{-0.3}) \times 10^{15} M_\odot h^{-1}$ and
$c_{\rm vir}=12.7\pm  1\pm 2.8$ (statistical followed by systematic
uncertainty at 68$\%$ confidence).\footnote{
In \citet{UB2008} cluster masses are
expressed in units of $10^{15}M_\odot$ with $h=0.7$. The systematic
uncertainty in $M_{\rm vir}$ is tightly correlated with that in $c_{\rm
vir}$ through the Einstein radius constraint by the ACS observations.
}
This properly reproduces the Einstein radius, which is 
tightly constrained by detailed strong-lens modeling
\citep{2005ApJ...621...53B,2006MNRAS.372.1425H,2007ApJ...668..643L}:
$\theta_{\rm E}\simeq 52\arcsec$ at $z_s=3.05$ (or $\theta_{\rm E}\simeq 
45\arcsec$ at a fiducial source redshift of $z_s=1$).
With the improved color selection for the red background sample
(see \S\ref{subsec:color}),
we have redone a joint fit to the ACS and Subaru lensing observations
using the 2D method of \citet{UB2008}: The refined constraints on the
NFW parameters are  $M_{\rm vir} =1.55^{+0.13}_{-0.12} \times
10^{15} M_\odot h^{-1}$ and  
$c_{\rm vir} =12.3^{+0.9}_{-0.8}$, 
yielding an Einstein radius of $50^{+6.5}_{-6.0}$ arcsec at $z_s=1.5$.
In the following, we will adopt this refined NFW profile as our primary 
mass model of A1689.

\section{Distributions of Mass and Hot Baryons}
\label{sec:comparison}
  
Here we aim to compare the projected distribution of mass and ICM in the
clusters using our Subaru weak lensing and AMiBA SZE maps.
To make a quantitative comparison, we first define 
the ``cluster shapes'' on weak lensing mass structure
by introducing a spin-2 halo ellipticity $e^{\rm halo}= e_1^{\rm
halo}+i e_2^{\rm halo}$, defined in terms of weighted quadrupole shape moments
$Q_{\alpha\beta}^{\rm halo}$ ($\alpha,\beta=1,2$), as
\begin{eqnarray}
\label{eq:haloshape}
e^{\rm halo}_\alpha (\theta_{\rm ap})&=& \left(
\frac{Q^{\rm halo}_{11}-Q^{\rm halo}_{22}}
{Q^{\rm halo}_{11}+Q^{\rm halo}_{22}}, 
 \frac{2Q^{\rm halo}_{12}}
      {Q^{\rm halo}_{11} + Q^{\rm halo}_{22}}
\right),\\
\label{eq:qij_halo}
Q^{\rm halo}_{\alpha\beta}(\theta_{\rm ap}) 
&=& \int_{\Delta\theta\le \theta_{\rm ap}}
\!d^2\theta\,  
 \Delta\theta_\alpha\Delta\theta_\beta \, \kappa(\btheta),
\end{eqnarray}
where $\theta_{\rm ap}$ is the circular aperture radius, 
and
$\Delta\theta_\alpha$ is the angular displacement vector from the
cluster center.
Similarly, the spin-2 halo
ellipticity for the SZE is defined using the cleaned SZE decrement map
$-\Delta I(\btheta) \propto y(\btheta)$ instead of $\kappa(\btheta)$ in
equation (\ref{eq:qij_halo}). 
The degree of halo ellipticity is quantified by the modulus of the
spin-2 
ellipticity, $|e^{\rm halo}|=\sqrt{(e_1^{\rm halo})^2+(e_2^{\rm
halo})^2}$, and  the orientation of halo is defined by the position
angle of the major axis, $\phi^{\rm halo} = \arctan(e_2^{\rm
halo}/e_1^{\rm halo})/2$.  
In order to avoid noisy shape
measurements, we introduce a lower limit of $\kappa(\btheta)>0$ and
$-\Delta I(\btheta)>0$ in equation (\ref{eq:qij_halo}).
Practical shape measurements are done using pixelized lensing and SZE
maps shown in Figure \ref{fig:maps}. The images are sufficiently
oversampled that 
the integral in equation (\ref{eq:qij_halo}) can be approximated by the
discrete sum.
Note, a comparison in terms of the shape parameters is optimal for the
present case where the paired AMiBA and weak lensing images have
different angular 
resolutions: $\theta_{\rm FWHM}\simeq 6\arcmin$ FWHM for AMiBA7, 
and $\theta_{\rm FWHM}\simeq 1.7\arcmin$
FWHM for Subaru weak lensing. When the aperture diameter
is larger than the resolution $\theta_{\rm FWHM}$, i.e.,
$\theta_{\rm ap} > \theta_{\rm FWHM}/2$, the halo shape parameters
can be reasonably defined and measured from the maps.

Now we measure as a
function of aperture radius $\theta_{\rm ap}$
the cluster ellipticity and orientation profiles 
for projected mass and ICM pressure as represented by the lensing
$\kappa$ and SZE decrement maps, respectively. 
For the Subaru weak lensing, the shape parameters are measured at
$\theta_{\rm ap}=[1,2,3, ..., 11]\times \theta_{\rm FWHM}$
($1.7\arcmin \simlt \theta_{\rm ap}\simlt 18.3\arcmin)$;
for the AMiBA SZE, 
$\theta_{\rm ap}=[1,2,3]\times \theta_{\rm FWHM}$
($6\arcmin \simlt \theta_{\rm ap}\simlt 18\arcmin)$. 
The level of uncertainty in the halo shape parameters is assessed 
by a Monte-Carlo error analysis assuming Gaussian errors for weak
lensing distortion and AMiBA visibility measurements 
\citep[for the Gaussianity of AMiBA data, see][]{Nishioka_AMiBA}.
For each cluster and dataset, 
we generate a set of 500 Monte Carlo simulations of Gaussian noise 
and propagate into final
uncertainties in the spin-2 halo ellipticity, $e^{\rm halo}$. 
Figure \ref{fig:eplot} displays, for the four clusters,
the resulting cluster ellipticity and orientation profiles on mass and
ICM structure 
as measured  from 
the Subaru weak lensing and AMiBA SZE maps, shown separately for the
ellipticity modulus $|e^{\rm halo}|$ and the orientation, $2\phi^{\rm
halo}$ (twice the position angle). 
Overall, a good agreement is found between the shapes of mass and
ICM structure up to large radii, in terms of both ellipticity and
orientation. In particular, our results on A2142 and A2390 show that  
the mass and pressure distributions trace each other well at all radii.
At a large radius of $\theta_{\rm ap}\simgt 10'$, 
the position angle of A2142 is $\phi^{\rm halo}\sim 50^\circ$.
For A2390, 
the position angle  is $\phi^{\rm halo}\sim 30^\circ$ at all radii.


\section{Cluster Gas Mass Fraction Profiles}

\subsection{Method}
\label{sec:method}

In modeling the clusters,
we consider two representative analytic models for describing 
the cluster DM and ICM distributions,  namely (1) 
the \citet[][hereafter KS01]{2001MNRAS.327.1353K}
model of the universal gas density/temperature profiles 
and (2) the isothermal $\beta$ model, where both are physically
motivated under the hypothesis of 
hydrostatic equilibrium and polytropic equation-of-state,
$P\propto \rho^\gamma$,
with an additional assumption about the spherical symmetry of the
system.

Joint AMiBA SZE and Subaru weak lensing observations probe cluster
structures on angular scales up to $\Delta\theta \sim
23\arcmin$.\footnote{The FWHM of the primary beam patten of the AMiBA is 
about $23\arcmin$, while the field-of-view of the Subaru/Suprime-Cam is
about $34\arcmin$}
At the median redshift $\bar{z}\simeq  0.2$ of our clusters, this maximum
angle covered by the data corresponds roughly to 
$r_{200}\approx 0.8 r_{\rm vir}$, 
except $r_{500}\approx 0.5 r_{\rm vir}$ for A2142 at $z=0.09$.
In order to better constrain the gas mass fraction
in the outer parts of the clusters, 
we adopt a prior that the gas density profile $\rho_{\rm gas}(r)$  
traces the underlying (total) mass density profile,
$\rho_{\rm tot}(r)$. Such a relationship is expected at large radii,
where non-gravitational processes, such as radiative cooling and star
formation, have not had a major effect on the structure of the
atmosphere so that the polytropic assumption remains valid (Lewis et
al. 2000). Clearly this results in the gas mass fraction,
$\rho_{\rm gas}(r)/\rho_{\rm tot}(r)$, tending to a constant at large
radius. 
In the context of the isothermal $\beta$ model, this simply means that
$\beta=2/3$.

In both models, for each cluster, the mass density profile $\rho_{\rm
tot}(r)$ is constrained solely by the Subaru weak lensing data 
(\S\ref{sec:subaru}),
the gas temperature profile $T_{\rm gas}(r)$ is normalized by 
the spatially-averaged X-ray temperature
(see Table \ref{tab:amiba_target}),
and the electron pressure profile  $P_e(r) = n_e(r) k_B T_e(r)$
is normalized by the AMiBA SZE data,
where $n_e(r)$ is the electron number density, and
$T_e(r)=T_{\rm gas}(r)$ is the electron temperature.
The gas density is then given by
$\rho_{\rm gas}(r)=\mu_e m_p n_e(r)$.

\subsection{Cluster Models}
\label{sec:model}

\subsubsection{NFW-Consistent Model of  Komatsu \& Seljak 2001}

The KS01 model describes the polytropic gas in hydrostatic equilibrium with
a gravitational potential described by the {\it universal} density
profile for collisionless CDM halos
proposed by 
\citet[][hereafter NFW]{1996ApJ...462..563N}.
See KS01, 
\citet[][hereafter KS02]{2002MNRAS.336.1256K},
and
\citet{Worrall+Birkinshaw2006}
for more detailed discussions.
High mass clusters with virial masses 
$M_{\rm vir}\simgt 10^{15}M_\odot/h$
are so massive that the virial temperature of the gas is too
high for efficient cooling and hence the cluster potential simply
reflects the dominant DM. This has been recently established by our
Subaru weak lensing study of several massive clusters 
\citep{BTU+05,BUM+2008,UB2008}.

In this model, the gas mass profile traces the
NFW profile in the outer region of the halo 
($r_{\rm vir}/2 \simlt r \simlt r_{\rm vir}$; see KS01),
satisfying the adopted prior of the constant gas mass fraction
$\rho_{\rm gas}(r)/\rho_{\rm tot}(r)$ at large radii.
This behavior is supported by cosmological hydrodynamic simulations 
\citep[e.g.,][]{Yoshikawa+2000},
and is recently found from the stacked SZE analysis of the
WMAP 3-year data 
\citep{Atrio+2008_WMAP3}.
The shape of the gas distribution functions, as well as the polytropic
index $\gamma_{\rm gas}$, can be fully specified by 
the halo virial mass, $M_{\rm vir}$, and the halo concentration, $c_{\rm
vir}=r_{\rm vir}/r_s$,  of the NFW profile.

In the following,
we use the form of the NFW profile to determine 
$r_{\rm vir}$, $r_{200}$, $r_{500}$, and $r_{2500}$.
Table \ref{tab:cluster} summarizes 
the NFW model parameters
derived from our lensing analysis for the four clusters
(see \S\ref{subsec:profile}).
For each cluster we also list the corresponding
$(r_{2500}, r_{500},r_{200},r_{\rm vir})$. 
For calculating $\gamma_{\rm gas}$ and the normalization
factor $\eta(0)$ for a structure constant ($B$ in equation [16] of
KS02), we follow the fitting formulae given by KS02, which are valid for 
halo concentration, $1< c_{\rm vir}<25$ (see
Table \ref{tab:cluster}). 
For our clusters, $\gamma_{\rm gas}$ is in the range of $1.15$ to
$1.20$. Following the prescription in KS01, we convert
the X-ray cluster temperature $T_X$ to the central gas
temperature $T_{\rm gas}(0)$ of the KS01 model.

\subsubsection{Isothermal $\beta$ Profile}
\label{subsubsec:beta}

The isothermal $\beta$ model provides an alternative consistent solution
of the hydrostatic equilibrium equation \citep{1999PThPS.133....1H},
assuming the ICM is isothermal and its density profile follows
$\rho_{\rm gas}(r)=\rho_{\rm gas}(0)[1+(r/r_c)^2]^{-3\beta/2}$
with the gas core radius $r_c$. At large radii, $r\gg r_c$, where both
of our SZE and weak lensing observations are sensitive, the total mass density  
follows $\rho_{\rm tot}(r) \propto r^{-2}$.
Thus we set $\beta=2/3$ to satisfy 
our assumption of  constant $\rho_{\rm gas}(r)/\rho_{\rm tot}(r)$ at
large radius.
We adopt the values of $r_c$ and $T_X$ listed in Table
\ref{tab:amiba_target}, taken from X-ray observations, 
and use $T_{\rm gas}(r)=T_X$ as the gas temperature for
this model.
At $r\gg r_c$, the
$\rho_{\rm tot}(r)$ profile can be approximated by that of a 
SIS
(see \S\ref{subsubsec:gt}) parametrized by the
isothermal
1D velocity dispersion $\sigma_v$ 
(see Table~\ref{tab:cluster}), 
constrained by the Subaru distortion data (see \S\ref{subsec:profile}).

Requiring hydrostatic balance gives an
isothermal temperature $T_{\rm SIS}$, equivalent to $\sigma_v$, as
\begin{equation}
\label{eq:tratio}
k_B T_{\rm SIS}\equiv \mu m_p \sigma_v^2 \frac{2}{3\beta}.
\end{equation}
For $\beta=2/3$, $k_B T_{\rm SIS} = \mu m_p \sigma_v^2$, 
which can be compared with the observed $T_X$ (Table
\ref{tab:amiba_target}).  
For our AMiBA-lensing cluster sample, we found 
X-ray to SIS temperature ratios
$T_X/T_{\rm SIS} = 0.82 \pm 0.03,  1.65 \pm 0.15, 0.94 \pm 0.05, 1.28\pm
0.15$ for A1689, A2142, A2261, and A2390, respectively. 
For A2261 and A2390, the inferred temperature ratios are consistent with
unity at 1--2$\sigma$.
For the merging cluster A2142, the observed spatially-averaged X-ray
temperature \citep[cooling-flow corrected; see][]{1998ApJ...504...27M}
is significantly higher than the lensing-derived temperature. This
temperature excess of $\sim 4\sigma$ could be explained by the effects
of merger boosts, as discussed in \citet{Okabe&Umetsu08}.
The temperature ratio $T_X/T_{\rm SIS}$ for A1689, on the other hand, is
significantly lower than unity. Recently, a similar level of 
discrepancy was also found in \cite{2008MNRAS.386.1092L}, who performed
a careful joint X-ray and lensing analysis of this cluster. 
A deprojected
3D temperature profile was obtained 
using a model-independent
approach to the Chandra X-ray emission measurements 
and the projected mass profile obtained from the joint strong/weak
lensing analysis of \citet{BTU+05}.
The projected temperature profile predicted from their joint analysis 
exceeds the observed temperature by $30\%$ 
at all radii, a level of discrepancy suggested from hydrodynamical 
simulations that find that denser, more X-ray luminous small-scale
structure can 
bias X-ray temperature measurements
downward at about the same level
\citep{2007ApJ...659..257K}.
If we accept this $+30\%$ correction
for $T_X$, the ratio $T_X/T_{\rm SIS} \rightarrow 1.07 \pm 0.04$ for
A1689, consistent with $\beta = 2/3$.

\subsection{AMiBA SZE Data}
\label{subsec:szedata}

We use our AMiBA data to constrain the remaining
normalization parameter for the $\rho_{\rm gas}(r)$ profile, 
$\rho_{\rm gas}(0)$.
The calibrated output of the AMiBA interferometer, after the
lag-to-visibility transformation
\citep{Wu_AMiBA}, 
is the complex visibility $V(\bu)$ as a function of 
baseline vector in units of wavelength, $\bu=\bd/\lambda$, 
given as the Fourier transform of the sky brightness distribution
$\Delta I(\btheta)$ 
attenuated by the antenna primary beam pattern $A(\btheta)$.

In targeted AMiBA observations at 94$\,$GHz, the sky signal $\Delta
I(\btheta)$ with respect to the background 
(i.e., atmosphere and the mean CMB intensity)
is dominated by the thermal SZE due to hot electrons in the
cluster, $\Delta I_{\rm SZE} = I_{\rm norm}\,g (\nu) y$ 
(see eq.~[\ref{eq:SZE}]).
The Comptonization parameter $y$ is expressed as a line-of-sight integral of
the thermal electron pressure (see eq.~[\ref{eq:y}]).
In the line-of-sight projection of equation (\ref{eq:y}), the cutoff
radius $r_{\rm max}$ needs to be specified.
We take
$r_{\rm max}\equiv \alpha_r r_{\rm vir}$ with a dimensionless
constant $\alpha_r$ which we set to $\alpha_r=2$. In the present study we
found the line-of-sight projection in equation (\ref{eq:y}) is insensitive to
the choice of $\alpha_r$ as long as $\alpha_r\simgt 1$.

A useful measure of the thermal SZE is the integrated Comptonization
parameter $Y(\theta)$, 
\begin{equation}
\label{eq:Y}
Y(\theta) = 2\pi\int_0^\theta\!d\theta'\,\theta' y(\theta'),
\end{equation}
which is proportional to
the SZE flux,
and is a measure of the thermal energy content in the ICM. 
The value of $Y$ is less
sensitive to the details of the model fitted than the central
Comptonization parameter $y_0 \equiv y(0)$, with the current
configuration of AMiBA.
If the $A(\btheta)y(\btheta)$ field  has reflection symmetry about the
pointing center, then the imaginary part of $V(\bu)$ vanishes, and the
sky signal is entirely contained in the real visibility flux.
If the $A(\btheta)y(\btheta)$ field is further azimuthally symmetric,
the real visibility flux is expressed
by the Hankel transform of order zero as
\begin{eqnarray}
\label{eq:V}
V^{Re}(u) &=&2\pi
I_{\rm norm} g(\nu_c) \int_0^\infty \!d\theta\, \theta A(\theta) y(\theta) 
J_0(2\pi u \theta)\nonumber\\
 &\equiv&
2\pi I_0  \int_0^\infty \!d\theta\, \theta A(\theta) \frac{y(\theta)}{y_0}
J_0(2\pi u \theta),
\end{eqnarray} 
where 
$I_0=I_{\rm norm}g(\nu_c)y_0$ is the central SZE intensity at
$\nu_c=94\,$GHz, 
$J_0(x)$ is the Bessel function of the first kind and
order zero, and 
$A(\theta)$ is well approximated by a circular Gaussian with  
${\rm FWHM}=1.22(\lambda/D)\simeq 23\arcmin$ at $\nu_c=94 {\rm GHz}$ with 
an antenna diameter of $D=60 {\rm cm}$ 
\citep{Wu_AMiBA}.
The observed imaginary flux can be
used to check for the effects of
primary CMB and radio source contamination
\citep[][]{Liu_AMiBA}.
From our AMiBA data we derive in the Fourier domain
azimuthally-averaged visibility profiles $\langle V(u)\rangle$ 
for individual clusters. 

We constrain the normalization $I_0$
from $\chi^2$ fitting to the $\langle V(u)\rangle$ profile
\citep{Liu_AMiBA}.
In order to convert  $I_0$ into the 
central Comptonization parameter, we take account of (i) the
relativistic correction $\delta_{\rm SZE}(\nu,T_{\rm gas})$
in the SZE spectral function $g(\nu)$ (see eq.~[\ref{eq:sze_gfunc}])
and (ii) corrections for contamination by discrete radio point
sources \citep{Liu_AMiBA}.
The level of 
contamination in $I_0$
from known discrete point sources has been estimated to be
about 
$(6-35)\%$
in our four clusters
\citep{Liu_AMiBA}.
In all cases, a net positive contribution of point sources was found
in our 2-patch differencing AMiBA observations (\S\ref{sec:amiba}), 
indicating that there are more radio sources
towards clusters than in the background
\citep{Liu_AMiBA}. 
Thus ignoring the point source correction would systematically bias the
SZE flux estimates, leading to an underestimate of $y_0$.
The relativistic correction to the thermal SZE 
is 6--7$\%$ in our $T_X$ range at $94\,$GHz.

Table \ref{tab:amiba} summarizes,
  for our two models,
the best-fitting parameter, $y_0$, 
and the $Y$-parameter
interior to a cylinder of radius $\theta=3\arcmin$ that roughly
matches the AMiBA synthesized beam with $6'$ FWHM. 
For each case,
both cluster models yield consistent values of $y_0$ and $Y(3')$
within $1\sigma$; 
in particular, the inferred values of $Y(3^\prime)$ for the two models
are in excellent agreement.
Following the procedure in \S
\ref{sec:method} we convert $y_0$ into the central gas mass density,
$\rho_{\rm gas}(0)$.

\subsection{Gas Mass Fraction Profiles}
\label{subsec:fg}

We derive cumulative gas fraction profiles, 
\begin{equation}
\label{eq:fgas}
f_{\rm gas}(<r) = \frac{M_{\rm
gas}(<r)} {M_{\rm tot}(<r)}, 
\end{equation}
for our cluster sample using two sets of
cluster models described in \S \ref{sec:model}, where $M_{\rm gas}(<r)$
and $M_{\rm tot}(<r)$ are the hot gas and total cluster masses contained
within a spherical radius $r$. 
In Table \ref{tab:fg} we list, for each of
the clusters, $M_{\rm gas}$ and $f_{\rm gas}$
within $r_{2500}, $$r_{500}$, and $r_{200}$ 
(see also Table \ref{tab:cluster}) calculated with the
two models.
Note that our total mass estimates do not require the
assumption of hydrostatic balance, but are determined based solely
on the weak lensing measurements. 
Gaussian error propagation was used to derive the errors on 
$M_{\rm gas}(<r)$ and $f_{\rm gas}(<r)$.
We propagate errors on the individual cluster parameters (Tables
\ref{tab:amiba_target} and \ref{tab:cluster}) by a Monte-Carlo method.
For A2142, the isothermal model increasingly overpredicts $f_{\rm gas}$
at all radii
$r>r_{2500}$, exceeding the cosmic baryon fraction
$f_b = \Omega_b/\Omega_m=0.171\pm 0.009$ 
\citep{Dunkley+2008_WMAP5}. For other clusters in our sample, both
cluster models yield consistent $f_{\rm gas}$ and $M_{\rm gas}$
measurements 
within the statistical uncertainties from the SZE and weak lensing data.

Our SZE/weak lensing-based measurements can be
compared with other X-ray and SZE measurements.
\citet{Grego+2001_SZE} derived gas fractions for a sample of 18
clusters 
 from 30$\,$GHz SZE observations with BIMA and OVRO 
in combination with
 published X-ray temperatures. They found 
$f_{\rm gas}(<r_{500})=0.140^{+0.041}_{-0.047} h_{70}^{-1}$ and
$0.053^{+0.139}_{-0.031} h_{70}^{-1}$
($h_{70}= h/0.7$) for A1689 and A2261, respectively, 
in agreement with our results.
For A2142,  
the $f_{\rm gas}$ and $M_{\rm gas}$
values inferred from the KS01 model are
in good agreement with those from the VSA SZE observations at 30$\,$GHz 
\citep{Lancaster+05_VSA},
$M_{\rm gas}(r_{500})=6.1^{+1.7}_{-1.8}\times 10^{13}M_{\odot}h^{-2}$
and $f_{\rm gas}(r_{500})=0.123^{+0.080}_{-0.050} h_{70}^{-1}$.
From a detailed analysis of Chandra X-ray data,
\citet{2006ApJ...640..691V} obtained 
$f_{\rm gas}(<r_{500}) = (0.141\pm 0.009) h_{72}^{-3/2}$
$(h_{72}= h/0.72)$ for A2390, in good
agreement with our results.

Furthermore, it is interesting to compare our results 
with 
the detailed joint X-ray and lensing analysis of A1689 
by \citet{2008MNRAS.386.1092L},
who derived deprojected profiles of 
$\rho_{\rm tot}(r)$, $\rho_{\rm gas}(r)$, and $T_{\rm gas}(r)$
assuming hydrostatic equilibrium,
using a model-independent approach to the Chandra X-ray emission
profile and the projected lensing mass profile of 
\citet{BTU+05}.
A steep 3D mass profile was obtained by this approach, with the inferred
concentration of $c_{\rm vir} =  12.2^{+0.9}_{-1.0}$, consistent with the
detailed lensing analysis of 
\citet{BTU+05} and \citet{UB2008},
whereas the observed X-ray temperature
profile falls short of the derived profile at all radii by a constant
factor of $\sim  30\%$ 
(see \S\ref{subsubsec:beta}).
With the pressure profile of Lemze et al.
we find $y_0 = (4.7\pm 0.3)\times 10^{-4}$,
which is in agreement with our KS01 prediction,
$y_0=(4.2\pm 1.0)\times 10^{-4}$ (Table
\ref{tab:amiba}). 
The integrated Comptonization parameter predicted by 
the Lemze et al. model is
$Y(3^\prime) =(3.0\pm 0.1)\times 10^{-10} $, 
which roughly agrees with the AMiBA measurement of 
$Y(3')=(2.5\pm 0.6)\times 10^{-10}$.
Alternatively,
adopting the observed temperature profile in the Lemze et
al. model reduces the predicted SZE signal by a factor of $\sim 30\%$,
yielding 
$y_0\simeq 3.3\times 10^{-4}$ and $Y(3')\simeq 2.1\times 10^{-10}$,
again in agreement with the AMiBA measurements. Therefore, more accurate
SZE measurements are required to further test and verify
this detailed cluster model.

We now use our data to find the average
gas fraction profile over our sample of
four hot X-ray clusters.
The weighted average of $M_{\rm vir}$ in our AMiBA-lensing sample is
$\langle M_{\rm vir}\rangle = (1.19\pm 0.08)\times
10^{15}M_{\odot}h^{-1}$, with a weighted-mean concentration of 
$\langle c_{\rm vir}\rangle =8.9\pm 0.6$.
The weighted average of the cluster virial radius is $\langle r_{\rm
vir}\rangle \simeq 1.95\,$Mpc$h^{-1}$.
At each radius 
we compute the sample-averaged gas fraction,
$\langle f_{\rm gas}(<r)\rangle$, 
weighted by the inverse square of the statistical $1\sigma$
uncertainty. 
In Figure \ref{fig:fgas} we show for the two models
the resulting $\langle f_{\rm gas}\rangle$ profiles 
as a function of radius in units of $r_{\rm vir}$,
along with the published 
results for other X-ray and SZE observations.
Here the uncertainties ({\it cross-hatched}) represent the standard error
($1\sigma$) of
the weighted mean at each radius point, including both the
statistical measurement uncertainties and cluster-to-cluster variance.  
Note A2142 has been excluded for the isothermal case (see
above). The averaged $\langle f_{\rm gas}\rangle$ profiles derived for
the isothermal and KS01 models are consistent within $1\sigma$
at all radii, and lie below the cosmic baryon fraction 
$f_b=0.171\pm 0.009$ constrained by the WMAP 5-year
data \citep{Dunkley+2008_WMAP5}. At $r=\langle r_{200}\rangle\simeq 0.79
\langle r_{\rm vir}\rangle$, the KS01  model gives 
\begin{equation}
\langle f_{\rm gas, 200}\rangle = 0.133 \pm 0.020 \pm 0.018
\end{equation}
where the first error is statistical, and the second is the standard
error due to cluster-to-cluster variance. This is marginally consistent
with $\langle f_{\rm gas,200} \rangle=0.109\pm 0.013$ obtained from the
averaged SZE profile of a sample of 193 X-ray clusters 
($T_X>3\, {\rm keV}$)
using the WMAP 3-year data \citep{Afshordi+2007_WMAP3}.
A similar value of $\langle f_{\rm gas,200}\rangle=0.11\pm 0.03$ was obtained
by \citet{Biviano+Salucci2006}
for a sample of 59 nearby clusters from the ESO Nearby Abell Cluster
Survey, where the total and ICM mass profile are determined by their
dynamical and X-ray analyses, respectively. 
At $r=\langle r_{500}\rangle\simeq 0.53 \langle r_{\rm vir}\rangle$, 
we have  
\begin{equation}
\langle f_{\rm gas, 500}\rangle = 0.126 \pm 0.019 \pm 0.016
\end{equation}
for the KS01 model, in good agreement with the Chandra X-ray
measurements for a subset of six $T_X>5\, {\rm keV}$ clusters
in  \citet{2006ApJ...640..691V}.
At $r=\langle r_{2500}\rangle\simeq 0.25 \langle r_{\rm vir}\rangle$, 
which is close to the resolution limit of AMiBA7,
we have for the KS01 model
\begin{equation}
\langle f_{\rm gas, 2500}\rangle = 0.105  \pm 0.015 \pm 0.012,
\end{equation}
again marginally consistent with the Chandra gas fraction measurements
in 26 X-ray luminous clusters with $T_X>5\,$keV
\citep{2004MNRAS.353..457A}.


\section{Discussion and Conclusions} 
\label{sec:con}

\begin{figure}[htb]
 \begin{center}
  \includegraphics[width=150mm,angle=0]{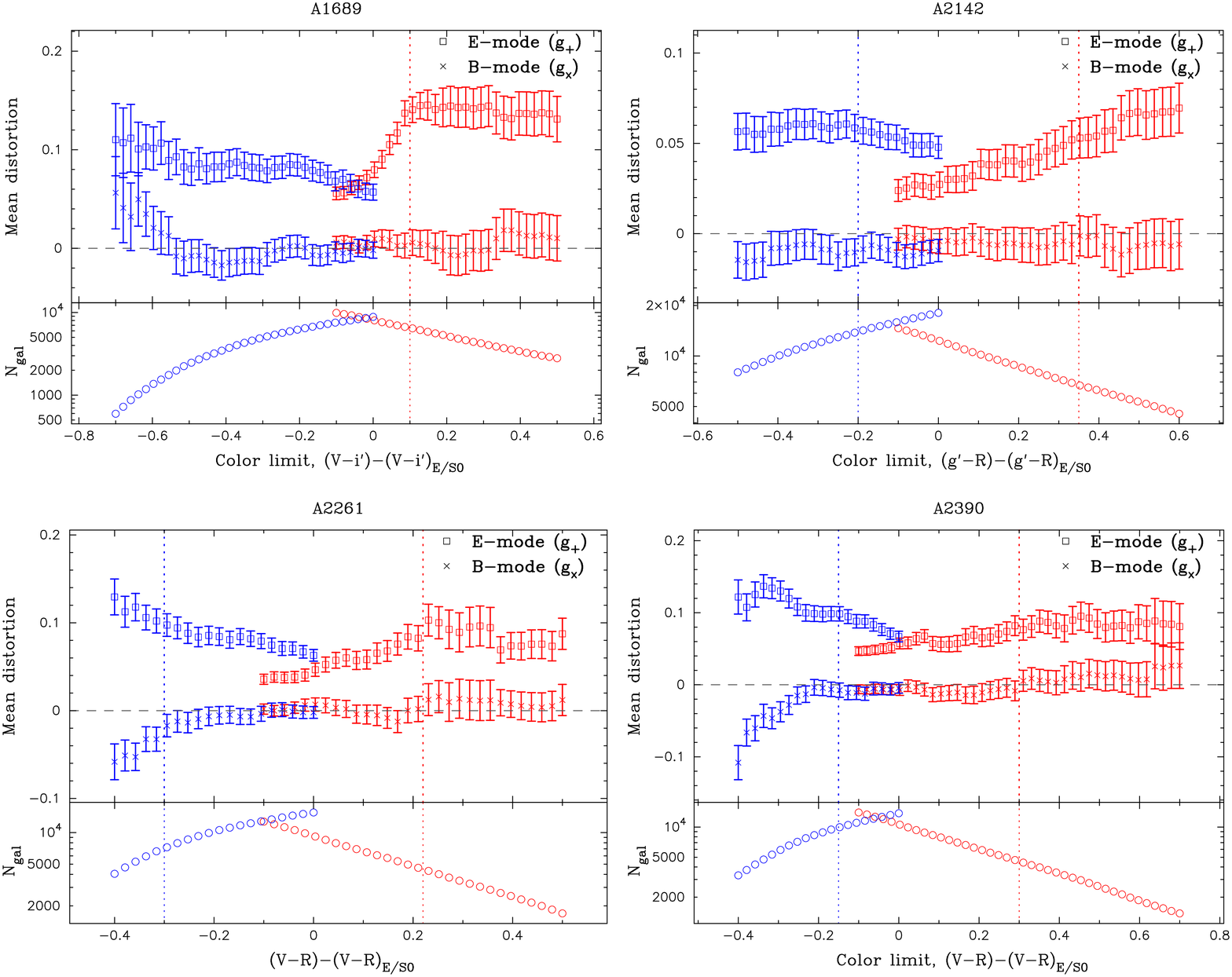}
 \end{center}
\caption{
{\it Top panels}:
Mean shape distortions $(g_+,g_\times)$
averaged over the entire cluster region
($1' < \theta  < 18'$) for the four clusters
done separately for the blue and red samples, in order to establish the
boundaries of the color distribution free of cluster members.
{\it Bottom panels}: Respective numbers of galaxies as a function of
 color-limit
in the red ({\it  right}) and the blue ({\it blue}) samples.
\label{fig:dilution}
}
\end{figure}

We have obtained secure, model-independent profiles of the lens
distortion and projected mass  (Figures \ref{fig:gprof} and \ref{fig:kprof})
by using the shape distortion measurements
from high-quality Subaru imaging,
for our AMiBA lensing sample of four high-mass clusters.
We utilized weak lensing dilution in deep Subaru color images to define
color-magnitude boundaries for blue/red galaxy samples,
where a reliable weak lensing signal can be 
derived, free of unlensed cluster members (Figure \ref{fig:dilution}). 
Cluster contamination
otherwise preferentially dilutes the inner lensing signal leading to
spuriously shallower profiles. 
With the observed lensing profiles we have examined 
cluster mass-density profiles dominated by DM.
For all of the clusters in our sample, the lensing profiles are well
described by the NFW profile predicted for collisionless CDM halos.

A qualitative comparison between our weak lensing and SZE data,
on scales $r\simgt 3'$ limited by the current AMiBA resolution,
shows
a good correlation between the distribution of mass (weak
lensing) and hot baryons (SZE) in massive cluster environments
(\S\ref{subsec:2dmap}), 
as
physically expected for high mass clusters with deep gravitational
potential wells (\S\ref{subsec:2dmap}).
We have also examined and compared, for the first time, the 
cluster ellipticity and orientation profiles on mass and ICM structure
in the Subaru weak lensing and AMiBA SZE observations, respectively.
For all of the four clusters, the mass and ICM distribution shapes are
in good agreement at all relevant radii in terms of both ellipticity and
orientation (Figure \ref{fig:eplot}).
In the context of the CDM model, 
the mass density, dominated by collisionless DM, 
is expected to have a more irregular and elliptical
distribution than the ICM density 
due to inherent triaxiality of CDM halos.
We do not see such a tendency in our lensing and SZE datasets, although
  our ability to find such effects is limited by the resolution of the
  current AMiBA SZE measurements.

We have obtained cluster gas fraction profiles (Figure \ref{fig:fgas})
for the AMiBA-lensing sample
($T_X>8\, {\rm keV}$) 
based on joint AMiBA SZE and Subaru weak lensing observations
(\S\ref{subsec:fg}).  
Our cluster gas fraction measurements are overall in good agreement with
previously-published values.
At $r=\langle r_{200}\rangle\simeq 0.79 \langle r_{\rm vir} \rangle$,
corresponding roughly to the maximum available radius in our joint
SZE/weak lensing data, the sample-averaged gas fraction is 
$\langle f_{\rm gas,200}\rangle=0.133 \pm 0.027$ 
for the NFW-consistent KS01 model,
representing 
the average over
our high-mass cluster
sample with a mean virial mass of
$\langle M_{\rm
 vir}\rangle =(1.2\pm 
 0.1)\times 10^{15}M_\odot h^{-1}$. 
When compared to the cosmic baryon fraction $f_b$, 
this indicates $\langle f_{\rm gas,200}\rangle/f_b=0.78\pm 0.16$, i.e.,
$(22\pm 16)\%$ of the baryons
is missing from the hot phase 
in our cluster sample
\citep[cf.][]{Afshordi+2007_WMAP3,Crain+2007}. 
This missing cluster baryon fraction is partially made up by
observed stellar and cold gas fractions of 
$\sim$ several $\%$ in our $T_X$ range  
\citep{Gonzalez+2005_ICL}.

Halo triaxiality
may affect our projected total and gas mass
measurements based on the assumption of spherical symmetry, producing an
orientation bias.
A degree of triaxiality 
is inevitable for collisionless
gravitationally collapsed structures. 
The likely effect
of triaxiality on the measurements of lensing properties has been
examined analytically 
\citep{2005ApJ...632..841O,2007MNRAS.380.1207S,2007MNRAS.380..149C},
and in numerical investigations 
\citep{2002ApJ...574..538J,2007ApJ...654..714H}.
The effect of triaxiality will be less for the collisional ICM, which
follows the gravitational potential and will be more
  spherical and more smoothly distributed than the total mass density
  distribution. 
For an unbiased measurement of the gas mass fractions, a large, homogeneous
sample of clusters would be needed to beat down the orientation bias. 

Possible biases in X-ray spectroscopic temperature
measurements \citep{Mazzotta+2004,2007ApJ...659..257K} may also affect our
gas fraction measurements
based on 
the overall normalization by the observed X-ray temperature.
This would need to be taken seriously into
account in future investigations with larger samples and higher
statistical precision. 

\begin{figure}[!htb]
 \begin{center}
    \includegraphics[width=150mm]{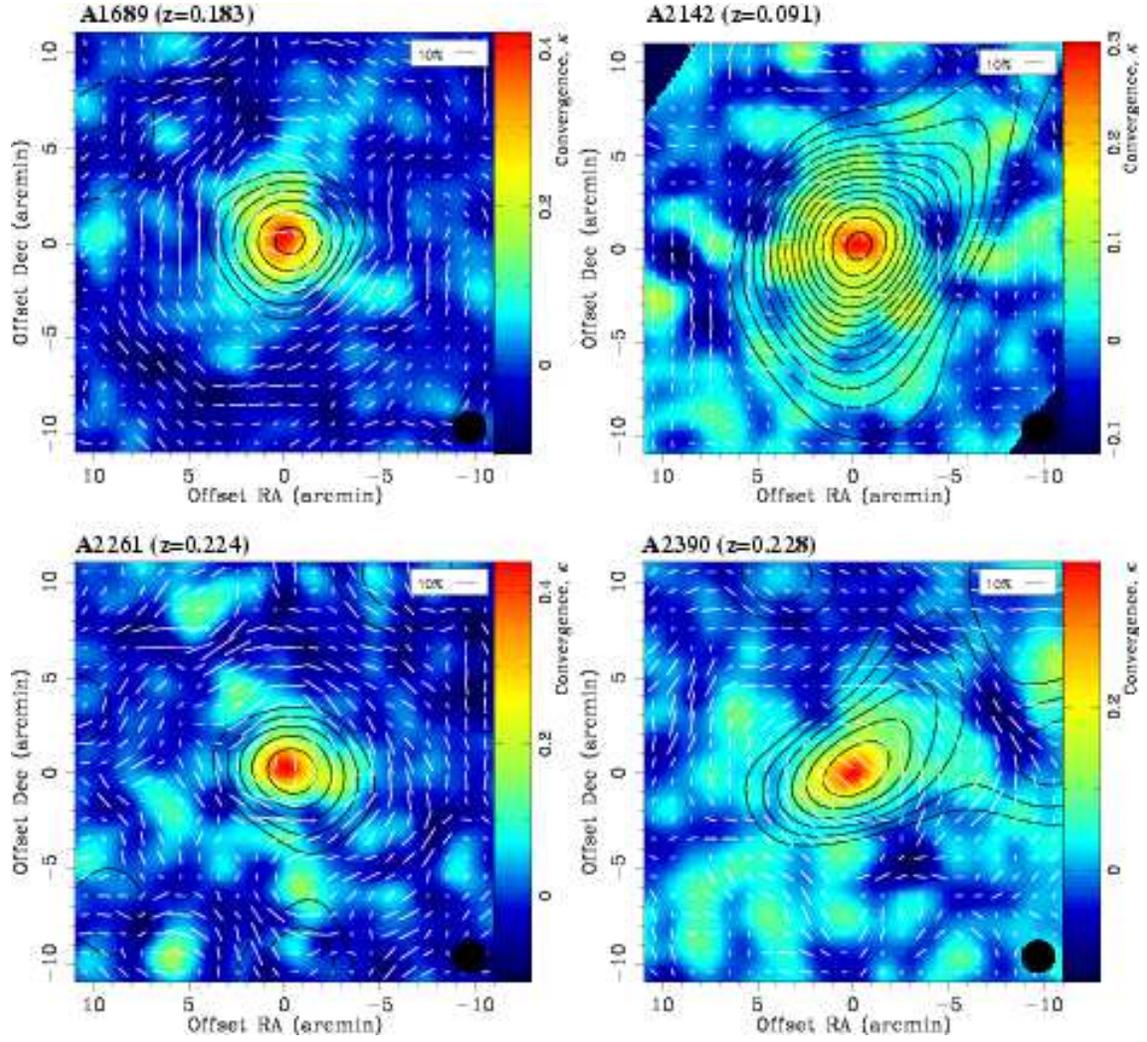}
 \end{center}
\caption{
\label{fig:maps}
Mass maps of the central $22^\prime \times
22^\prime$ of four AMiBA/Subaru clusters reconstructed from Subaru 
 weak lensing data, with the gravitational shear field of background
 galaxies overlaid; $10\%$ ellipticity is indicated top right, and
the resolution characterized by Gaussian FWHM is shown bottom right.
Also overlaid are contours of the SZE flux densities at $94\,$GHz,
 observed with the
 7-element AMiBA, given in units of $1\sigma$ reconstruction error.
The resolution of AMiBA, given in Gaussian FWHM, is $6'$
For  all four clusters
the distribution of the SZE signal is well correlated with
the projected mass distribution, indicating that 
the hot gas in the clusters traces well the underlying gravitational
 potential dominated by unseen dark matter.
The dark blue regions in the mass map of A2142 are outside the Subaru
 observations.
}
\end{figure}

\begin{figure}[htb]
 \begin{center}
   \includegraphics[width=150mm, angle=0]{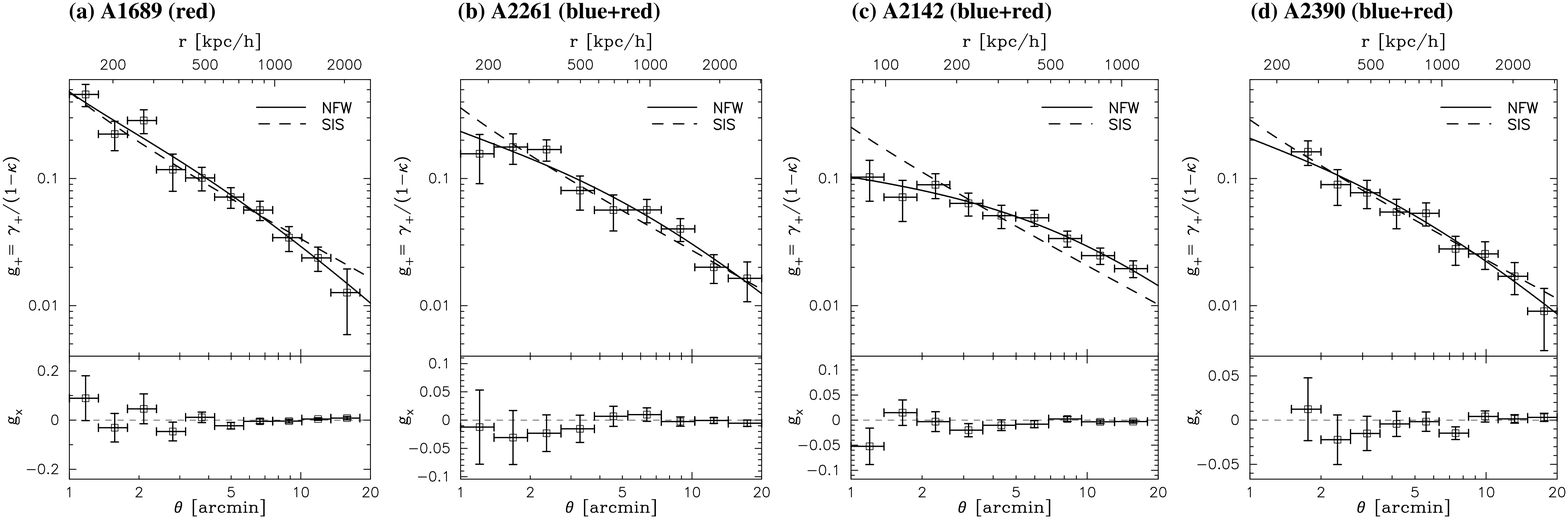}
 \end{center}
\caption{
Azimuthally-averaged radial profiles of the tangential reduced shear
 $g_+$ ({\it upper panels}) for the four clusters
based on the combined red and blue background samples.
The solid and dashed curves show the best-fitting NFW and SIS profiles
 for each cluster.
Shown below is the  $45^\circ$ rotated ($\times$) component, $g_\times$.
\label{fig:gprof}
}
\end{figure}

\begin{figure}[htb]
 \begin{center}
  \includegraphics[width=150mm, angle=0]{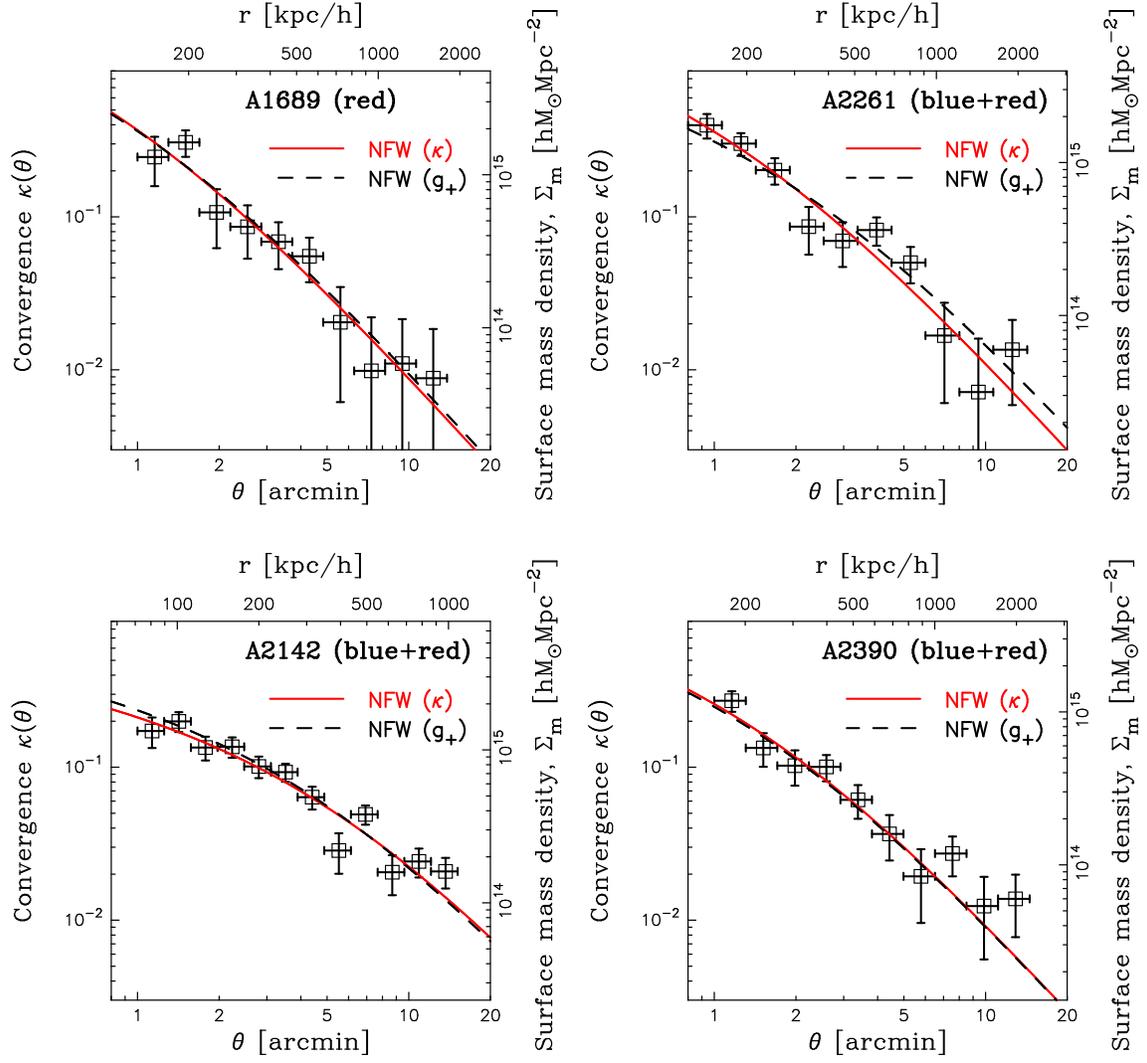}
 \end{center}
\caption{
Model-independent radial profiles
of the lensing convergence
 $\kappa(\theta)=\Sigma_m(\theta)/\Sigma_{\rm crit}$ for 
the four clusters derived from a variant of the non-linear aperture-mass
 densitometry. For each cluster, the best-fitting NFW model for the
 $\kappa$ profile is shown with a solid line.  The dashed curve shows the
 best-fitting NFW model for the $g_+$ profile in Figure
 \ref{fig:gprof}. 
\label{fig:kprof}
}
\end{figure}

Our joint analysis of high quality Subaru 
weak lensing and AMiBA SZE observations
allows for a detailed study of individual clusters.
The cluster A2142 shows complex mass substructure 
\citep{Okabe&Umetsu08},  
and displays a shallower density profile with $c_{\rm vir}\sim 5$,
consistent with detailed X-ray observations which imply
recent interaction. Due to its low-$z$ and low $c_{\rm vir}$,
the curvature in the lensing profiles is highly pronounced, so that a
SIS profile for A2142 is strongly ruled out by the Subaru distortion
data alone (\S\ref{subsubsec:gt}). 
For this cluster, our AMiBA SZE map shows an extended
structure in the ICM distribution, elongated along the
northwest-southeast direction. 
This direction of elongation in the SZE halo is in good
agreement with the cometary X-ray appearance seen by Chandra
\citep{2000ApJ...541..542M,Okabe&Umetsu08}. 
In addition, 
an extended structure showing some excess SZE can be seen 
in the northwest region of the cluster. 
A joint
weak-lensing, optical-photometric, and X-ray analysis 
\citep{Okabe&Umetsu08} 
revealed northwest mass substructure in this SZE excess region,
located ahead of 
the northwest edge of the central gas core seen in X-rays.
The northwest mass  
substructure is also seen in our weak lensing mass map (Figure
\ref{fig:maps}) based on the much improved color 
selection for the background sample.
A slight excess
of cluster sequence galaxies associated with the northwest
substructure is also found in \cite{Okabe&Umetsu08}, while no X-ray
counterpart is seen in the Chandra and
XMM-Newton images \citep{Okabe&Umetsu08}. 
Good consistency found between the SZE and weak lensing
maps is encouraging, and may suggest that the northwest excess SZE is a
pressure increase in the ICM associated with the moving northwest
substructure. 
Clearly further improvements in both sensitivity and
resolution are needed if SZE data are to attain a significant
detection of the 
excess structure in the northwest region.
Nonetheless,
this demonstrates the potential of SZE observations as a powerful
tool for measuring the distribution of ICM in cluster outskirts
where the X-ray emission measure ($\propto n_e^2$) is 
rapidly decreasing.
This also demonstrates the potential of AMiBA, and 
the power of multiwavelength cluster analysis 
for probing the distribution of mass and baryons in
clusters. 
A further detailed
multiwavelength analysis of A2142 will be of great importance for
further understanding of the cluster merger dynamics and associated
physical processes of the intracluster gas.

For A2390 we obtain a highly elliptical mass 
distribution at all radii from both weak and strong lensing
\citep{1998ApJ...499L.115F}.
The elliptical mass distribution agrees well with the shape seen by AMiBA in 
the thermal SZE (Figures \ref{fig:maps} and \ref{fig:eplot}).
Our joint lensing, SZE, and X-ray modeling  leads to a relatively high
gas mass fraction for this cluster, $f_{\rm gas,500}\sim 0.15$
for the NFW-consistent case, which is in good agreement with careful
X-ray work by \citet{2006ApJ...640..691V}, 
$f_{\rm gas,500}=(0.141\pm 0.009) h_{72}^{-3/2}$.

We have refined for A1689
the statistical constraints on the NFW mass model of
\citet{UB2008}, 
with our improved color selection for the red background sample, where
all possible lensing measurements are combined to achieve the maximum
possible lensing precision,
 $M_{\rm vir}=(1.55^{+0.13}_{-0.12})\times 10^{15} M_\odot h^{-1}$ 
and $c_{\rm vir}= 12.3^{+0.9}_{-0.8}$ (quoted are statistical errors at
68$\%$ confidence level), confirming again the high concentration found
by detailed lensing work
\citep{BTU+05,2006MNRAS.372.1425H,2007ApJ...668..643L,UB2008}. 
The AMiBA SZE measurements at 94$\,$GHz support the compact structure in
the ICM distribution for this cluster (Figure \ref{fig:maps}).
Recently, good consistency was found 
between high-quality multiwavelength datasets available for this cluster
\citep{2008MNRAS.386.1092L,Lemze_spec}. 
\citet{2008MNRAS.386.1092L} performed a joint analysis of
Chandra X-ray, ACS strong lensing, and Subaru weak
lensing measurements, and derived an improved mass profile in a
model-independent way. Their NFW fit to the derived mass profile yields
a virial mass of 
$M_{\rm vir}=(1.58\pm 0.15)\times 10^{15}M_\odot h^{-1}$ and 
a high concentration of $c_{\rm vir}=12.2^{+0.9}_{-1.0}$, both of which
are in excellent agreement with our full lensing constraints.
More recently, \citet{Lemze_spec} further extended their multiwavelength
analysis to combine their 
X-ray/lensing measurements with two dynamical datasets from
VLT/VIRMOS spectroscopy and Subaru/Suprime-Cam imaging.
Their joint lensing, X-ray, and dynamical analysis provides a tight
constraint on the cluster virial mass: $1.5 < M_{\rm
vir}/(10^{15}M_\odot h^{-1}) < 1.65$. Their purely dynamical analysis
constrains the concentration parameter to be $c_{\rm vir} > 9.8$
for A1689, in agreement with our independent lensing analysis and the
joint X-ray/lensing analysis of \citet{2008MNRAS.386.1092L}.
We remark that 
NFW fits to the Subaru outer profiles alone give consistent
but somewhat higher concentrations, $c_{\rm vir}\sim 15$ (Table
\ref{tab:model}; see also Umetsu \& Broadhurst 2008 and Broadhurst et
al. 2008). This slight discrepancy can be explained by  
the mass density slope at large radii ($\theta\simgt 5'$) for A1689 being
slightly steeper than the NFW profile where the asymptotic decline tends
to $\rho_{\rm NFW}(r)\propto r^{-3}$ 
 \citep[see][]{BTU+05,Medezinski+07,2008MNRAS.386.1092L,UB2008,Lemze_spec}.
Recent detailed modeling by \citet{Saxton+Wu2008} suggests such a
steeper outer density profile in stationary, self-gravitating halos
composed of adiabatic DM and radiative gas components.
For accurate measurements of the outermost lensing profile, a wider
optical/near-infrared wavelength coverage is required to improve the
contamination-free selection of background galaxies, including blue
background galaxies, behind this rich cluster.

\begin{figure}[!htb]
 \begin{center}
    \includegraphics[width=140mm]{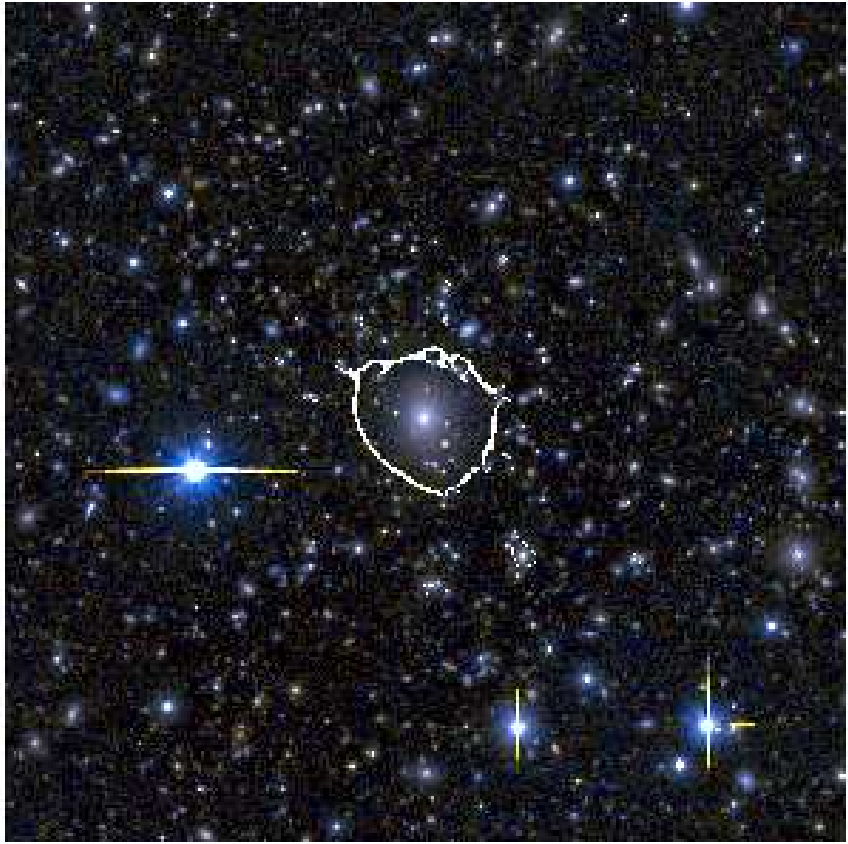}
 \end{center}
\caption{
\label{fig:a2261}
Subaru
 $V+R_{\rm c}$ pseudo-color image of the central $6.7'\times 6.7'$
 ($2{\rm K}\times 2{\rm K}$ pixels) region of
 the cluster A2261 at $z_d=0.226$. Overlaid is the tangential critical
 curve predicted 
 for a background source at $z_s\sim 1.5$ based on strong lensing
 modeling of multiply-lensed images and tangential arcs registered in
 deep Subaru $VR_{\rm c}$ and CFHT/WIRCam $JHK_S$
 images. 
 The effective radius of the tangential  
 critical curve defines the Einstein
 radius, $\theta_{\rm E}\approx 40'$ at $z_s\sim 1.5$. 
}
\end{figure}

\begin{figure}[!htb]
 \begin{center}
    \includegraphics[width=160mm,angle=0]{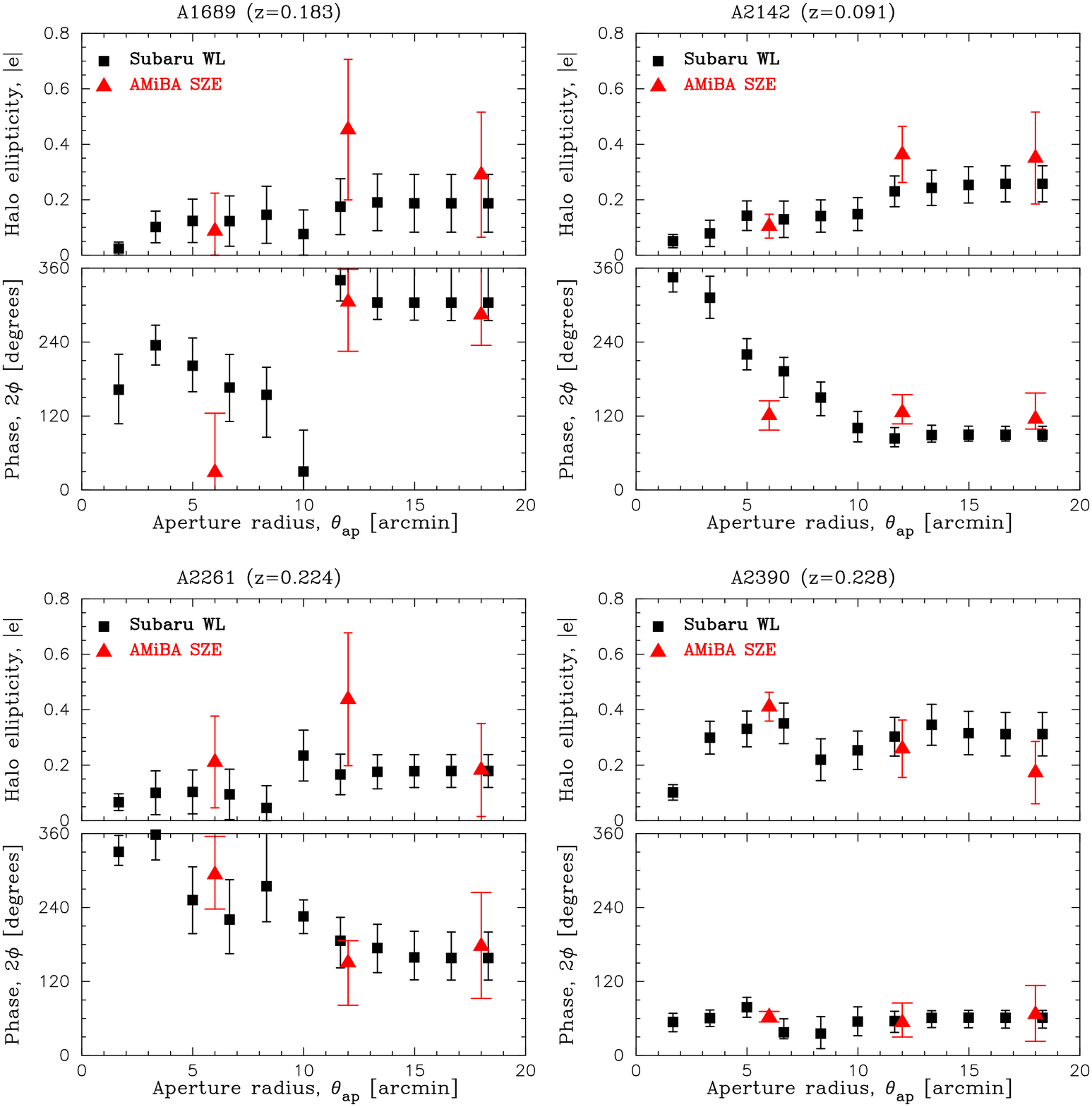}
 \end{center}
\caption{
\label{fig:eplot}
Cluster ellipticity and orientation 
profiles on mass and ICM structure
as a function of aperture radius $\theta_{\rm ap}$,
measured  from 
the Subaru weak lensing and AMiBA SZE maps shown in Figure
 \ref{fig:maps}.
For each cluster, the top panel shows the halo ellipticity profile
$|e^{\rm halo}|(\theta_{\rm ap})$,
and the bottom panel shows the orientation profile $2\phi(\theta_{\rm
 ap})$, where $\phi^{\rm halo}$ represents the position angle of the
 major axis as measured from weighted quadrupole shape moments. 
} 
\end{figure}

\begin{figure}[!htb]
 \begin{center}
    \includegraphics[width=160mm,angle=0]{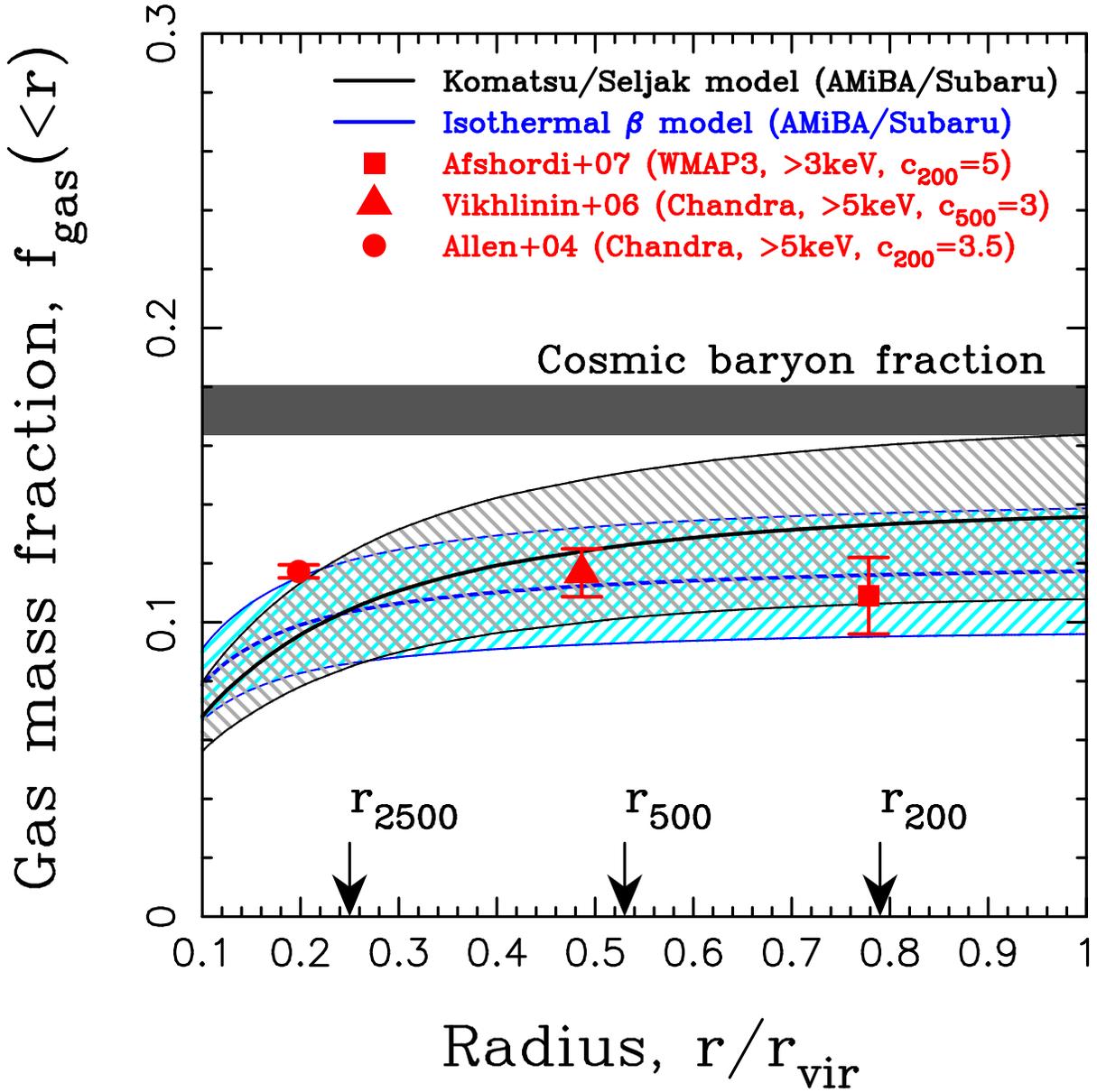}
 \end{center}
\caption{
\label{fig:fgas}
Gas mass fraction profiles $\langle f_{\rm gas}(<r)\rangle = \langle
 M_{\rm gas}(<r)/M_{\rm tot}(<r)\rangle$ averaged over the
 sample of four hot ($T_X>8\, {\rm keV}$) clusters (A1689, A2142, A2261,
 A2390) obtained from joint AMiBA SZE and Subaru weak lensing observations,
 shown for the NFW-consistent Komatsu \& Seljak 2001 model ({\it
 black}) and the isothermal $\beta$ model with $\beta=2/3$ ({\it
 blue}), along with published results ({\it square}, {\it triangle}, and
 {\it circle}) from other X-ray and SZE observations. The isothermal
 results exclude the cluster A2142 (see \S \ref{subsec:fg}).
 For each model, the {\it cross-hatched} region represents
 $1\sigma$ uncertainties for
 the weighted mean at each radius point, including both the
 statistical measurement uncertainties and cluster-to-cluster variance.  
 The black horizontal bar shows the constraints on the cosmic baryon
 fraction from the WMAP 5-year data.
}
\end{figure}

Our Subaru observations have established that A2261 
is very similar to A1689 in terms of both weak and strong lensing
properties:
Our preliminary strong lens modeling 
reveals many tangential arcs and multiply-lensed images around A2261,  
with an effective Einstein radius $\theta_{\rm E}\sim 40\arcsec$ at
$z\sim 1.5$ (Figure \ref{fig:a2261}),  
which, when combined with our weak lensing measurements, implies a 
mass profile well fitted by an NFW model with a concentration
$c_{\rm vir}\sim 10$, similar to A1689
\citep{UB2008},
and considerably higher than theoretically expected for the 
standard $\Lambda$CDM model,
where $c_{\rm vir}\sim 5$ is predicted for the most massive
relaxed clusters with $M_{\rm vir}\simgt 10^{15}M_\odot$ 
\citep{2001MNRAS.321..559B,2007MNRAS.381.1450N,Duffy+2008}.

Such a high concentration is also seen in several other
well-studied massive clusters from careful lensing work
\citep{2003A&A...403...11G,
2003ApJ...598..804K,
BTU+05,2007ApJ...668..643L,2008MNRAS.386.1092L,BUM+2008}.    
The orientation bias due to halo triaxiality can potentially affect the
projected lensing measurements, and hence the lensing-based
concentration measurement \citep[e.g.,][]{2005ApJ...632..841O}.
A statistical bias in favor of prolate structure pointed to the observer
is unavoidable at some level, as this orientation boosts the projected
surface mass density and hence the lensing signal. 
In the context of the $\Lambda$CDM model, this leads to an increase of
$\sim 18\%$ in the mean value of the lensing-based concentrations 
\citep{2007ApJ...654..714H}. A larger bias of $\sim 30$ up to
50$\%$ is expected for CDM halos
selected by the presence of large gravitational arcs
\citep{2007ApJ...654..714H,Oguri+Blandford}.
Our cluster sample is identified by their being X-ray/SZE strong, with
the added requirement of the availability of high-quality
multi-band Subaru/Suprime-Cam imaging (see \S\ref{subsec:targets}).
Hence, it is unlikely that the four clusters are all particularly
triaxial with long axes pointing to the observer. 
Indeed, in the context of $\Lambda$CDM,
the highly elliptical mass distribution of A2390 would
suggest that its major axis is 
not far from the sky plane,
and that its true concentration is higher than the projected
measurement $c_{\rm vir}\simeq 7$.

A chance projection of structure along the line-of-sight
may also influence the lensing-based cluster parameter determination.
It can locally boost the surface mass density, and hence can
affect in a non-local manner (see eq. [\ref{eq:gt2kappa}])
the tangential distortion measurement that is sensitive to the
total interior mass, 
if this physically
unassociated mass structure is contained within the measurement radius.
For the determination of the NFW concentration parameter,
it can lead to either an under or over-estimate of the concentration
depending on the apparent position of the projected structure with
respect to the cluster center.
When the projected structure is well isolated from the cluster center,
one way to overcome this problem is to utilize the convergence profile to
examine the cluster mass profile, by locally masking out the
contribution of known foreground/background structure (see
\S\ref{subsubsec:kappa} for the case of A2261).

The ongoing upgrade of AMiBA to 13-elements with 1.2m antennas
will improve its spatial resolution and dynamic range, 
and the 13-element AMiBA (AMiBA13) will be sensitive to structures on
scales down to 
$2\arcmin$, matching the angular scales probed by ground-based weak
lensing observations \citep{Umetsu2004_MPLA}.
For our initial target clusters, joint constraints with AMiBA7 and
AMiBA13 data will complement the baseline coverage, which will further
improve our multiwavelength analysis of the relation between 
mass and hot baryons in  the clusters.
A joint analysis of complementary high-resolution
lensing, SZE, and X-ray observations will be of great interest to
address the issue of halo triaxiality and further improve the
constraints on cluster density profiles \citep{2007MNRAS.380.1207S}.
The AMiBA upgrade will also make the instrument faster by a factor of $\sim
60$ in pointed observations. 
Our constraints can be further improved in the
near future by observing a larger sample with AMiBA13.
A detailed comparison between X-ray based and SZE/weak lensing-based gas
fraction measurements 
will enable us to test the degree of clumpiness 
($\langle n_e^2\rangle/\langle n_e\rangle^2$) 
and of hydrostatic balance in hot cluster gas.
The high angular resolution ($2'$) of AMiBA13 
combined with dynamically-improved imaging capabilities will allow
for direct tests of the gas pressure profile in deep single pointed
observations
\citep{Molnar_AMiBA}.


\acknowledgments
We thank the anonymous referee for providing useful comments.
We are grateful to N. Okabe, M. Takada, and Y. Rephaeli
for valuable discussions.
We thank the Ministry of Education, the National Science Council, and
the Academia Sinica for their support 
of this project. We thank the Smithsonian Astrophysical Observatory for
hosting the AMiBA project staff at
the SMA Hilo Base Facility. We thank the NOAA for
locating the AMiBA project on their site on Mauna Loa.
We thank the Hawaiian people for allowing astronomers
to work on their mountains in order to study the Universe.
We thank all the members of the AMiBA team for their hard work.
The work is partially supported by the National Science Council of Taiwan
under the grant NSC95-2112-M-001-074-MY2.
Support from the STFC for MB is also acknowledged.



\begin{thebibliography}{115}
\expandafter\ifx\csname natexlab\endcsname\relax\def\natexlab#1{#1}\fi

\bibitem[{{Afshordi} {et~al.}(2007){Afshordi}, {Lin}, {Nagai}, \&
  {Sanderson}}]{Afshordi+2007_WMAP3}
{Afshordi}, N., {Lin}, Y.-T., {Nagai}, D., \& {Sanderson}, A.~J.~R. 2007,
  \mnras, 378, 293

\bibitem[{{Allen} {et~al.}(2001){Allen}, {Ettori}, \&
  {Fabian}}]{2001MNRAS.324..877A}
{Allen}, S.~W., {Ettori}, S., \& {Fabian}, A.~C. 2001, \mnras, 324, 877

\bibitem[{{Allen} {et~al.}(2008){Allen}, {Rapetti}, {Schmidt}, {Ebeling},
  {Morris}, \& {Fabian}}]{Allen+2008_fgas}
{Allen}, S.~W., {Rapetti}, D.~A., {Schmidt}, R.~W., {Ebeling}, H., {Morris},
  R.~G., \& {Fabian}, A.~C. 2008, \mnras, 383, 879

\bibitem[{{Allen} {et~al.}(2004){Allen}, {Schmidt}, {Ebeling}, {Fabian}, \&
  {van Speybroeck}}]{2004MNRAS.353..457A}
{Allen}, S.~W., {Schmidt}, R.~W., {Ebeling}, H., {Fabian}, A.~C., \& {van
  Speybroeck}, L. 2004, \mnras, 353, 457

\bibitem[{{Allen} {et~al.}(2002){Allen}, {Schmidt}, \&
  {Fabian}}]{2002MNRAS.334L..11A}
{Allen}, S.~W., {Schmidt}, R.~W., \& {Fabian}, A.~C. 2002, \mnras, 334, L11

\bibitem[{{Atrio-Barandela} {et~al.}(2008){Atrio-Barandela}, {Kashlinsky},
  {Kocevski}, \& {Ebeling}}]{Atrio+2008_WMAP3}
{Atrio-Barandela}, F., {Kashlinsky}, A., {Kocevski}, D., \& {Ebeling}, H. 2008,
  \apjl, 675, L57

\bibitem[{{Bartelmann}(1996)}]{1996A&A...313..697B}
{Bartelmann}, M. 1996, \aap, 313, 697

\bibitem[{{Bartelmann} {et~al.}(1996){Bartelmann}, {Narayan}, {Seitz}, \&
  {Schneider}}]{1996ApJ...464L.115B}
{Bartelmann}, M., {Narayan}, R., {Seitz}, S., \& {Schneider}, P. 1996, \apjl,
  464, L115+

\bibitem[{{Bartelmann} \& {Schneider}(2001)}]{BS2001}
{Bartelmann}, M., \& {Schneider}, P. 2001, \physrep, 340, 291

\bibitem[{{Benson} {et~al.}(2004){Benson}, {Church}, {Ade}, {Bock}, {Ganga},
  {Henson}, \& {Thompson}}]{Benson2004_SuZIE}
{Benson}, B.~A., {Church}, S.~E., {Ade}, P.~A.~R., {Bock}, J.~J., {Ganga},
  K.~M., {Henson}, C.~N., \& {Thompson}, K.~L. 2004, \apj, 617, 829

\bibitem[{{Bertin} \& {Arnouts}(1996)}]{1996A&AS..117..393B}
{Bertin}, E., \& {Arnouts}, S. 1996, \aaps, 117, 393

\bibitem[{{Bialek} {et~al.}(2001){Bialek}, {Evrard}, \& {Mohr}}]{Bialek+2001}
{Bialek}, J.~J., {Evrard}, A.~E., \& {Mohr}, J.~J. 2001, \apj, 555, 597

\bibitem[{{Birkinshaw}(1999)}]{1999PhR...310...97B}
{Birkinshaw}, M. 1999, \physrep, 310, 97

\bibitem[{{Boehringer} {et~al.}(1998){Boehringer}, {Tanaka}, {Mushotzky},
  {Ikebe}, \& {Hattori}}]{Boehringer+1998_A2390}
{Boehringer}, H., {Tanaka}, Y., {Mushotzky}, R.~F., {Ikebe}, Y., \& {Hattori},
  M. 1998, \aap, 334, 789

\bibitem[{{Broadhurst} {et~al.}(2005{\natexlab{a}}){Broadhurst},
  {Ben{\'{\i}}tez}, {Coe}, {Sharon}, {Zekser}, {White}, {Ford}, {Bouwens},
  {Blakeslee}, {Clampin}, {Cross}, {Franx}, {Frye}, {Hartig}, {Illingworth},
  {Infante}, {Menanteau}, {Meurer}, {Postman}, {Ardila}, {Bartko}, {Brown},
  {Burrows}, {Cheng}, {Feldman}, {Golimowski}, {Goto}, {Gronwall}, {Herranz},
  {Holden}, {Homeier}, {Krist}, {Lesser}, {Martel}, {Miley}, {Rosati},
  {Sirianni}, {Sparks}, {Steindling}, {Tran}, {Tsvetanov}, \&
  {Zheng}}]{2005ApJ...621...53B}
{Broadhurst}, T., {Ben{\'{\i}}tez}, N., {Coe}, D., {Sharon}, K., {Zekser}, K.,
  {White}, R., {Ford}, H., {Bouwens}, R., {Blakeslee}, J., {Clampin}, M.,
  {Cross}, N., {Franx}, M., {Frye}, B., {Hartig}, G., {Illingworth}, G.,
  {Infante}, L., {Menanteau}, F., {Meurer}, G., {Postman}, M., {Ardila}, D.~R.,
  {Bartko}, F., {Brown}, R.~A., {Burrows}, C.~J., {Cheng}, E.~S., {Feldman},
  P.~D., {Golimowski}, D.~A., {Goto}, T., {Gronwall}, C., {Herranz}, D.,
  {Holden}, B., {Homeier}, N., {Krist}, J.~E., {Lesser}, M.~P., {Martel},
  A.~R., {Miley}, G.~K., {Rosati}, P., {Sirianni}, M., {Sparks}, W.~B.,
  {Steindling}, S., {Tran}, H.~D., {Tsvetanov}, Z.~I., \& {Zheng}, W.
  2005{\natexlab{a}}, \apj, 621, 53

\bibitem[{{Broadhurst} {et~al.}(2005{\natexlab{b}}){Broadhurst}, {Takada},
  {Umetsu}, {Kong}, {Arimoto}, {Chiba}, \& {Futamase}}]{BTU+05}
{Broadhurst}, T., {Takada}, M., {Umetsu}, K., {Kong}, X., {Arimoto}, N.,
  {Chiba}, M., \& {Futamase}, T. 2005{\natexlab{b}}, \apjl, 619, L143

\bibitem[{{Broadhurst} {et~al.}(2008){Broadhurst}, {Umetsu}, {Medezinski},
  {Oguri}, \& {Rephaeli}}]{BUM+2008}
{Broadhurst}, T., {Umetsu}, K., {Medezinski}, E., {Oguri}, M., \& {Rephaeli},
  Y. 2008, \apjl, 685, L9
 
 
\bibitem[{{Biviano} \& {Salucci}(2006)}]{Biviano+Salucci2006}
{Biviano}, A. \& {Salucci}, P. 2006, \aap, 452, 75 

\bibitem[{{Bullock} {et~al.}(2001){Bullock}, {Kolatt}, {Sigad}, {Somerville},
  {Kravtsov}, {Klypin}, {Primack}, \& {Dekel}}]{2001MNRAS.321..559B}
{Bullock}, J.~S., {Kolatt}, T.~S., {Sigad}, Y., {Somerville}, R.~S.,
  {Kravtsov}, A.~V., {Klypin}, A.~A., {Primack}, J.~R., \& {Dekel}, A. 2001,
  \mnras, 321, 559

\bibitem[{{Capak} {et~al.}(2007){Capak}, {Aussel}, {Ajiki}, {McCracken},
  {Mobasher}, {Scoville}, {Shopbell}, {Taniguchi}, {Thompson}, {Tribiano},
  {Sasaki}, {Blain}, {Brusa}, {Carilli}, {Comastri}, {Carollo}, {Cassata},
  {Colbert}, {Ellis}, {Elvis}, {Giavalisco}, {Green}, {Guzzo}, {Hasinger},
  {Ilbert}, {Impey}, {Jahnke}, {Kartaltepe}, {Kneib}, {Koda}, {Koekemoer},
  {Komiyama}, {Leauthaud}, {Lefevre}, {Lilly}, {Liu}, {Massey}, {Miyazaki},
  {Murayama}, {Nagao}, {Peacock}, {Pickles}, {Porciani}, {Renzini}, {Rhodes},
  {Rich}, {Salvato}, {Sanders}, {Scarlata}, {Schiminovich}, {Schinnerer},
  {Scodeggio}, {Sheth}, {Shioya}, {Tasca}, {Taylor}, {Yan}, \&
  {Zamorani}}]{Capak+07_COSMOS}
{Capak}, P., {Aussel}, H., {Ajiki}, M., {McCracken}, H.~J., {Mobasher}, B.,
  {Scoville}, N., {Shopbell}, P., {Taniguchi}, Y., {Thompson}, D., {Tribiano},
  S., {Sasaki}, S., {Blain}, A.~W., {Brusa}, M., {Carilli}, C., {Comastri}, A.,
  {Carollo}, C.~M., {Cassata}, P., {Colbert}, J., {Ellis}, R.~S., {Elvis}, M.,
  {Giavalisco}, M., {Green}, W., {Guzzo}, L., {Hasinger}, G., {Ilbert}, O.,
  {Impey}, C., {Jahnke}, K., {Kartaltepe}, J., {Kneib}, J.-P., {Koda}, J.,
  {Koekemoer}, A., {Komiyama}, Y., {Leauthaud}, A., {Lefevre}, O., {Lilly}, S.,
  {Liu}, C., {Massey}, R., {Miyazaki}, S., {Murayama}, T., {Nagao}, T.,
  {Peacock}, J.~A., {Pickles}, A., {Porciani}, C., {Renzini}, A., {Rhodes}, J.,
  {Rich}, M., {Salvato}, M., {Sanders}, D.~B., {Scarlata}, C., {Schiminovich},
  D., {Schinnerer}, E., {Scodeggio}, M., {Sheth}, K., {Shioya}, Y., {Tasca},
  L.~A.~M., {Taylor}, J.~E., {Yan}, L., \& {Zamorani}, G. 2007, \apjs, 172, 99

\bibitem[{{Capak} {et~al.}(2004){Capak}, {Cowie}, {Hu}, {Barger}, {Dickinson},
  {Fernandez}, {Giavalisco}, {Komiyama}, {Kretchmer}, {McNally}, {Miyazaki},
  {Okamura}, \& {Stern}}]{Capak+04_HDFN}
{Capak}, P., {Cowie}, L.~L., {Hu}, E.~M., {Barger}, A.~J., {Dickinson}, M.,
  {Fernandez}, E., {Giavalisco}, M., {Komiyama}, Y., {Kretchmer}, C.,
  {McNally}, C., {Miyazaki}, S., {Okamura}, S., \& {Stern}, D. 2004, \aj, 127,
  180

\bibitem[{{Carlstrom} {et~al.}(2002){Carlstrom}, {Holder}, \&
  {Reese}}]{Carlstrom+2002_SZE}
{Carlstrom}, J.~E., {Holder}, G.~P., \& {Reese}, E.~D. 2002, \araa, 40, 643

\bibitem[{{Challinor} \& {Lasenby}(1998)}]{Challinor+Lasenby1998_SZE}
{Challinor}, A., \& {Lasenby}, A. 1998, \apj, 499, 1

\bibitem[{{Chen} {et~al.}(2008){Chen}, {AMiBA 1}, \& {AMiBA 2}}]{Chen_AMiBA}
{Chen}, M.-T et al.. 2008, \apj, submitted

\bibitem[{{Clowe} {et~al.}(2006){Clowe}, {Brada{\v c}}, {Gonzalez},
  {Markevitch}, {Randall}, {Jones}, \& {Zaritsky}}]{2006ApJ...648L.109C}
{Clowe}, D., {Brada{\v c}}, M., {Gonzalez}, A.~H., {Markevitch}, M., {Randall},
  S.~W., {Jones}, C., \& {Zaritsky}, D. 2006, \apjl, 648, L109

\bibitem[{{Clowe} {et~al.}(2000){Clowe}, {Luppino}, {Kaiser}, \&
  {Gioia}}]{2000ApJ...539..540C}
{Clowe}, D., {Luppino}, G.~A., {Kaiser}, N., \& {Gioia}, I.~M. 2000, \apj, 539,
  540

\bibitem[{{Corless} \& {King}(2007)}]{2007MNRAS.380..149C}
{Corless}, V.~L., \& {King}, L.~J. 2007, \mnras, 380, 149

\bibitem[{{Crain} {et~al.}(2007){Crain}, {Eke}, {Frenk}, {Jenkins}, {McCarthy},
  {Navarro}, \& {Pearce}}]{Crain+2007}
{Crain}, R.~A., {Eke}, V.~R., {Frenk}, C.~S., {Jenkins}, A., {McCarthy}, I.~G.,
  {Navarro}, J.~F., \& {Pearce}, F.~R. 2007, \mnras, 377, 41

\bibitem[{{Crittenden} {et~al.}(2002){Crittenden}, {Natarajan}, {Pen}, \&
  {Theuns}}]{2002ApJ...568...20C}
{Crittenden}, R.~G., {Natarajan}, P., {Pen}, U.-L., \& {Theuns}, T. 2002, \apj,
  568, 20


\bibitem[{{Duffy} {et~al.}(2008){Duffy}, {Schaye}, {Kay}, \& {Dalla
  Vecchia}}]{Duffy+2008}
{Duffy}, A.~R., {Schaye}, J., {Kay}, S.~T., \& {Dalla Vecchia}, C. 2008,
  \mnras, 390, L64

\bibitem[{{Dunkley} {et~al.}(2008){Dunkley}, {Komatsu}, {Nolta}, {Spergel},
  {Larson}, {Hinshaw}, {Page}, {Bennett}, {Gold}, {Jarosik}, {Weiland},
  {Halpern}, {Hill}, {Kogut}, {Limon}, {Meyer}, {Tucker}, {Wollack}, \&
  {Wright}}]{Dunkley+2008_WMAP5}
{Dunkley}, J., {Komatsu}, E., {Nolta}, M.~R., {Spergel}, D.~N., {Larson}, D.,
  {Hinshaw}, G., {Page}, L., {Bennett}, C.~L., {Gold}, B., {Jarosik}, N.,
  {Weiland}, J.~L., {Halpern}, M., {Hill}, R.~S., {Kogut}, A., {Limon}, M.,
  {Meyer}, S.~S., {Tucker}, G.~S., {Wollack}, E., \& {Wright}, E.~L. 2008,
  ApJS, submitted (arXiv:0803.0586)

\bibitem[{{Erben} {et~al.}(2001){Erben}, {Van Waerbeke}, {Bertin}, {Mellier},
  \& {Schneider}}]{2001A&A...366..717E}
{Erben}, T., {Van Waerbeke}, L., {Bertin}, E., {Mellier}, Y., \& {Schneider},
  P. 2001, \aap, 366, 717

\bibitem[{{Fahlman} {et~al.}(1994){Fahlman}, {Kaiser}, {Squires}, \&
  {Woods}}]{1994ApJ...437...56F}
{Fahlman}, G., {Kaiser}, N., {Squires}, G., \& {Woods}, D. 1994, \apj, 437, 56

\bibitem[{{Frye} \& {Broadhurst}(1998)}]{1998ApJ...499L.115F}
{Frye}, B., \& {Broadhurst}, T. 1998, \apjl, 499, L115+

\bibitem[{{Fukugita} {et~al.}(1998){Fukugita}, {Hogan}, \&
  {Peebles}}]{Fukugita+1998}
{Fukugita}, M., {Hogan}, C.~J., \& {Peebles}, P.~J.~E. 1998, \apj, 503, 518

\bibitem[{{Gavazzi} {et~al.}(2003){Gavazzi}, {Fort}, {Mellier}, {Pell{\'o}}, \&
  {Dantel-Fort}}]{2003A&A...403...11G}
{Gavazzi}, R., {Fort}, B., {Mellier}, Y., {Pell{\'o}}, R., \& {Dantel-Fort}, M.
  2003, \aap, 403, 11

\bibitem[{{Gonzalez} {et~al.}(2005){Gonzalez}, {Zabludoff}, \&
  {Zaritsky}}]{Gonzalez+2005_ICL}
{Gonzalez}, A.~H., {Zabludoff}, A.~I., \& {Zaritsky}, D. 2005, \apj, 618, 195

\bibitem[{{Grego} {et~al.}(2001{\natexlab{a}}){Grego}, {Carlstrom}, {Reese},
  {Holder}, {Holzapfel}, {Joy}, {Mohr}, \& {Patel}}]{Grego2001_SZE}
{Grego}, L., {Carlstrom}, J.~E., {Reese}, E.~D., {Holder}, G.~P., {Holzapfel},
  W.~L., {Joy}, M.~K., {Mohr}, J.~J., \& {Patel}, S. 2001{\natexlab{a}}, \apj,
  552, 2

\bibitem[{{Grego} {et~al.}(2001{\natexlab{b}}){Grego}, {Carlstrom}, {Reese},
  {Holder}, {Holzapfel}, {Joy}, {Mohr}, \& {Patel}}]{Grego+2001_SZE}
---. 2001{\natexlab{b}}, \apj, 552, 2

\bibitem[{{Halkola} {et~al.}(2006){Halkola}, {Seitz}, \&
  {Pannella}}]{2006MNRAS.372.1425H}
{Halkola}, A., {Seitz}, S., \& {Pannella}, M. 2006, \mnras, 372, 1425

\bibitem[{{Halverson} {et~al.}(2008){Halverson}, {Lanting}, {Ade}, {Basu},
  {Bender}, {Benson}, {Bertoldi}, {Cho}, {Chon}, {Clarke}, {Dobbs}, {Ferrusca},
  {Guesten}, {Holzapfel}, {Kovacs}, {Kennedy}, {Kermish}, {Kneissl}, {Lee},
  {Lueker}, {Mehl}, {Menten}, {Muders}, {Nord}, {Pacaud}, {Plagge},
  {Reichardt}, {Richards}, {Schaaf}, {Schilke}, {Schuller}, {Schwan},
  {Spieler}, {Tucker}, {Weiss}, \& {Zahn}}]{Halverson2008_APEX}
{Halverson}, N.~W., {Lanting}, T., {Ade}, P.~A.~R., {Basu}, K., {Bender},
  A.~N., {Benson}, B.~A., {Bertoldi}, F., {Cho}, H.~., {Chon}, G., {Clarke},
  J., {Dobbs}, M., {Ferrusca}, D., {Guesten}, R., {Holzapfel}, W.~L., {Kovacs},
  A., {Kennedy}, J., {Kermish}, Z., {Kneissl}, R., {Lee}, A.~T., {Lueker}, M.,
  {Mehl}, J., {Menten}, K.~M., {Muders}, D., {Nord}, M., {Pacaud}, F.,
  {Plagge}, T., {Reichardt}, C., {Richards}, P.~L., {Schaaf}, R., {Schilke},
  P., {Schuller}, F., {Schwan}, D., {Spieler}, H., {Tucker}, C., {Weiss}, A.,
  \& {Zahn}, O. 2008, \apj, submitted (arXiv:0807.4208)

\bibitem[{{Hamana} {et~al.}(2003){Hamana}, {Miyazaki}, {Shimasaku}, {Furusawa},
  {Doi}, {Hamabe}, {Imi}, {Kimura}, {Komiyama}, {Nakata}, {Okada}, {Okamura},
  {Ouchi}, {Sekiguchi}, {Yagi}, \& {Yasuda}}]{2003ApJ...597...98H}
{Hamana}, T., {Miyazaki}, S., {Shimasaku}, K., {Furusawa}, H., {Doi}, M.,
  {Hamabe}, M., {Imi}, K., {Kimura}, M., {Komiyama}, Y., {Nakata}, F., {Okada},
  N., {Okamura}, S., {Ouchi}, M., {Sekiguchi}, M., {Yagi}, M., \& {Yasuda}, N.
  2003, \apj, 597, 98

\bibitem[{{Hattori} {et~al.}(1999){Hattori}, {Kneib}, \&
  {Makino}}]{1999PThPS.133....1H}
{Hattori}, M., {Kneib}, J., \& {Makino}, N. 1999, Progress of Theoretical
  Physics Supplement, 133, 1

\bibitem[{{Hennawi} {et~al.}(2007){Hennawi}, {Dalal}, {Bode}, \&
  {Ostriker}}]{2007ApJ...654..714H}
{Hennawi}, J.~F., {Dalal}, N., {Bode}, P., \& {Ostriker}, J.~P. 2007, \apj,
  654, 714

\bibitem[{{Heymans} {et~al.}(2006){Heymans}, {Van Waerbeke}, {Bacon}, {Berge},
  {Bernstein}, {Bertin}, {Bridle}, {Brown}, {Clowe}, {Dahle}, {Erben}, {Gray},
  {Hetterscheidt}, {Hoekstra}, {Hudelot}, {Jarvis}, {Kuijken}, {Margoniner},
  {Massey}, {Mellier}, {Nakajima}, {Refregier}, {Rhodes}, {Schrabback}, \&
  {Wittman}}]{2006MNRAS.368.1323H}
{Heymans}, C., {Van Waerbeke}, L., {Bacon}, D., {Berge}, J., {Bernstein}, G.,
  {Bertin}, E., {Bridle}, S., {Brown}, M.~L., {Clowe}, D., {Dahle}, H.,
  {Erben}, T., {Gray}, M., {Hetterscheidt}, M., {Hoekstra}, H., {Hudelot}, P.,
  {Jarvis}, M., {Kuijken}, K., {Margoniner}, V., {Massey}, R., {Mellier}, Y.,
  {Nakajima}, R., {Refregier}, A., {Rhodes}, J., {Schrabback}, T., \&
  {Wittman}, D. 2006, \mnras, 368, 1323

\bibitem[{{Ho} {et~al.}(2008){Ho}, {AMiBA 1}, \& {AMiBA 2}}]{Ho_AMiBA}
{Ho}, P.~T.~P. et al. 2008, ApJ, submitted (arXiv:0810.1871)

\bibitem[{{Hoekstra} {et~al.}(1998){Hoekstra}, {Franx}, {Kuijken}, \&
  {Squires}}]{1998ApJ...504..636H}
{Hoekstra}, H., {Franx}, M., {Kuijken}, K., \& {Squires}, G. 1998, \apj, 504,
  636

\bibitem[{{Huang} {et~al.}(2008){Huang}, {AMiBA 1}, \& {AMiBA 2}}]{Huang_AMiBA}
{Huang}, C.-W.~L. et al. 2008, in preparation

\bibitem[{{Itoh} {et~al.}(1998){Itoh}, {Kohyama}, \& {Nozawa}}]{Itoh+1998_SZE}
{Itoh}, N., {Kohyama}, Y., \& {Nozawa}, S. 1998, \apj, 502, 7

\bibitem[{{Jain} {et~al.}(2000){Jain}, {Seljak}, \&
  {White}}]{2000ApJ...530..547J}
{Jain}, B., {Seljak}, U., \& {White}, S. 2000, \apj, 530, 547

\bibitem[{{Jing} \& {Suto}(2002)}]{2002ApJ...574..538J}
{Jing}, Y.~P., \& {Suto}, Y. 2002, \apj, 574, 538

\bibitem[{{Kaiser}(1995)}]{1995ApJ...439L...1K}
{Kaiser}, N. 1995, \apjl, 439, L1

\bibitem[{{Kaiser} \& {Squires}(1993)}]{1993ApJ...404..441K}
{Kaiser}, N., \& {Squires}, G. 1993, \apj, 404, 441

\bibitem[{{Kaiser} {et~al.}(1995){Kaiser}, {Squires}, \&
  {Broadhurst}}]{1995ApJ...449..460K}
{Kaiser}, N., {Squires}, G., \& {Broadhurst}, T. 1995, \apj, 449, 460

\bibitem[{{Kawahara} {et~al.}(2007){Kawahara}, {Suto}, {Kitayama}, {Sasaki},
  {Shimizu}, {Rasia}, \& {Dolag}}]{2007ApJ...659..257K}
{Kawahara}, H., {Suto}, Y., {Kitayama}, T., {Sasaki}, S., {Shimizu}, M.,
  {Rasia}, E., \& {Dolag}, K. 2007, \apj, 659, 257

\bibitem[{{Kneib} {et~al.}(2003){Kneib}, {Hudelot}, {Ellis}, {Treu}, {Smith},
  {Marshall}, {Czoske}, {Smail}, \& {Natarajan}}]{2003ApJ...598..804K}
{Kneib}, J.-P., {Hudelot}, P., {Ellis}, R.~S., {Treu}, T., {Smith}, G.~P.,
  {Marshall}, P., {Czoske}, O., {Smail}, I., \& {Natarajan}, P. 2003, \apj,
  598, 804

\bibitem[{{Kneissl} {et~al.}(2001){Kneissl}, {Jones}, {Saunders}, {Eke},
  {Lasenby}, {Grainge}, \& {Cotter}}]{AMI2001}
{Kneissl}, R., {Jones}, M.~E., {Saunders}, R., {Eke}, V.~R., {Lasenby}, A.~N.,
  {Grainge}, K., \& {Cotter}, G. 2001, \mnras, 328, 783

\bibitem[{{Koch} {et~al.}(2008{\natexlab{a}}){Koch}, {AMiBA 1}, \& {AMiBA
  2}}]{Koch_mount}
{Koch}, P. et al. 2008{\natexlab{a}}, \apjs, submitted

\bibitem[{{Koch} {et~al.}(2008{\natexlab{b}}){Koch}, {AMiBA 1}, \& {AMiBA
  2}}]{Koch_AMiBA}
---. 2008{\natexlab{b}}, in preparation

\bibitem[{{Koch} {et~al.}(2006){Koch}, {Raffin}, {Proty Wu}, \& {et
  al.}}]{Koch2006_0.6m}
{Koch}, P., {Raffin}, P.~A., {Proty Wu}, J.-H., \& {et al.} 2006, in ESA
  Special Publication, Vol. 626, The European Conference on Antennas and
  Propagation: EuCAP 2006

\bibitem[{{Komatsu} {et~al.}(2008){Komatsu}, {Dunkley}, {Nolta}, {Bennett},
  {Gold}, {Hinshaw}, {Jarosik}, {Larson}, {Limon}, {Page}, {Spergel},
  {Halpern}, {Hill}, {Kogut}, {Meyer}, {Tucker}, {Weiland}, {Wollack}, \&
  {Wright}}]{2008arXiv0803.0547K}
{Komatsu}, E., {Dunkley}, J., {Nolta}, M.~R., {Bennett}, C.~L., {Gold}, B.,
  {Hinshaw}, G., {Jarosik}, N., {Larson}, D., {Limon}, M., {Page}, L.,
  {Spergel}, D.~N., {Halpern}, M., {Hill}, R.~S., {Kogut}, A., {Meyer}, S.~S.,
  {Tucker}, G.~S., {Weiland}, J.~L., {Wollack}, E., \& {Wright}, E.~L. 2008,
  \apjs, submitted (arXiv:0803.0547)

\bibitem[{{Komatsu} \& {Seljak}(2001)}]{2001MNRAS.327.1353K}
{Komatsu}, E., \& {Seljak}, U. 2001, \mnras, 327, 1353

\bibitem[{{Komatsu} \& {Seljak}(2002)}]{2002MNRAS.336.1256K}
---. 2002, \mnras, 336, 1256

\bibitem[{{Kravtsov} {et~al.}(2005){Kravtsov}, {Nagai}, \&
  {Vikhlinin}}]{2005ApJ...625..588K}
{Kravtsov}, A.~V., {Nagai}, D., \& {Vikhlinin}, A.~A. 2005, \apj, 625, 588

\bibitem[{{Lancaster} {et~al.}(2005){Lancaster}, {Genova-Santos}, {Falc{\`o}n},
  {Grainge}, {Guti{\`e}rrez}, {Kneissl}, {Marshall}, {Pooley}, {Rebolo},
  {Rubi{\~n}o-Martin}, {Saunders}, {Waldram}, \& {Watson}}]{Lancaster+05_VSA}
{Lancaster}, K., {Genova-Santos}, R., {Falc{\`o}n}, N., {Grainge}, K.,
  {Guti{\`e}rrez}, C., {Kneissl}, R., {Marshall}, P., {Pooley}, G., {Rebolo},
  R., {Rubi{\~n}o-Martin}, J.-A., {Saunders}, R.~D.~E., {Waldram}, E., \&
  {Watson}, R.~A. 2005, \mnras, 359, 16

\bibitem[{{Lemze} {et~al.}(2008{\natexlab{a}}){Lemze}, {Barkana}, {Broadhurst},
  \& {Rephaeli}}]{2008MNRAS.386.1092L}
{Lemze}, D., {Barkana}, R., {Broadhurst}, T.~J., \& {Rephaeli}, Y.
  2008{\natexlab{a}}, \mnras, 386, 1092

\bibitem[{{Lemze} {et~al.}(2008{\natexlab{b}}){Lemze}, {Broadhurst},
  {Rephaeli}, {Barkana}, {Czoske}, \& {Umetsu}}]{Lemze_spec}
{Lemze}, D., {Broadhurst}, T., {Rephaeli}, Y., {Barkana}, R., {Czoske}, O., \&
  {Umetsu}, K. 2008{\natexlab{b}}, submitted to ApJ (arXiv:0810.3129)

\bibitem[{{Limousin} {et~al.}(2007){Limousin}, {Richard}, {Jullo}, {Kneib},
  {Fort}, {Soucail}, {El{\'{\i}}asd{\'o}ttir}, {Natarajan}, {Ellis}, {Smail},
  {Czoske}, {Smith}, {Hudelot}, {Bardeau}, {Ebeling}, {Egami}, \&
  {Knudsen}}]{2007ApJ...668..643L}
{Limousin}, M., {Richard}, J., {Jullo}, E., {Kneib}, J.-P., {Fort}, B.,
  {Soucail}, G., {El{\'{\i}}asd{\'o}ttir}, {\'A}., {Natarajan}, P., {Ellis},
  R.~S., {Smail}, I., {Czoske}, O., {Smith}, G.~P., {Hudelot}, P., {Bardeau},
  S., {Ebeling}, H., {Egami}, E., \& {Knudsen}, K.~K. 2007, \apj, 668, 643

\bibitem[{{Lin} {et~al.}(2008){Lin}, {AMiBA 1}, \& {AMiBA 2}}]{Lin_AMiBA}
{Lin}, K.-Y. et al., \apj, submitted

\bibitem[{{Liu} {et~al.}(2008){Liu}, {AMiBA 1}, \& {AMiBA 2}}]{Liu_AMiBA}
{Liu}, G.-C. et al., in preparation

\bibitem[{{Mandelbaum} {et~al.}(2008){Mandelbaum}, {Seljak}, \&
  {Hirata}}]{Mandelbaum+2008}
{Mandelbaum}, R., {Seljak}, U., \& {Hirata}, C.~M. 2008, Journal of Cosmology
  and Astro-Particle Physics, 8, 6


\bibitem[{{Markevitch}(1998)}]{1998ApJ...504...27M} 
{Markevitch}, M. 1998, \apj, 504, 27

\bibitem[{{Markevitch} {et~al.}(1998){Markevitch}, {Forman}, {Sarazin}, \&
  {Vikhlinin}}]{Markevitch+1998}
{Markevitch}, M., {Forman}, W.~R., {Sarazin}, C.~L., \& {Vikhlinin}, A. 1998,
  \apj, 503, 77

\bibitem[{{Markevitch} {et~al.}(2002){Markevitch}, {Gonzalez}, {David},
  {Vikhlinin}, {Murray}, {Forman}, {Jones}, \& {Tucker}}]{2002ApJ...567L..27M}
{Markevitch}, M., {Gonzalez}, A.~H., {David}, L., {Vikhlinin}, A., {Murray},
  S., {Forman}, W., {Jones}, C., \& {Tucker}, W. 2002, \apjl, 567, L27

\bibitem[{{Markevitch} {et~al.}(2000){Markevitch}, {Ponman}, {Nulsen}, {Bautz},
  {Burke}, {David}, {Davis}, {Donnelly}, {Forman}, {Jones}, {Kaastra},
  {Kellogg}, {Kim}, {Kolodziejczak}, {Mazzotta}, {Pagliaro}, {Patel}, {Van
  Speybroeck}, {Vikhlinin}, {Vrtilek}, {Wise}, \& {Zhao}}]{2000ApJ...541..542M}
{Markevitch}, M., {Ponman}, T.~J., {Nulsen}, P.~E.~J., {Bautz}, M.~W., {Burke},
  D.~J., {David}, L.~P., {Davis}, D., {Donnelly}, R.~H., {Forman}, W.~R.,
  {Jones}, C., {Kaastra}, J., {Kellogg}, E., {Kim}, D.-W., {Kolodziejczak}, J.,
  {Mazzotta}, P., {Pagliaro}, A., {Patel}, S., {Van Speybroeck}, L.,
  {Vikhlinin}, A., {Vrtilek}, J., {Wise}, M., \& {Zhao}, P. 2000, \apj, 541,
  542

\bibitem[{{Mason} {et~al.}(2001){Mason}, {Myers}, \&
  {Readhead}}]{Mason2001_Hubble}
{Mason}, B.~S., {Myers}, S.~T., \& {Readhead}, A.~C.~S. 2001, \apjl, 555, L11

\bibitem[{{Mason} {et~al.}(2003){Mason}, {Pearson}, {Readhead}, {Shepherd},
  {Sievers}, {Udomprasert}, {Cartwright}, {Farmer}, {Padin}, {Myers}, {Bond},
  {Contaldi}, {Pen}, {Prunet}, {Pogosyan}, {Carlstrom}, {Kovac}, {Leitch},
  {Pryke}, {Halverson}, {Holzapfel}, {Altamirano}, {Bronfman}, {Casassus},
  {May}, \& {Joy}}]{Mason2003_CBI}
{Mason}, B.~S., {Pearson}, T.~J., {Readhead}, A.~C.~S., {Shepherd}, M.~C.,
  {Sievers}, J., {Udomprasert}, P.~S., {Cartwright}, J.~K., {Farmer}, A.~J.,
  {Padin}, S., {Myers}, S.~T., {Bond}, J.~R., {Contaldi}, C.~R., {Pen}, U.,
  {Prunet}, S., {Pogosyan}, D., {Carlstrom}, J.~E., {Kovac}, J., {Leitch},
  E.~M., {Pryke}, C., {Halverson}, N.~W., {Holzapfel}, W.~L., {Altamirano}, P.,
  {Bronfman}, L., {Casassus}, S., {May}, J., \& {Joy}, M. 2003, \apj, 591, 540

\bibitem[{{Maughan} {et~al.}(2008){Maughan}, {Jones}, {Forman}, \& {Van
  Speybroeck}}]{2008ApJS..174..117M}
{Maughan}, B.~J., {Jones}, C., {Forman}, W., \& {Van Speybroeck}, L. 2008,
  \apjs, 174, 117

\bibitem[{{Mazzotta} {et~al.}(2004){Mazzotta}, {Rasia}, {Moscardini}, \&
  {Tormen}}]{Mazzotta+2004}
{Mazzotta}, P., {Rasia}, E., {Moscardini}, L., \& {Tormen}, G. 2004, \mnras,
  354, 10

\bibitem[{{Medezinski} {et~al.}(2007){Medezinski}, {Broadhurst}, {Umetsu},
  {Coe}, {Ben{\'{\i}}tez}, {Ford}, {Rephaeli}, {Arimoto}, \&
  {Kong}}]{Medezinski+07}
{Medezinski}, E., {Broadhurst}, T., {Umetsu}, K., {Coe}, D., {Ben{\'{\i}}tez},
  N., {Ford}, H., {Rephaeli}, Y., {Arimoto}, N., \& {Kong}, X. 2007, \apj, 663,
  717

\bibitem[{{Miyazaki} {et~al.}(2002){Miyazaki}, {Komiyama}, {Sekiguchi},
  {Okamura}, {Doi}, {Furusawa}, {Hamabe}, {Imi}, {Kimura}, {Nakata}, {Okada},
  {Ouchi}, {Shimasaku}, {Yagi}, \& {Yasuda}}]{2002PASJ...54..833M}
{Miyazaki}, S., {Komiyama}, Y., {Sekiguchi}, M., {Okamura}, S., {Doi}, M.,
  {Furusawa}, H., {Hamabe}, M., {Imi}, K., {Kimura}, M., {Nakata}, F., {Okada},
  N., {Ouchi}, M., {Shimasaku}, K., {Yagi}, M., \& {Yasuda}, N. 2002, \pasj,
  54, 833

\bibitem[{{Molnar} {et~al.}(2008){Molnar}, {AMiBA 1}, \& {AMiBA
  2}}]{Molnar_AMiBA}
{Molnar}, S.~M. et al. 2008, in preparation

\bibitem[{{Mroczkowski} {et~al.}(2008){Mroczkowski}, {Bonamente}, {Carlstrom},
  {Culverhouse}, {Greer}, {Hawkins}, {Hennessy}, {Joy}, {Lamb}, {Leitch},
  {Loh}, {Maughan}, {Marrone}, {Miller}, {Nagai}, {Muchovej}, {Pryke}, {Sharp},
  \& {Woody}}]{SZA2008}
{Mroczkowski}, T., {Bonamente}, M., {Carlstrom}, J.~E., {Culverhouse}, T.~L.,
  {Greer}, C., {Hawkins}, D., {Hennessy}, R., {Joy}, M., {Lamb}, J.~W.,
  {Leitch}, E.~M., {Loh}, M., {Maughan}, B., {Marrone}, D.~P., {Miller}, A.,
  {Nagai}, D., {Muchovej}, S., {Pryke}, C., {Sharp}, M., \& {Woody}, D. 2008,
  \apj, submitted (arXiv:0809.5077)

\bibitem[{{Myers} {et~al.}(1997){Myers}, {Baker}, {Readhead}, {Leitch}, \&
  {Herbig}}]{Myers+1997}
{Myers}, S.~T., {Baker}, J.~E., {Readhead}, A.~C.~S., {Leitch}, E.~M., \&
  {Herbig}, T. 1997, \apj, 485, 1

\bibitem[{{Navarro} {et~al.}(1996){Navarro}, {Frenk}, \&
  {White}}]{1996ApJ...462..563N}
{Navarro}, J.~F., {Frenk}, C.~S., \& {White}, S.~D.~M. 1996, \apj, 462, 563

\bibitem[{{Navarro} {et~al.}(1997){Navarro}, {Frenk}, \&
  {White}}]{1997ApJ...490..493N}
---. 1997, \apj, 490, 493

\bibitem[{{Neto} {et~al.}(2007){Neto}, {Gao}, {Bett}, {Cole}, {Navarro},
  {Frenk}, {White}, {Springel}, \& {Jenkins}}]{2007MNRAS.381.1450N}
{Neto}, A.~F., {Gao}, L., {Bett}, P., {Cole}, S., {Navarro}, J.~F., {Frenk},
  C.~S., {White}, S.~D.~M., {Springel}, V., \& {Jenkins}, A. 2007, \mnras, 381,
  1450

\bibitem[{{Nishioka} {et~al.}(2008){Nishioka}, {AMiBA 1}, \& {AMiBA
  2}}]{Nishioka_AMiBA}
{Nishioka}, H. et al. 2008, \apj, accepted (arXiv:0811.1675)

\bibitem[{{Oguri} \& {Blandford}(2008)}]{Oguri+Blandford}
{Oguri}, M., \& {Blandford}, R.~D. 2008, preprint (arXiv:0808.0192O)

\bibitem[{{Oguri} {et~al.}(2005){Oguri}, {Takada}, {Umetsu}, \&
  {Broadhurst}}]{2005ApJ...632..841O}
{Oguri}, M., {Takada}, M., {Umetsu}, K., \& {Broadhurst}, T. 2005, \apj, 632,
  841

\bibitem[{{Okabe} \& {Umetsu}(2008)}]{Okabe&Umetsu08}
{Okabe}, N., \& {Umetsu}, K. 2008, \pasj, 60, 345

\bibitem[{{Padin} {et~al.}(2001){Padin}, {Cartwright}, {Mason}, {Pearson},
  {Readhead}, {Shepherd}, {Sievers}, {Udomprasert}, {Holzapfel}, {Myers},
  {Carlstrom}, {Leitch}, {Joy}, {Bronfman}, \& {May}}]{Padin2001_CBI_1st}
{Padin}, S., {Cartwright}, J.~K., {Mason}, B.~S., {Pearson}, T.~J., {Readhead},
  A.~C.~S., {Shepherd}, M.~C., {Sievers}, J., {Udomprasert}, P.~S.,
  {Holzapfel}, W.~L., {Myers}, S.~T., {Carlstrom}, J.~E., {Leitch}, E.~M.,
  {Joy}, M., {Bronfman}, L., \& {May}, J. 2001, \apjl, 549, L1

\bibitem[{{Padin} {et~al.}(2002){Padin}, {Shepherd}, {Cartwright}, {Keeney},
  {Mason}, {Pearson}, {Readhead}, {Schaal}, {Sievers}, {Udomprasert},
  {Yamasaki}, {Holzapfel}, {Carlstrom}, {Joy}, {Myers}, \&
  {Otarola}}]{Padin2002_CBI}
{Padin}, S., {Shepherd}, M.~C., {Cartwright}, J.~K., {Keeney}, R.~G., {Mason},
  B.~S., {Pearson}, T.~J., {Readhead}, A.~C.~S., {Schaal}, W.~A., {Sievers},
  J., {Udomprasert}, P.~S., {Yamasaki}, J.~K., {Holzapfel}, W.~L., {Carlstrom},
  J.~E., {Joy}, M., {Myers}, S.~T., \& {Otarola}, A. 2002, \pasp, 114, 83

\bibitem[{{Pearson} {et~al.}(2003){Pearson}, {Mason}, {Readhead}, {Shepherd},
  {Sievers}, {Udomprasert}, {Cartwright}, {Farmer}, {Padin}, {Myers}, {Bond},
  {Contaldi}, {Pen}, {Prunet}, {Pogosyan}, {Carlstrom}, {Kovac}, {Leitch},
  {Pryke}, {Halverson}, {Holzapfel}, {Altamirano}, {Bronfman}, {Casassus},
  {May}, \& {Joy}}]{Pearson2003_CBI}
{Pearson}, T.~J., {Mason}, B.~S., {Readhead}, A.~C.~S., {Shepherd}, M.~C.,
  {Sievers}, J.~L., {Udomprasert}, P.~S., {Cartwright}, J.~K., {Farmer}, A.~J.,
  {Padin}, S., {Myers}, S.~T., {Bond}, J.~R., {Contaldi}, C.~R., {Pen}, U.-L.,
  {Prunet}, S., {Pogosyan}, D., {Carlstrom}, J.~E., {Kovac}, J., {Leitch},
  E.~M., {Pryke}, C., {Halverson}, N.~W., {Holzapfel}, W.~L., {Altamirano}, P.,
  {Bronfman}, L., {Casassus}, S., {May}, J., \& {Joy}, M. 2003, \apj, 591, 556


\bibitem[{{Press} {et~al.}(1992)}]{press1992}
Press, W.~H., Teukolsky, S.~A., Vetterling, W.~T.,
Flannery, B.~P., 1992, Numerical Recipes in FORTRAN: The Art of
Scientific Computing, Third Edition, Cambridge University
Press.


\bibitem[{{Reese} {et~al.}(2002){Reese}, {Carlstrom}, {Joy}, {Mohr}, {Grego},
  \& {Holzapfel}}]{Reese2002_SZE}
{Reese}, E.~D., {Carlstrom}, J.~E., {Joy}, M., {Mohr}, J.~J., {Grego}, L., \&
  {Holzapfel}, W.~L. 2002, \apj, 581, 53
 
\bibitem[{{Rephaeli}(1995)}]{1995ARA&A..33..541R} 
{Rephaeli}, Y. 1995, \araa, 33, 541

\bibitem[{{Rines} \& {Geller}(2008)}]{Rines+Geller2008}
{Rines}, K., \& {Geller}, M.~J. 2008, \aj, 135, 1837

\bibitem[{{Sanderson} {et~al.}(2003){Sanderson}, {Ponman}, {Finoguenov},
  {Lloyd-Davies}, \& {Markevitch}}]{Sanderson+2003_fgas}
{Sanderson}, A.~J.~R., {Ponman}, T.~J., {Finoguenov}, A., {Lloyd-Davies},
  E.~J., \& {Markevitch}, M. 2003, \mnras, 340, 989

\bibitem[{{Sasaki}(1996)}]{Sasaki1996}
{Sasaki}, S. 1996, \pasj, 48, L119

\bibitem[{{Saxton} \& {Wu}(2008)}]{Saxton+Wu2008}
{Saxton}, C.~J., \& {Wu}, K. 2008, \mnras in press (arXiv:0809.3795)

\bibitem[{{Schneider} \& {Seitz}(1995)}]{1995A&A...294..411S}
{Schneider}, P., \& {Seitz}, C. 1995, \aap, 294, 411

\bibitem[{{Sereno}(2007)}]{2007MNRAS.380.1207S}
{Sereno}, M. 2007, \mnras, 380, 1207

\bibitem[{{Spergel} {et~al.}(2007){Spergel}, {Bean}, {Dor{\'e}}, {Nolta},
  {Bennett}, {Dunkley}, {Hinshaw}, {Jarosik}, {Komatsu}, {Page}, {Peiris},
  {Verde}, {Halpern}, {Hill}, {Kogut}, {Limon}, {Meyer}, {Odegard}, {Tucker},
  {Weiland}, {Wollack}, \& {Wright}}]{2007ApJS..170..377S}
{Spergel}, D.~N., {Bean}, R., {Dor{\'e}}, O., {Nolta}, M.~R., {Bennett}, C.~L.,
  {Dunkley}, J., {Hinshaw}, G., {Jarosik}, N., {Komatsu}, E., {Page}, L.,
  {Peiris}, H.~V., {Verde}, L., {Halpern}, M., {Hill}, R.~S., {Kogut}, A.,
  {Limon}, M., {Meyer}, S.~S., {Odegard}, N., {Tucker}, G.~S., {Weiland},
  J.~L., {Wollack}, E., \& {Wright}, E.~L. 2007, \apjs, 170, 377

\bibitem[{{Umetsu} \& {Broadhurst}(2008)}]{UB2008}
{Umetsu}, K., \& {Broadhurst}, T. 2008, \apj, 684, 177

\bibitem[{{Umetsu} {et~al.}(2004){Umetsu}, {Chiueh}, {Lin}, {Wu}, \&
  {Tseng}}]{Umetsu2004_MPLA}
{Umetsu}, K., {Chiueh}, T., {Lin}, K.-Y., {Wu}, J.-M., \& {Tseng}, Y.-H. 2004,
  Modern Physics Letters A, 19, 1027

\bibitem[{{Umetsu} {et~al.}(1999){Umetsu}, {Tada}, \&
  {Futamase}}]{1999PThPS.133...53U}
{Umetsu}, K., {Tada}, M., \& {Futamase}, T. 1999, Progress of Theoretical
  Physics Supplement, 133, 53

\bibitem[{{Umetsu} {et~al.}(2007){Umetsu}, {Takada}, \&
  {Broadhurst}}]{2007MPLA...22.2099U}
{Umetsu}, K., {Takada}, M., \& {Broadhurst}, T. 2007, Modern Physics Letters A,
  22, 2099

\bibitem[{{Umetsu} {et~al.}(2005){Umetsu}, {Wu}, {Chiueh}, \&
  {Birkinshaw}}]{2005astro.ph..6065U}
{Umetsu}, K., {Wu}, J.-M., {Chiueh}, T., \& {Birkinshaw}, M. 2005,
	       preprint (arXiv:astro-ph/0607542)

\bibitem[{{Vikhlinin} {et~al.}(2006){Vikhlinin}, {Kravtsov}, {Forman}, {Jones},
  {Markevitch}, {Murray}, \& {Van Speybroeck}}]{2006ApJ...640..691V}
{Vikhlinin}, A., {Kravtsov}, A., {Forman}, W., {Jones}, C., {Markevitch}, M.,
  {Murray}, S.~S., \& {Van Speybroeck}, L. 2006, \apj, 640, 691

\bibitem[{{Watson} {et~al.}(2003){Watson}, {Carreira}, {Cleary}, {Davies},
  {Davis}, {Dickinson}, {Grainge}, {Guti{\'e}rrez}, {Hobson}, {Jones},
  {Kneissl}, {Lasenby}, {Maisinger}, {Pooley}, {Rebolo}, {Rubi{\~n}o-Martin},
  {Rusholme}, {Saunders}, {Savage}, {Scott}, {Slosar}, {Sosa Molina}, {Taylor},
  {Titterington}, {Waldram}, \& {Wilkinson}}]{Watson2003_VSA}
{Watson}, R.~A., {Carreira}, P., {Cleary}, K., {Davies}, R.~D., {Davis}, R.~J.,
  {Dickinson}, C., {Grainge}, K., {Guti{\'e}rrez}, C.~M., {Hobson}, M.~P.,
  {Jones}, M.~E., {Kneissl}, R., {Lasenby}, A., {Maisinger}, K., {Pooley},
  G.~G., {Rebolo}, R., {Rubi{\~n}o-Martin}, J.~A., {Rusholme}, B., {Saunders},
  R.~D.~E., {Savage}, R., {Scott}, P.~F., {Slosar}, A., {Sosa Molina}, P.~J.,
  {Taylor}, A.~C., {Titterington}, D., {Waldram}, E., \& {Wilkinson}, A. 2003,
  \mnras, 341, 1057

\bibitem[{{Worrall} \& {Birkinshaw}(2006)}]{Worrall+Birkinshaw2006}
{Worrall}, D.~M., \& {Birkinshaw}, M. 2006, in Lecture Notes in Physics, Berlin
  Springer Verlag, Vol. 693, Physics of Active Galactic Nuclei at all Scales,
  ed. D.~{Alloin}, 39--+

\bibitem[{{Wright} \& {Brainerd}(2000)}]{2000ApJ...534...34W}
{Wright}, C.~O., \& {Brainerd}, T.~G. 2000, \apj, 534, 34

\bibitem[{{Wu} {et~al.}(2008{\natexlab{a}}){Wu}, {AMiBA 1}, \& {AMiBA
  2}}]{Wu_AMiBA}
{Wu}, J.~H.~P. et al. 2008, \apj, submitted (arXiv:0810.1015)

\bibitem[{{Wu} {et~al.}(2008{\natexlab{b}}){Wu}, {Chiueh}, {Huang}, {Liao},
  {Wang}, {Altimirano}, {Chang}, {Chang}, {Chang}, {Chen}, {Chereau}, {Han},
  {Ho}, {Huang}, {Hwang}, {Jiang}, {Koch}, {Kubo}, {Li}, {Lin}, {Liu},
  {Martin-Cocher}, {Molnar}, {Nishioka}, {Raffin}, {Umetsu}, {Kesteven},
  {Wilson}, {Birkinshaw}, \& {Lancaster}}]{Wu_AMiBA_MPLA}
{Wu}, J.-H.~P., {Chiueh}, T.-H., {Huang}, C.-W., {Liao}, Y.-W., {Wang}, F.-C.,
  {Altimirano}, P., {Chang}, C.-H., {Chang}, S.-H., {Chang}, S.-W., {Chen},
  M.-T., {Chereau}, G., {Han}, C.-C., {Ho}, P.~T.~P., {Huang}, Y.-D., {Hwang},
  Y.-J., {Jiang}, H., {Koch}, P., {Kubo}, D., {Li}, C.-T., {Lin}, K.-Y., {Liu},
  G.-C., {Martin-Cocher}, P., {Molnar}, S., {Nishioka}, H., {Raffin}, P.,
  {Umetsu}, K., {Kesteven}, M., {Wilson}, W., {Birkinshaw}, M., \& {Lancaster},
  K. 2008{\natexlab{b}}, Modern Physics Letters A, 23, 1675

\bibitem[{{Yagi} {et~al.}(2002){Yagi}, {Kashikawa}, {Sekiguchi}, {Doi},
  {Yasuda}, {Shimasaku}, \& {Okamura}}]{2002AJ....123...66Y}
{Yagi}, M., {Kashikawa}, N., {Sekiguchi}, M., {Doi}, M., {Yasuda}, N.,
  {Shimasaku}, K., \& {Okamura}, S. 2002, \aj, 123, 66

\bibitem[{{Yoshikawa} {et~al.}(2000){Yoshikawa}, {Jing}, \&
  {Suto}}]{Yoshikawa+2000}
{Yoshikawa}, K., {Jing}, Y.~P., \& {Suto}, Y. 2000, \apj, 535, 593

\bibitem[{{Zhang} {et~al.}(2002){Zhang}, {Pen}, \&
  {Wang}}]{2002ApJ...577..555Z}
{Zhang}, P., {Pen}, U.-L., \& {Wang}, B. 2002, \apj, 577, 555

\end{thebibliography}




\begin{deluxetable}{ccc|cc|ccc}
 \centering
\tabletypesize{\footnotesize}
\tablecolumns{8}
\tablecaption{
 \label{tab:amiba_target}
Target clusters and AMiBA/X-ray properties
} 
\tablewidth{0pt} 
\tablehead{ 
 \multicolumn{1}{c}{Cluster} &
 \multicolumn{1}{c}{$z$} &
 \multicolumn{1}{c|}{1$\,$arcmin\tablenotemark{a}} &
 \multicolumn{2}{c|}{AMiBA7\tablenotemark{b}} &
 \multicolumn{3}{c}{X-ray\tablenotemark{c}} 
\\
 \colhead{} &
 \colhead{} &
 \colhead{} &
 \multicolumn{1}{|c}{SZE flux} &
 \multicolumn{1}{c}{Image FWHM} &
 \multicolumn{1}{|c}{$T_X$} &
 \multicolumn{1}{c}{$\theta_c$} &
 \multicolumn{1}{c}{Refs}
\\
 \colhead{} &
 \colhead{} &
 \multicolumn{1}{c}{(${\rm kpc}h^{-1}$)} &
 \multicolumn{1}{|c}{(${\rm mJy}$)} &
 \multicolumn{1}{c|}{(arcmin)} &
  \multicolumn{1}{c}{(keV)} &
 \multicolumn{1}{c}{(arcmin)} &
 \colhead{} 
} 
\startdata  
  A1689  & 0.183 & 129.6 & $-168\pm 28$ & 5.7 & $9.66^{+0.22}_{-0.20}$ &
 $0.44\pm 0.01$ & 3\\
 A2142  & 0.091 & 71.4 & $-316\pm 23$ & 9.0 & $9.7\pm 1.0$ & $3.14\pm
 0.22$ & 1, 4, 5\\
 A2261  & 0.224 & 151.8 & $-90\pm 17$ & 5.8 & $8.82^{+0.37}_{-0.32}$ &
 $0.26\pm 0.02 $ & 3\\
 A2390  & 0.228 & 153.3 & $-158\pm 24$ & 8.0 &$10.1\pm 1.1$ &
 $0.47 \pm 0.05$ & 2 

\enddata
\tablecomments{Uncertainties are $68\%$ confidence.}
\tablenotetext{a}{Physical scale in kpc$\,h^{-1}$ units corresponding to
 $1\arcmin$ at the cluster redshift.}
\tablenotetext{b}{SZE properties from AMiBA7 at 94$\,$GHz: cluster peak
 SZE flux (mJy) and angular size ($\arcmin$) in FWHM measured from the cleaned
 image \citep[][]{Wu_AMiBA}.}
\tablenotetext{c}{Published X-ray properties: X-ray temperature (keV),
 X-ray core radius (kpc$\,h^{-1}$), and references. For A2142, $T_X$ and
 $\theta_c$ are  
 taken from Ref.~[1], and Refs.~[4,5], respectively.
 For A2390 a $10\%$ error is assumed for $(T_X,\beta)$, for which no
 error estimate was presented in the original reference.}
\tablerefs{ 
 [1] \cite{Markevitch+1998};
 [2] \cite{Boehringer+1998_A2390};
 [3] \cite{Reese2002_SZE}; 
 [4] \cite{Sanderson+2003_fgas};
 [5] \cite{Lancaster+05_VSA}. 
 }
\end{deluxetable}



\begin{deluxetable}{cccccccc}
\tablecolumns{8}
\tablecaption{
 \label{tab:lensdata}
Subaru weak lensing data and background galaxy sample
} 
\tablewidth{0pt} 
\tablehead{ 
 \multicolumn{1}{c}{Cluster} &
 \multicolumn{1}{c}{Filters} &
 \multicolumn{1}{c}{Seeing\tablenotemark{a}} &
 \multicolumn{1}{c}{$n_g$\tablenotemark{b}} &
 \multicolumn{1}{c}{B/R\tablenotemark{c}} &
 \multicolumn{1}{c}{$\langle D_{ds}/D_s\rangle$\tablenotemark{d}} &
 \multicolumn{1}{c}{$z_{s,D}$\tablenotemark{e}} &
 \multicolumn{1}{c}{$\sigma_\kappa$\tablenotemark{f}}
\\
 \colhead{} &
 \colhead{} &
 \multicolumn{1}{c}{(arcsec)} &
 \multicolumn{1}{c}{(${\rm arcmin}^{-2}$)} &
 \colhead{} &
 \colhead{} &
 \colhead{} &
 \colhead{}
}
\startdata  
 A1689  & $Vi'$ & 0.88 & 8.8 & 0 & $0.70\pm 0.02$ & 
   $0.70^{+0.06}_{-0.05}$ & 0.029 \\
 A2142  & $g'R_{\rm c}$ & 0.55 & 30.4 & 2.1 & $0.88\pm 0.04$ & 
   $0.95^{+0.79}_{-0.30}$ & 0.021\\
 A2261  & $VR_{\rm c}$ & 0.65 & 13.8 & 1.5 &  $0.72\pm 0.04$ & 
 $ 0.98^{+0.24}_{-0.16}$ & 0.032\\
 A2390  & $VR_{\rm c}$ & 0.70 & 20.7 &2.1 &   $0.72\pm 0.04$ &
   $1.00^{+0.25}_{-0.16}$ & 0.026 
\enddata
\tablenotetext{a}{Seeing FWHM in the final co-added image in the redder
 band.}
\tablenotetext{b}{Surface number density of blue+red galaxies.}
\tablenotetext{c}{Fraction of blue to red galaxies in the 
blue+red background sample.}
\tablenotetext{d}{Distance ratio averaged over the redshift distribution
 of the blue+red sample.}
\tablenotetext{e}{Effective source redshift (see eq. [\ref{eq:zD}])
corresponding to the mean depth $\langle D_{ds}/D_s\rangle$.}
\tablenotetext{f}{RMS noise level in the reconstructed $\kappa$ map.}
\end{deluxetable}


\begin{deluxetable}{c|ccc|cccc|cccc}
\tabletypesize{\scriptsize}
\tablecolumns{12}
\tablecaption{
\label{tab:model}
Summary of best-fit mass models from Subaru distortion data
} 
\tablewidth{0pt} 
\tablehead{ 
 \multicolumn{1}{c}{Cluster} &
 \multicolumn{7}{|c}{Tangential reduced shear, $g_+$} &
 \multicolumn{4}{|c}{Lensing convergence, $\kappa$} \\
 \colhead{} &
 \multicolumn{3}{|c}{SIS}  &
 \multicolumn{4}{c}{NFW}   & 
 \multicolumn{4}{|c}{NFW}  \\
 \colhead{} &
 \multicolumn{1}{|c}{$\sigma_v$} & 
 \multicolumn{1}{c}{$\chi^2/{\rm dof}$} &
 \multicolumn{1}{c}{$\theta_{\rm E}$} &
 \multicolumn{1}{|c}{$M_{\rm vir}$}  & 
 \multicolumn{1}{c}{$c_{\rm vir}$}  & 
 \multicolumn{1}{c}{$\chi^2/{\rm dof}$} & 
 \multicolumn{1}{c|}{$\theta_{\rm E}$} &
 \multicolumn{1}{|c}{$M_{\rm vir}$}  & 
 \multicolumn{1}{c}{$c_{\rm vir}$}  & 
 \multicolumn{1}{c}{$\chi^2/{\rm dof}$} &  
 \multicolumn{1}{c}{$\theta_{\rm E}$} 
\\
 \colhead{} &
 \multicolumn{1}{|c}{$({\rm km} s^{-1})$} & 
 \multicolumn{1}{c}{} &
 \multicolumn{1}{c}{($\arcsec$)} &
 \multicolumn{1}{|c}{$(10^{15}M_\odot/h)$}  & 
 \multicolumn{1}{c}{}  & 
 \multicolumn{1}{c}{} & 
 \multicolumn{1}{c}{($\arcsec$)} &
 \multicolumn{1}{|c}{$(10^{15}M_\odot/h)$}  & 
 \multicolumn{1}{c}{}  & 
 \multicolumn{1}{c}{}  &
 \multicolumn{1}{c}{($\arcsec$)} 
}
\startdata 
 A1689   & $1403\pm 41$& 11/9 & $47\pm 3$&
 $1.09^{+0.18}_{-0.16}$&$15.6^{+4.8}_{-3.3}$ & 7.3/8 & $47^{+15}_{-14}$&
 $1.05^{+0.18}_{-0.15}$& $15.8^{+14.2}_{-8.0}$&  5.3/8 &
 $46^{+26}_{-31}$\\ 
 A2142   & $970\pm 27$& 39/8 & $25\pm 1$ & 
 $1.07^{+0.22}_{-0.16}$& $5.6^{+0.9}_{-0.8}$ & 2.1/7 & $1.2^{+2.9}_{-0.9}$& 
 $1.06^{+0.19}_{-0.16}$& $4.9^{+1.2}_{-1.0}$& 20/10 &
 $0.5^{+2.3}_{-0.4}$\\ 
 A2261   & $1276\pm 43$& 8.7/8 & $37\pm 3$&
 $1.35^{+0.26}_{-0.22}$&$6.4^{+1.9}_{-1.4}$ &7.7/7  &  $20^{+16}_{-11}$&
 $1.26^{+0.20}_{-0.17}$&$10.2^{+7.1}_{-3.5}$ & 9.8/8 & $37^{+25}_{-19}$\\
 A2390   & $1139\pm 38$& 3.8/8& $30\pm 2$&
 $0.90^{+0.15}_{-0.14}$& $6.9^{+2.3}_{-1.5}$& 3.8/7 & $15^{+13}_{-8}$&
 $0.92^{+0.15}_{-0.12}$& $7.3^{+6.9}_{-2.9}$& 8.1/8 & $17^{+26}_{-14}$
\enddata
\tablecomments{A flat prior of $c_{\rm vir}\le 30$ is assumed for the
 halo concentration of the NFW model. The Einstein radius $\theta_{\rm E}$ is
 calculated for a background source at $z_s=1.5$, corresponding roughly
 to the mean depth of blue+red background galaxies.
} 
\end{deluxetable}

 
\begin{deluxetable}{c|cc|c|cccc} 
\tabletypesize{\scriptsize}
\tablecolumns{8}
\tablecaption{ 
 \label{tab:cluster}
Cluster mass models for gas mass fraction measurements
} 
\tablewidth{0pt} 
\tablehead{ 
 \multicolumn{1}{c|}{Cluster} &
 \multicolumn{2}{c|}{NFW model} &
 \multicolumn{1}{c|}{SIS model} &
 \multicolumn{1}{c}{$r_{2500}$} &
 \multicolumn{1}{c}{$r_{500}$} &
 \multicolumn{1}{c}{$r_{200}$} &
 \multicolumn{1}{c}{$r_{\rm vir}$} 
\\
 \colhead{} &
 \multicolumn{1}{|c}{$M_{\rm vir}$} &
 \multicolumn{1}{c|}{$c_{\rm vir}$} &
 \multicolumn{1}{c|}{$\sigma_v$} &
 \colhead{} &
 \colhead{} &
 \colhead{} 
\\
 \colhead{} &
 \multicolumn{1}{|c}{($10^{15}M_\odot h^{-1}$)} &
 \multicolumn{1}{c|}{} &
 \multicolumn{1}{c|}{(${\rm km s}^{-1}$)} &
 \multicolumn{1}{c}{(${\rm Mpc}h^{-1}$)} & 
 \multicolumn{1}{c}{(${\rm Mpc}h^{-1}$)} & 
 \multicolumn{1}{c}{(${\rm Mpc}h^{-1}$)} & 
 \multicolumn{1}{c}{(${\rm Mpc}h^{-1}$)} 
}
\startdata 
 A1689\tablenotemark{a}  & 
 $1.55^{+0.13}_{-0.12}$ &
 $12.3^{+0.9}_{-0.8}$ & 
 $1403\pm 41$ &
 $0.57\pm 0.01$ &
 $1.16\pm 0.02$ &
 $1.70\pm 0.04$ &
 $2.13\pm 0.05$
\\
 A2142  & 
 $1.07^{+0.22}_{-0.16}$ &
 $5.6^{+0.9}_{-0.8}$ &  
 $970\pm 27$ &
 $0.43\pm 0.02$ &
 $0.99\pm 0.04$ &
 $1.51\pm 0.07$ & 
 $1.98\pm 0.10$
\\
 A2261\tablenotemark{b}  & 
 $1.25^{+0.17}_{-0.16}$ &
 $11.1^{+2.2}_{-1.9}$ & 
 $1276\pm 43$ &
 $0.52\pm 0.02$ &
 $1.06\pm 0.04$ &
 $1.56\pm 0.06$ &
 $1.94\pm 0.07$
\\
 A2390  & 
 $0.90^{+0.15}_{-0.14}$ & 
 $6.9^{+2.3}_{-1.5}$ &
 $1139\pm 38$  & 
 $0.42\pm 0.03$ &
 $0.92\pm 0.04$ &
 $1.38\pm 0.06$ &
 $1.73\pm 0.07$
\enddata
\tablenotetext{a}{The NFW model is constrained by a joint fit to
ACS strong lensing and Subaru distortion+magnification data, presented in Umetsu \& Broadhurst (2008), but with our improved color selection of the red background sample for Subaru distortion measurements (\S\ref{subsubsec:kappa}).}
\tablenotetext{b}{The NFW model is constrained by a joint fit to
the inner Einstein-radius constraint and the outer
Subaru $\kappa$ profile (\S\ref{subsubsec:kappa}).
  }
\end{deluxetable}


\begin{deluxetable}{c|cc|cc}
\tablecolumns{7}
\tablecaption{
 \label{tab:amiba}
AMiBA visibility analysis
} 
\tablewidth{0pt}  
\tablehead{ 
 \multicolumn{1}{c}{Cluster} & 
 \multicolumn{2}{|c}{KS01} &
 \multicolumn{2}{|c}{isothermal $\beta (=2/3$)} \\
 \colhead{} &
 \multicolumn{1}{|c}{$y_0$} & 
 \multicolumn{1}{c}{$Y(3')$} &
 \multicolumn{1}{|c}{$y_0$} &
 \multicolumn{1}{c}{$Y(3')$} \\
 \colhead{} &
 \multicolumn{1}{|c}{$(10^{-4})$} &
 \multicolumn{1}{c}{$(10^{-10})$} &
 \multicolumn{1}{|c}{$(10^{-4})$} &
 \multicolumn{1}{c}{$(10^{-10})$}  
}
\startdata 
 A1689 &
 $4.15\pm 1.00$ &  
 $2.5^{+0.6}_{-0.6}$ &
 $4.31\pm 1.10$& 
 $2.6^{+0.6}_{-0.6}$  
\\
 A2142  & 
 $2.29\pm 0.28$ &
 $3.5^{+0.5}_{-0.5}$ &
 $2.00\pm 0.25$ &
 $4.0^{+0.5}_{-0.5}$ 
 \\
 A2261  & 
 $3.00\pm 0.84$ & 
 $1.5^{+0.5}_{-0.4}$ &
 $4.25\pm 1.22$ &
 $1.6^{+0.5}_{-0.4}$ 
 \\
 A2390  & 
  $2.87\pm 0.61$ &
  $1.9^{+0.6}_{-0.5}$ &
  $3.40 \pm 0.72$  &
  $2.1^{+0.8}_{-0.5}$
\enddata
\tablecomments{
The effects of radio point source contamination in the thermal SZE  have
 been 
 corrected for \citep[see][]{Liu_AMiBA}. The relativistic correction to
 the SZE is also taken into account.
}
\end{deluxetable}


\begin{deluxetable}{c|cccccc|cccccc}
\tabletypesize{\tiny}
\tablecolumns{11}
\tablecaption{
 \label{tab:fg}
Cluster gas properties derived from the AMiBA/Subaru data
} 
\tablewidth{0pt} 
\tablehead{ 
 \multicolumn{1}{c}{Cluster} &
 \multicolumn{6}{|c}{KS01 + NFW} &
 \multicolumn{6}{|c}{isothermal $\beta (=2/3)$ + SIS}\\
 \colhead{} &
 \multicolumn{1}{|c}{$M_{\rm gas,2500}$} &
 \multicolumn{1}{c}{$M_{\rm gas,500}$} &
 \multicolumn{1}{c}{$M_{\rm gas,200}$} &
  \multicolumn{1}{c}{$f_{\rm gas,2500}$} &
 \multicolumn{1}{c}{$f_{\rm gas,500}$} &
 \multicolumn{1}{c}{$f_{\rm gas,200}$} &
 \multicolumn{1}{|c}{$M_{\rm gas,2500}$} &
 \multicolumn{1}{c}{$M_{\rm gas,500}$} & 
 \multicolumn{1}{c}{$M_{\rm gas,200}$} &
  \multicolumn{1}{c}{$f_{\rm gas,2500}$} &
 \multicolumn{1}{c}{$f_{\rm gas,500}$} &
 \multicolumn{1}{c}{$f_{\rm gas,200}$} 
\\
 \colhead{} &
 \multicolumn{3}{|c}{($10^{13}M_\odot h^{-2}$)} &
 \colhead{} & 
 \colhead{} &
 \colhead{} &
 \multicolumn{3}{|c}{($10^{13}M_\odot h^{-2}$)} &
 \colhead{} &
 \colhead{} &
 \colhead{} 
}
\startdata
A1689 &
$ 4.4^{+ 1.1}_{- 2.2}$ &
$ 8.8^{+ 2.3}_{- 2.2}$ &
$11.5^{+ 3.0}_{- 3.0}$ &
$0.098^{+0.025}_{-0.026}$ &
$0.115^{+0.029}_{-0.029}$ &
$0.119^{+0.031}_{-0.030}$ &
$ 3.5^{+ 0.9}_{- 0.8}$ &
$ 7.8^{+ 2.0}_{- 1.8}$ &
$11.8^{+ 3.0}_{- 2.7}$ &
$0.100^{+0.024}_{-0.023}$ &
$0.108^{+0.026}_{-0.025}$ &
$0.111^{+0.027}_{-0.026}$ \\
A2142 &
$ 2.3^{+ 0.4}_{- 1.3}$ &
$ 7.2^{+ 1.5}_{- 1.3}$ &
$11.2^{+ 2.6}_{- 2.2}$ &
$0.128^{+0.036}_{-0.025}$ &
$0.169^{+0.046}_{-0.034}$ &
$0.183^{+0.049}_{-0.037}$ &
 -- &
 -- &
 -- &
 -- &
 -- &
 -- \\ 
A2261 &
$ 3.0^{+ 0.9}_{- 2.1}$ &
$ 6.3^{+ 1.9}_{- 2.1}$ &
$ 8.4^{+ 2.7}_{- 2.8}$ &
$0.087^{+0.030}_{-0.028}$ &
$0.103^{+0.036}_{-0.033}$ &
$0.108^{+0.040}_{-0.035}$ &
$ 2.5^{+ 0.7}_{- 0.7}$ &
$ 5.4^{+ 1.5}_{- 1.5}$ &
$ 8.1^{+ 2.3}_{- 2.3}$ &
$0.097^{+0.030}_{-0.028}$ &
$0.103^{+0.031}_{-0.030}$ &
$0.105^{+0.032}_{-0.030}$ \\
A2390 &
$ 2.3^{+ 0.7}_{- 1.8}$ &
$ 6.1^{+ 2.4}_{- 1.8}$ &
$ 8.8^{+ 4.0}_{- 2.7}$ &
$0.122^{+0.059}_{-0.037}$ &
$0.153^{+0.075}_{-0.049}$ &
$0.164^{+0.084}_{-0.053}$ &
$ 2.2^{+ 0.6}_{- 0.6}$ &
$ 5.6^{+ 1.5}_{- 1.5}$ &
$ 8.8^{+ 2.4}_{- 2.4}$ &
$0.125^{+0.035}_{-0.034}$ &
$0.145^{+0.041}_{-0.041}$ &
$0.151^{+0.042}_{-0.043}$
\enddata
\tablecomments{
 The derived gas fractions $f_{\rm gas}$
 scale with the Hubble parameter $h$as $f_{\rm  gas}\propto h^{-1}$ 
 ($h=0.7$ adopted here).
 Confidence intervals are quoted at the $1\sigma$ ($68\%$) level.
 Here we exclude the results from the isothermal
 model for A2142 which overpredicts $f_{\rm gas}
 $at all relevant radii ($r>r_{2500}$)
 compared with the
 cosmic baryon fraction, $f_b=\Omega_b/\Omega_m=0.171\pm 0.009$.
}
\end{deluxetable} 

\clearpage





\clearpage

\appendix

\section{One-Dimensional Mass Reconstruction from Distortion Data}
  
Following the method developed by \citet{UB2008},
we derive an expression for the
discrete convergence profile  
using a non-linear extension of 
weak lensing aperture densitometry.

\subsection{Non-Linear Aperture Mass Densitometry}
\label{app:1dmass}

For a shear-based estimation of the cluster mass profile
we use a variant of weak lensing aperture densitometry, or
the so-called $\zeta$-statistic 
\citep{1994ApJ...437...56F,2000ApJ...539..540C}
of the form:
\begin{eqnarray}
\label{eq:zeta}
\zeta_{\rm c}(\theta)
&\equiv &
2\int_{\theta}^{\theta_{\rm inn}}\!d\ln\theta' 
\gamma_+(\theta')\nonumber\\
&&+ \frac{2}{1-(\theta_{\rm inn}/\theta_{\rm out})^2}\int_{\theta_{\rm
inn}}^{\theta_{\rm out}}\! d\ln\theta'  
\gamma_+(\theta')
\nonumber\\
&=& 
\bar{\kappa}(\theta) - \bar{\kappa}
(\theta_{\rm inn}<\vartheta <\theta_{\rm out}),
\end{eqnarray}
where 
$\kappa(\theta)$ is the azimuthal average of 
the convergence field $\kappa(\btheta)$ at radius $\theta$,
$\bar{\kappa}(\theta)$ is the average convergence interior to 
 radius $\theta$,
 $\theta_{\rm inn}$ and $\theta_{\rm out}$ are the inner and
outer radii of the annular background region in which the mean
background contribution, 
$\bar{\kappa}_b\equiv
\bar{\kappa}(\theta_{\rm inn}<\vartheta <\theta_{\rm out})$,
is defined;
$\gamma_+(\theta) = \bar{\kappa}(\theta)-\kappa(\theta)$ 
is an azimuthal average of the tangential
component of the gravitational shear at radius $\theta$
\citep{1994ApJ...437...56F},
which is observable in the weak lensing limit: $\gamma_+(\theta)\approx
\langle g_+(\theta)\rangle$.
This cumulative mass estimator
subtracts  from the mean convergence 
$\bar{\kappa}( \theta)$
a constant 
$ \bar{\kappa}_b$
for all apertures $\theta$ in the measurements, 
thus removing any DC component in the control
region $\theta = [\theta_{\rm inn}, \theta_{\rm out}]$.
Note that the $\bar{\kappa}_b$ is 
a non-observable free parameter. 
This degree of freedom can be used to
fix the outer boundary condition,
and hence to derive a convergence profile $\kappa(\theta)$.

In the non-linear regime $\gamma_+(\theta)$ is not a direct observable.
Therefore, non-linear corrections need to be taken into account 
in the mass reconstruction process \citep{UB2008}.
In the subcritical regime (i.e., outside the critical curves),
$\gamma_+(\theta)$ can be expressed in terms of 
the averaged tangential reduced shear as
$\langle g_+(\theta) \rangle \approx
\gamma_+(\theta)/[1-\kappa(\theta)]$
assuming a quasi-circular symmetry in the projected mass distribution
\citep{BTU+05,2007MPLA...22.2099U}.
This non-linear equation (\ref{eq:zeta})
for $\zeta_{\rm c}(\theta)$ can be solved by an iterative procedure:
Since the weak lensing limit ($\kappa,|\gamma|,|g|\ll 1$) holds in the
background region $\theta_{\rm inn}\le \theta \le \theta_{\rm
max}$, we have the following iterative equation for $\zeta_{\rm
c}(\theta)$: 
\begin{eqnarray}
\zeta_{\rm c}^{(k+1)}(\theta)
&\approx&
2\int_{\theta}^{\theta_{\rm inn}}\!d\ln\theta' 
\langle g_+(\theta)\rangle [1-\kappa^{(k)}(\theta)] \nonumber\\
&+& \frac{2}{1-(\theta_{\rm inn}/\theta_{\rm out})^2}\int_{\theta_{\rm
inn}}^{\theta_{\rm out}}\! d\ln\theta'  
\langle g_+(\theta')\rangle,
\end{eqnarray} 
where $\zeta_{\rm c}^{(k+1)}$ 
represents the 
aperture densitometry in the $(k+1)$th 
step of the iteration
$(k=0,1,2,...,N_{\rm iter})$; 
the $\kappa^{(k+1)}$ is
calculated from $\zeta_{\rm c}^{(k+1)}$ using equation (\ref{eq:zeta2kappa}).
This iteration is preformed by
starting with $\kappa^{(0)}=0$ for all radial bins, and repeated 
until convergence is reached at all radial bins. For a fractional
tolerance of $1\times 10^{-5}$, this iteration procedure converges
within $N_{\rm iter} \sim 10$ iterations.
We compute errors for $\zeta_{\rm c}$ and $\kappa$ 
with the linear approximation.

\subsection{Discretized Estimator for the Lensing Convergence}
\label{app:kappad}        
  
In the continuous limit,
the averaged convergence $\bar\kappa(\theta)$
and the convergence $\kappa(\theta)$ 
are related by
\begin{eqnarray}
\bar{\kappa}(\theta)&=&\frac{2}{\theta^2}\int_0^{\theta}
\!d\ln\theta'\theta'^2\kappa(\theta'),\\
\kappa(\theta) &=&
 \frac{1}{2\theta^2}\frac{d(\theta^2\bar{\kappa})}{d\ln\theta}.
\end{eqnarray}
For a given set of annular radii $\theta_m$
$(m=1,2,...,N)$,
discretized estimators can be written in the following way:
\begin{eqnarray}
\label{eq:avkappa_d}
\bar\kappa_m &\equiv&
\bar{\kappa}(\theta_m)=
\frac{2}{\theta_m^2}\sum_{l=1}^{m-1}
\Delta\ln\theta_l
\bar\theta_l^2
\kappa(\bar\theta_l),\\
\label{eq:kappa_d}
\kappa_l&\equiv&
\kappa(\bar\theta_l) =
\alpha^l_2\bar\kappa_{l+1}-\alpha^l_1\bar\kappa_l
\ \ \ \ \ (l=1,2,...,N-1),
\end{eqnarray}
where 
\begin{equation} 
\alpha_1^l = \frac{1}{2\Delta\ln\theta_l} 
\left( 
  \frac{\theta_{l}}{ \overline{\theta}_l }
\right)^2, \, \,  
\alpha_2^l = \frac{1}{2\Delta\ln\theta_l} 
\left(\frac{\theta_{l+1}}{ \overline{\theta}_l }\right)^2,
\end{equation}
with
$\Delta\ln\theta_l \equiv (\theta_{l+1}-\theta_l)/\bar\theta_l$
and $\bar\theta_l$
being the area-weighted center of the $l$th
annulus defined by $\theta_l$ and $\theta_{l+1}$;
in the continuous limit, we have
\begin{eqnarray}
\bar\theta_l
&\equiv& 
2\int_{\theta_l}^{\theta_{l+1}}\!d\theta'\theta'^2/
(\theta_{l+1}^2-\theta_{l}^2)\nonumber\\ 
&=&
\frac{2}{3}
\frac{\theta_{l}^2+\theta_{l+1}^2+\theta_{l}\theta_{l+1}}
{ \theta_{l}+\theta_{l+1} }. 
\end{eqnarray} 
 
The technique of the aperture densitometry
allows us
to measure the azimuthally averaged convergence $\bar\kappa(\theta)$
up to an additive constant $\bar{\kappa}_b$, corresponding to 
the mean convergence in the outer background annulus
with inner and outer radii of $\theta_{\rm inn}$ and $\theta_{\rm out}$,
respectively:
\begin{equation}
\label{eq:z2k}
\bar\kappa(\theta)=
\zeta_{\rm c}(\theta)
+\bar\kappa_b.
\end{equation}
Substituting equation (\ref{eq:z2k}) into equation (\ref{eq:kappa_d})
yields the desired expression 
as
\begin{equation}
\label{eq:zeta2kappa}
\kappa(\theta_l)=\alpha^{l}_2\zeta_{\rm c}(\theta_{l+1})
-\alpha^l_1\zeta_{\rm c}(\theta_l)
+ (\alpha_2^l-\alpha_1^l)\bar\kappa_b.
\end{equation}
Finally,
the error covariance matrix of $\kappa_l$ is expressed as
\begin{eqnarray}
C_{kl} \equiv \langle \delta\kappa_k \delta\kappa_l \rangle
&=& 
\alpha_2^k \alpha_2^l C^{\zeta}_{k+1,l+1}
+  
\alpha_1^k \alpha_1^l C^{\zeta}_{k,l}\nonumber\\
&-&
\alpha_1^k \alpha_2^l C^{\zeta}_{k,l+1}
-
\alpha_2^k \alpha_1^l C^{\zeta}_{k+1,l},
\end{eqnarray}
where $C^{\zeta}_{kl}\equiv \langle \delta\zeta_k\delta\zeta_l\rangle$
is the bin-to-bin error covariance matrix of the aperture densitometry
measurements which is calculated by propagating the rms errors
$\sigma_+(\theta_l)$ for the tangential shear measurement.


\end{document}